\documentclass[longauth]{aa} 
\usepackage[dvipsnames,svgnames]{xcolor}
\usepackage{graphicx}
\usepackage{multirow}
\usepackage{adjustbox}
\usepackage{amsmath}
\usepackage{amssymb}
\usepackage{rotating}
\usepackage{xspace}
\usepackage{multicol}
\bibpunct{(}{)}{;}{a}{}{,} 
\usepackage[colorlinks=true,allcolors=Blue]{hyperref}

\usepackage{txfonts}
\newcommand{\kms}{\ensuremath{\mathrm{km}\,\mathrm{s}^{-1}}}
\newcommand*{\Mjup}{\ensuremath{M_{\textrm{Jup}}}\xspace}
\newcommand*{\mjup}{\ensuremath{M_{\textrm {Jup}}}\xspace}

\newcommand*{\Msun}{\ensuremath{M_\odot}}
\newcommand*{\msun}{\ensuremath{M_\odot}}

\newcommand{\bpic}{$\beta$\,Pictoris\,}
\newcommand{\mds}{$\mu ^2$~Sco\,}
\newcommand{\bcen}{b Centauri\,}

\newcommand*{\teff}{\ensuremath{T_{\mathrm{eff}}}}
\newcommand{\degree}{$^{\circ}$}

\def\DvtwoD{\ensuremath{v_{\mathrm{2D}}}\xspace}
\def\DvthreeD{\ensuremath{v_{\mathrm{3D}}}\xspace}
\def\pcapt{\ensuremath{p_{\mathrm{capt}}\xspace}}
\def\rmin{\ensuremath{r_{\mathrm{min}}\xspace}}

\defcitealias{parker22}{PD22}
\defcitealias{Janson.2021_beast}{J21}
\defcitealias{alecian13}{A13}

\begin{document}
   \title{Giant planets population around B stars from the first part of the BEAST survey\thanks{Based on data obtained with the ESO/VLT SPHERE instrument under programs 1101.C-0258(A/B/C/D).}}
    \author{P.~Delorme\inst{\ref{IPAG}}
     \and A.~Chomez\inst{\ref{LESIA},\ref{IPAG}} 
     \and V.~Squicciarini\inst{\ref{LESIA},\ref{INAF}}
     \and M.~Janson \inst{\ref{IfA}}
     \and O.~Flasseur\inst{\ref{CRAL}}
     \and O.~Schib\inst{\ref{bern}}
     \and R.~Gratton \inst{\ref{INAF}} 
     \and A.-M.~Lagrange\inst{\ref{LESIA},\ref{IPAG}} 
     \and M.~Langlois\inst{\ref{CRAL}} 
     \and L.~Mayer \inst{\ref{UZH}}
     \and R.~Helled \inst{\ref{UZH}}
     \and S.~Reffert \inst{\ref{Heid}}
     \and F.~Kiefer\inst{\ref{LESIA}}
     \and B.~Biller \inst{\ref{Edinburg}}
     \and G.~Chauvin  \inst{\ref{OCA},\ref{IPAG}} 
     \and C.~Fontanive \inst{\ref{UdeM}}
     \and Th.~Henning \inst{\ref{mpia}}
     \and M.~Kenworthy \inst{\ref{Leiden}}  
     \and G.-D.~Marleau \inst{\ref{mpia},\ref{duisburg},\ref{tuebingen},\ref{bern}}
     \and D.~Mesa\inst{\ref{INAF}}
     \and M.~R.~Meyer \inst{\ref{UM}}
     \and C.~Mordasini\inst{\ref{bern}}
     \and S.~C.~Ringqvist \inst{\ref{IfA}}
     \and M.~Samland \inst{\ref{IfA}}
     \and A.~Vigan  \inst{\ref{LAM}}
    \and G.~Viswanath \inst{\ref{IfA}}  
}

\institute{
\label{IPAG}
Univ. Grenoble Alpes, CNRS-INSU, Institut de Planetologie et d'Astrophysique de Grenoble (IPAG) UMR 5274, Grenoble, F-38041, France;\\
\email{philippe.delorme@univ-grenoble-alpes.fr}
\and
\label{LESIA}
LESIA, Observatoire de Paris, Universit\'{e} PSL, CNRS, 5 Place Jules Janssen, 92190 Meudon, France 
\and
\label{INAF}
INAF – Osservatorio Astronomico di Padova; Vicolo
dell’Osservatorio 5, I-35122 Padova, Italy
\and
\label{IfA}
Institutionen för astronomi, Stockholms universitet; AlbaNova universitetscentrum, SE-106 91 Stockholm, Sweden
\and
\label{CRAL}
Centre de Recherche Astrophysique de Lyon (CRAL) UMR 5574, CNRS, Univ. de Lyon, Univ. Claude Bernard Lyon 1, ENS de Lyon, F-69230 Saint-Genis-Laval, France
\and
\label{bern}
Physikalisches Institut, Universit\"{a}t Bern, Gesellschaftsstr.~6, 3012 Bern, Switzerland
\and
\label{OCA}
Université Côte d’Azur, OCA, CNRS, Lagrange, France
\and
\label{PIXYL}
Pixyl S.A. La Tronche, France
\and
\label{ETH_Zurich}
ETH Zurich, Institute for Particle Physics and Astrophysics, Wolfgang-Pauli-Strasse 27, CH-8093 Zurich, Switzerland 
\and
\label{mpia}
Max-Planck-Institut f\"ur Astronomie, K\"onigstuhl 17, 69117 Heidelberg, Germany
\and
\label{duisburg}
Fakult\"at f\"ur Physik, Universit\"at Duisburg-Essen, Lotharstra\ss{}e 1, 47057 Duisburg, Germany
\and
\label{tuebingen}
Instit\"ut f\"ur Astronomie und Astrophysik, Universit\"at T\"ubingen, Auf der Morgenstelle 10, 72076 T\"ubingen, Germany
\and
\label{UM}
Department of Astronomy, University of Michigan; 1085 S. University Ave, Ann Arbor MI 48109, USA
\and
\label{Heid}
Landessternwarte, Zentrum für Astronomie der Universität Heidelberg, Königstuhl 12, 69117 Heidelberg, Germany
\and
\label{OSUG}
Université Grenoble Alpes, CNRS, Observatoire des Sciences de l’Univers de Grenoble (OSUG), Grenoble, France
\and
\label{UZH}
University of Zurich, Department of Astrophysics, Winterthurerstr. 190
CH-8057, Zurich, Switzerland
\and
\label{UdeM}
Trottier Institute for Research on Exoplanets, Université de Montréal, Montréal, H3C 3J7, Québec, Canada
\and
\label{Leiden}
Leiden Observatory, Leiden University, Leiden, The Netherlands
\and
\label{Edinburg}
Scottish Universities Physics Alliance, Institute for Astronomy, University of Edinburgh, Royal Observatory, Blackford Hill, Edinburgh EH9 3HJ, UK
\and
\label{LAM}
Aix-Marseille Univ., CNRS, CNES, LAM, Marseille, France
}

   \date{\today}

   
   \abstract
   {Exoplanets form from circumstellar protoplanetary discs whose fundamental properties (notably their extent, composition, mass, temperature and lifetime) depend on the host star properties, such as their mass and luminosity. B-stars are among the most massive stars and their protoplanetary discs test extreme conditions for exoplanet formation.
   }
   {This paper investigates the frequency of giant planet companions around young B-stars (median age of 16 Myr) in the Scorpius-Centaurus association, the closest association containing a large population of B-stars.
   }
   {We systematically search for massive exoplanets with the high-contrast direct imaging instrument SPHERE using the data from the BEAST survey, that targets an homogeneous sample of young B-stars from the wide Sco-Cen association. We derive accurate detection limits in case of non-detections.
   }
   {We found evidence in previous papers for two substellar companions around 42 stars. The masses of these companions are straddling the $\sim$13 Jupiter mass deuterium burning limit but their mass ratio with respect to their host star is close to that of Jupiter. We derive a frequency of such massive planetary mass companions around B stars of $11_{-5}^{+7}$\%, accounting for the survey sensitivity.
   }
   {The discoveries of substellar companions \bcen b and \mds B happened after only few stars in the survey had been observed, raising the possibility that massive Jovian planets might be common around B-stars. However our statistical analysis show that the occurrence rate of such planets is similar around B-stars and around solar-type stars of similar age, while B-star companions exhibit low mass ratios and larger semi-major axis.
   }


\titlerunning{Giant planets population around B stars from BEAST survey}

\maketitle
\nolinenumbers
\section{Introduction}

Though exoplanets have been found around a wide variety of stellar hosts, notably around very low mass stars (from Earth-mass planets \citep[e.g. TRAPPIST\,b-f]{Gillon.2016} to giants, \citep[e.g. 2M1207b]{Chauvin.2004}) their presence around stars more massive than a few Solar masses has not been thoroughly investigated as of now. Their intrinsic rarity, large radii, masses, strong activity and scarce emission or absorption line density makes them  unsuited for radial velocity surveys and poor targets for transit surveys, while their intense brightness offers unfavourable contrast for direct imaging. \citet{Reffert.2015,Wolthoff.2022} have shown from radial velocity surveys of GK giants (which are the evolved counterparts of young A and late B stars) that short separation giant exoplanets are more frequent with increasing stellar masses up to around 2 \msun~and then decrease. The prospects on the theoretical side are not more promising for giant exoplanet formation around massive stars, with the strong UV flux of B-type stars quickly photo-evaporating their protoplanetary discs and thus making such massive stars an unfavorable environment for planet formation by core accretion. The first large direct imaging surveys conducted with the new generation AO instruments, SHINE \citep[using SPHERE,][]{Vigan.2021} and GPIES \citep[using GPI,][]{Nielsen.2019} target mostly solar type stars and found that while massive giant planets did exists around such stars they were relatively rare.
However direct imaging has recently shown that  massive stars do harbor planetary mass companions at large separations \citep{Janson.2021_bcen,Squicciarini.2022,Chomez.2023b}. Though most of them are very massive (11--15 \Mjup) in absolute mass, these companions have mass-ratios relative to their host stars that are below 0.2\%, lower than those of most imaged exoplanets and close to that of Jupiter. Interestingly these companions have been found at very large separations, hundreds of au from their host stars. Though \citet{Squicciarini.2022} shows that the expected stellar irradiation of the protoplanetary disc at the present-day separation ($\approx290$~au) of $\mu^2$ Sco b is similar to that of early Jupiter, the exoplanets discovered by the YSES survey \citep{Bohn.2020a,Bohn.2021}, targeting solar-type stars of the Sco-Cen young association show that these lower mass stars do also harbour very large separation giant planets.

The B-star Exoplanet Abundance Study (BEAST) is a direct imaging high-contrast survey with the extreme adaptive optics instrument SPHERE, targeting 85 B-type stars in the young Scorpius-Centaurus (Sco-Cen) region with the aim to detect giant planets at wide separations and constrain their occurrence rate and physical properties. The individual discoveries of \bcen b and \mds B, achieved during an early stage of survey (respectively first and 7th stars observed over the 85 stars in the sample) raised the possibility that massive Jovian planets might be relatively common around B-stars. However, a dedicated statistical analysis is necessary to confirm or refute such a hypothesis and firmly establish whether large separation planetary-mass companions around massive stars are common or not and whether they are distinct in any ways from the population that is known around solar--mass stars.

We present here the statistical analysis of the first half of the survey that already has a complete follow-up of all identified candidates. Numbering 34 stars after the removal of 8 outlying objects that did not respect the initial selection criteria, this homogeneous statistical sample is twice larger, and much deeper, than the previous largest direct imaging survey for exoplanets around B stars, the 18-strong sample studied by \citet{Janson.2011}. 


The sample used for this study is described in Section 2. Section 3 present the observations and the data analysis. Section 4 presents the companions identified around the stars of our sample. Section 5 describes the detection limits achieved around these stars to assess the completeness of the survey. Section 6 explores the statistics of the giant planet population as observed by our survey, and explores the plausible pathways to account for the formation of the companions we detected.

\section{Sample description}
\label{sec:sample}
\subsection{Initial BEAST sample}
Starting from the census of B-type members of Sco-Cen, defined as those stars with a kinematic membership probability to the association according to \citet{rizzuto11}, the target list of BEAST was assembled \citep[][hereafter \citetalias{Janson.2021_beast}]{Janson.2021_beast} by considering every B-star belonging to Sco-Cen, with no known stellar companion with separation in the $0.1"-6"$ range, no previous deep observations with SPHERE and declination outside of the $-24.6^\circ \pm 3.0^\circ$ range, due to telescope tracking limitations.
The final roster of 85 stars, aged between 2 and 27 Myr, spans a wide range of stellar masses $[2.4-16] M_\odot$, corresponding to spectral types from B0 to B9.5. 

No prior indications of the possible presence of substellar companions was available at the time of sample definition; additionally, no selection and/or ranking was operated based on a-priori estimations of planet detectability. In other words, neither the target selection nor the first-epoch observational strategy is affected by observational biases caused by assumptions on the underlying substellar companion population whose properties we aim at constraining. The same argument applies to the follow-up within the standard BEAST program, which follows up all identified point sources with equal priority, regardless of their estimated properties.

To further ensure the sample studied in this paper is free from observational biases, we decided to strictly adhere to a first-epoch criterion for the selection of the sample. We thus restricted the analysis to the first $N$ stars imaged in the survey, exploiting any additional follow-up epoch regardless of its date. The value of $N$ was optimized to simultaneously 1) allow for an even division of the sample between the intermediate and the final analysis, and 2) to minimize the fraction of stars with only a single epoch. The resulting value of $N=42$ corresponds to stars with a first epoch observed no later than April 9th, 2021. Only five among those stars have not yet been re-observed to date. Four of them show no candidate companion while the detection limits of the fifth will be carefully handled in Section~\ref{subsec:contrast_SES} to ensure statistical consistency.

\subsection{Revised stellar host properties}
\label{subsec:revised_prop}

We decided to reassess the main properties (namely age and mass) of the entire BEAST stellar sample compared to the original analysis presented in \citetalias{Janson.2021_beast}, so as to account for several theoretical as well as observational advancements occurred in the last few years: 1) the code underlying \citetalias{Janson.2021_beast}'s analysis is now a robust tool known as \textsc{madys} \citep{madys}; 2) a new version of non-rotating, solar-metallicity PARSEC isochrones, covering the full pre-MS evolution of $[0.09,14]$ \Msun~stars, is now available \citep{nguyen22}; 3) as a consequence of 1) and 2), stellar parameters based on the direct isochrone fitting of their photometry could be derived; 4) the astrometric solution for several stars in the sample is significantly improved thanks to {\it Gaia} DR3 \citep{gaia_dr3} compared to the previous DR2 solution \citep{gaia_dr2}; 5) the availability of the Hipparcos-{\it Gaia} proper motion catalog \citep{kervella22} allows for a more accurate identification of groups of comoving stars to our science targets.  In addition to this, the updated kinematic information was used to reassess the membership of BEAST stars to the association through BANYAN $\Sigma$ \citep{gagne18}.

The first step of the process involved the derivation of stellar ages in a similar fashion as in \citetalias{Janson.2021_beast}. \textsc{madys} is a tool aimed at deriving stellar or substellar astrophysical parameters (mass, age, effective temperature, etc) for any list of input objects, by comparing suitable photometric measurements and theoretical (sub)stellar models. For the purpose of this work, optical and near infrared photometry from {\it Gaia} DR3 \citep{gaia_dr3} and 2MASS \citep{2MASS_MAG} -- corrected by extinction, estimated by integration of the 3D reddening map by \citet{leike20} along the line of sight -- was compared to the above mentioned PARSEC isochrones.

The following four age indicators, ordered by ascending spatial scale, were employed: individual isochronal ages (individual ages); the ages of small (10s-100s stars) kinematic groups (cms) sharing a similar motion with BEAST targets\footnote{In order to remove systematic uncertainties related to the modeling of low-mass stars \citep[see discussion in][and references therein]{madys}, only stars with $ M > 0.85$ \Msun were considered.}, following the method developed in \citetalias{Janson.2021_beast} (cms ages); the 2D age map of Sco-Cen derived by \citep{pecaut16}, describing the variation of the mean association age as a function of galactic coordinates (map ages); the ages of the three classical Sco-Cen subgroups (Upper Scorpius, Upper Centaurus-Lupus, Lower Centaurus-Crux) routinely adopted in the literature \citep[subgroup ages; see, e.g.,][]{pecaut16}.

In most cases the different indicators turned out to be compatible with one another within their respective uncertainties; priority was assigned to cms ages (or, if not available, to map ages) otherwise. Unusually young individual ages are particularly informative as they might be indicative, as will be shown below, of unresolved multiplicity.

Based on the derived ages, the second part of the process used \textsc{madys} in its age-constrained mode to derive the values of stellar mass that were more compatible with the observed photometry.  We assume a minimum relative uncertainty of 10\% on the derived mass values to account for systematic uncertainties such as the amount of stellar rotation and the choice of the evolutionary model (see appendix ~\ref{appendix:mass_error}).

\begin{figure}[t!]
    \centering
    \includegraphics[width=\linewidth]{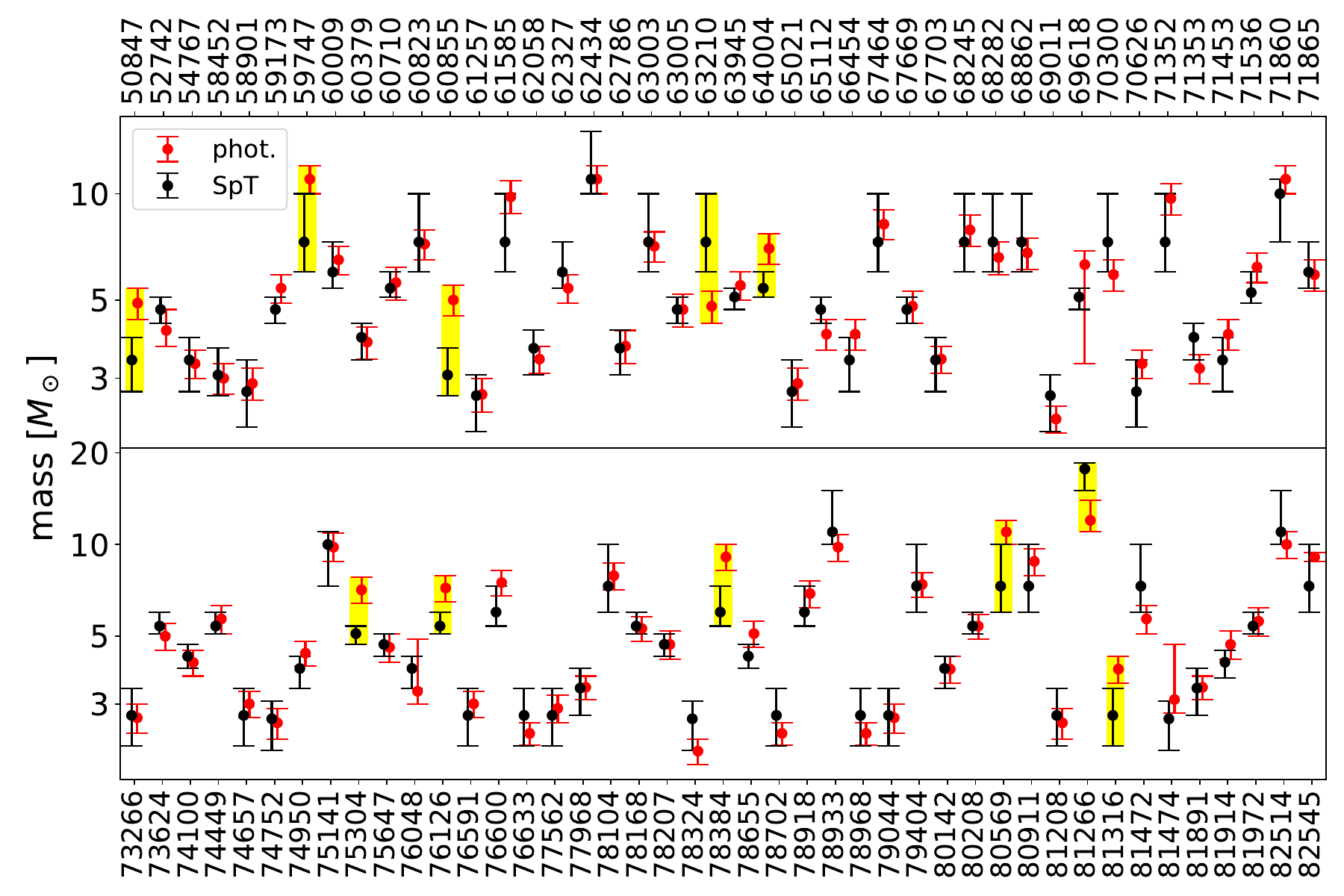}
    \caption{Comparison between spectral-type-based mass estimates (black) and photometric mass estimates (red). Stars where the two estimates are not compatible are highlighted through yellow boxes.}
    \label{fig:spt_iso_masses}
\end{figure}

The underlying assumption behind the method is that all stars are single; in other words, the contribution to the total measured flux from possible stellar companions is always assumed to be negligible\footnote{According to PARSEC isochrones, a $\sim 2~\msun$ companion to a 15 Myr, $5~\msun$ star (i.e., a mass ratio $q \sim 0.4$) is needed to cause a $\sim 10\%$ flux offset in the $G$ band-- approximately corresponding to 0.1 mag -- in the system's photometry. Multiple systems with a smaller $q$ are well approximated by the single-star model. This typical threshold $q$ value is between 0.4 and 0.6 for the age and mass range of BEAST stars.}. Even though intermediate-separation binaries were removed during the sample design, the large fraction of tighter ($a<10$~au) multiple systems with a B-type primary \citep[$\sim$40-70\%, increasing with mass;][]{moe17} actually calls for extreme caution and for additional checks. 
The nuisance caused by unresolved multiplicity was exactly the reason why stellar mass determination in \citetalias{Janson.2021_beast} was based on spectral type rather than on luminosity. Indeed, spectral types are weakly affected by this problem; however, the precision of spectral type determination is usually not as good as the one reached by absolute photometry. Hence, we decided to independently re-derive mass estimates based on spectral types with the goal of comparing them to photometry-based determinations 
to unveil significant differences which might be due to unresolved multiplicity. After collecting spectral types (SpT) from Simbad \citep{simbad}, and assuming an error of $\pm 1$ subclass  for SpT>B3 and $\pm 0.5$ for SpT$\leq$ B3\footnote{Approximately corresponding to a 10\% error on \teff.}, we converted them to stellar masses by means of the latest version (2021) of the empirical tables by \citet{pecaut13}, based on a large collection of $5-30$ Myr stars. The two mass estimates were found to be incompatible within their respective uncertainties in 11 cases (Figure~\ref{fig:spt_iso_masses}).

In order to verify if multiplicity was the reason for the observed discrepancies, we undertook a thorough search in the literature for high-contrast imaging, radial velocity, interferometry and photometric indication of multiplicity; moreover, the {\it Gaia}PMEX tool (Kiefer et al., accepted) was used to identify unseen companions inducing astrometric wobbles in Hipparcos \citep{hipparcos2} and {\it Gaia} data. We were able to identify 57 stellar companions to 52/85 stars (61\% of the sample), and to place meaningful detection limits in the case of non detections. Unresolved companions could be the reason for the observed differences between the two mass determinations in  9 cases; in one instance the discrepancy might be related to a combination of binarity and an inaccurate spectral type classification, while in a single case it is not clear whether the difference be related to parallax, spectral type or unknown multiplicity (see Appendix ~\ref{appendix:mass_error}).

The results of this analysis are provided in Table~\ref{tab:star_table}. Due to the heterogeneity of our binary data and to the manifold kinds of degeneracies (e.g.: mass vs semimajor axis for astrometry) hampering a precise companion mass determination, it was not possible to determine primary and companion masses for the entire sample in a uniform and precise way. For the sake of uniformity, we decided to always employ masses derived from the single-star model throughout this work. The combination of archival and future data  -- future {\it Gaia} releases, radial velocity monitoring and interferometric campaigns -- will allow to perform a systematic determination of the secondary masses in the coming years.

\begin{table}[!htbp]
\caption{Sample of stars used for this study's statistical analysis.}
\label{tab:sample}
\centering
\begin{tabular}{cccc}
\hline\hline
Name & Age (Myr) & Mass (\msun) & Sp. type \\
\hline
HIP58452 & $20_{-5}^{+5}$ & $3.0 \pm 0.3$ &B8.5V \\
HIP58901 & $13_{-3}^{+3}$ & $2.9 \pm 0.3$ &B9V \\
HIP60009 & $12_{-2}^{+4}$ & $6.5 \pm 0.6$ &B2.5V \\
HIP60379 & $13_{-3}^{+3}$ & $3.8 \pm 0.4$ &B7V \\
HIP60823$^{*1}$ & $17_{-3}^{+3}$ & $7.2 \pm 0.7$ &B2V \\ 
HIP61585$^{*1}$ & $15_{-3}^{+3}$ & $9.8_{-1.0}^{+1.1}$ &B2V \\
HIP62058 & $15_{-3}^{+3}$ & $3.4 \pm 0.3$ &B\\ 
HIP62327$^{*1}$  & $17_{-3}^{+3}$ & $5.4 \pm 0.5$ &B2.5V \\ 
HIP63003 & $16_{-3}^{+3}$ & $7.1 \pm 0.7$ &B2V \\
HIP65112 & $15_{-3}^{+3}$ & $4.0 \pm 0.4$ &B5V \\
HIP66454 & $17_{-3}^{+3}$ & $4.0 \pm 0.4$ &B8V \\
HIP67464$^{*1}$  & $18_{-5}^{+3}$ & $8.2 \pm 0.8$ &B2V \\ 
HIP67669$^{*1}$ & $17_{-3}^{+3}$ & $4.8 \pm 0.5$ &B5V \\
HIP67703 & $16_{-3}^{+3}$ & $3.4 \pm 0.3$ &B8V \\
HIP68245 & $17_{-3}^{+3}$ & $7.9 \pm 0.8$ &B2V \\
HIP68282$^{*1}$  & $21_{-2}^{+4}$ & $6.6 \pm 0.7$ &B2V \\ 
HIP69011 & $17_{-3}^{+3}$ & $2.3 \pm 0.2$ &B9.5V \\
HIP70300 & $17_{-5}^{+3}$ & $5.9 \pm 0.6$ &B2V \\
HIP70626 & $15_{-3}^{+3}$ & $3.3 \pm 0.3$ &B9V \\
HIP71352 & $15_{-3}^{+3}$ & $9.7 \pm 1.0$ &B2V \\
HIP71536$^{*2}$  & $14_{-4}^{+4}$ & $6.2 \pm 0.6$ &B3.5V \\
HIP71865$^{*2}$  & $14_{-3}^{+3}$ & $5.9 \pm 0.6$ &B2.5V \\
HIP74950$^{*1}$  & $18_{-3}^{+3}$ & $4.4 \pm 0.4$ &B7V \\ 
HIP76591 & $17_{-3}^{+3}$ & $3.0 \pm 0.3$ &B9V \\
HIP76600$^{*1}$  & $13_{-3}^{+4}$ & $7.5 \pm 0.7$ &B2.5V \\ 
HIP76633 & $7_{-2}^{+3}$ & $2.4 \pm 0.2$ &B9V \\
HIP77562 & $17_{-3}^{+3}$ & $2.9 \pm 0.3$ &B9V \\
HIP77968 & $17_{-2}^{+5}$ & $3.4 \pm 0.3$ &B8V \\
HIP78207 & $0.5_{-0.2}^{+0.5}$ & $4.7 \pm 0.5$ &B5V \\
HIP79044$^{*1}$ & $17_{-3}^{+3}$ & $2.7 \pm 0.3$ &B9V \\
HIP80911 & $16_{-4}^{+4}$ & $8.8 \pm 0.9$ &B2V \\
HIP82514$^{*1}$  & $20_{-4}^{+4}$ & $10.0 \pm 1.0$ &B1V \\
HIP82545 & $20_{-4}^{+4}$ & $9.1 \pm 0.3$ &B2V \\ \hline
\end{tabular}
\tablefoot{$^*$ Star is a short separation ($<$0.1") binary. Binarity source : $^{*1}$, from \citet{Gratton.2023}; $^{*2}$, from \citet{Rizzuto.2013}
Spectral types are from Simbad. }
\end{table}

\subsection{Sample selected for this study}
The membership analysis mentioned in Section~\ref{subsec:revised_prop} allowed us to determine that several stars from our initial sample (HIP52742, HIP54767, HIP60855, HIP65021, HIP76126, HIP81474) have a low chance of being members of the Sco-Cen association.  Of these, HIP52742 and HIP65021 are part of the 42-strong intermediary sample that we study here. Although they are still relatively young B-stars, we decided to remove them from our study, so that our sample is built only from Sco-Cen members, sharing the same environmental conditions associated with this relatively dense young star-forming region.
 
Another selection bias comes from binarity, since the BEAST survey has been built with the explicit selection of B stars from Sco-Cen with no stellar binary companions between 0.1" to 6". However, the BEAST observations themselves have uncovered some of the original sample stars as such intermediate separation stellar binaries. These objects are biasing the sample because their stellar companions are expected to have a strong impact on planet formation and orbital stability in the separation ranges probed by our data. To be able to derive the unbiased frequency of B stars with no stellar companions within 0.1" to 6" that host at least one planetary mass companion, we therefore need to remove any BEAST stars with such intermediate separation stellar companion from our statistical sample or, if the bias is small enough, try to correct it.

Our BEAST high-contrast imaging data has revealed that 8 stars within our initial sample were previously unknown binary stars with a stellar companion within 0.1" to 6", and are thus outside of the specification of our initial sample. These stars are identified in Table~\ref{tab:excluded_binaries} and were found by \citet{Gratton.2023}, and confirmed bound by our analysis of BEAST data.  Among them HIP52742 has already been rejected because it is not a Sco-Cen member. 
Six out of the remaining seven have bound stellar companions at relatively large separation, with projected semi-major axis (hereafter sma) ranging from 0.26" to 4.16"  that would strongly bias our sample because they would make planetary orbit unstable in large part of the parameter range we probe. Indeed, according to \citet{Holman.1999}, for a moderate eccentricity of the binary system, exoplanets cannot have stable orbits  if their sma is not at least $\sim$3 times larger than the sma of the stellar companion, and this dynamical instability arises fast, within 10\,000 stellar companion orbits.  We therefore also removed these six stars (HIP50847, HIP59173, HIP62434  HIP63005, HIP63945, HIP74100) from our initial sample.

One last system, HIP60009 is a tight stellar binary, with projected sma of 0.146". We decided to keep it in our final sample because the bias it causes on our sample selection is moderate, and we tried to minimize it as follows. We used the \citet{Holman.1999} stability criterion, and for separation smaller than three times the binary sma, we manually set the completeness of these observations to zero for any companion mass to account for the fact that these two observations could never detect any exoplanet companions in this range where their orbit would be quickly unstable. This is equivalent to removing them from our sample only within the separation range from which we know putative exoplanet companions cannot have stable orbits. We acknowledge that projected separation do not translate directly into sma and that the \citet{Holman.1999} stability criterion value of three times the sma is an approximation, but since the final exoplanet frequency is negligibly  affected by varying its value  by plus or minus 50\%  (the effect on the median of the posterior of exoplanet frequency is at the 0.1\% level, much smaller than our formal error bars), we are confident this bias correction is acceptable.

The 34 stars that we kept in the sample and use in the following statistical study are shown in Table~\ref{tab:sample}, together with their revised properties. Even if we did not use the stars we excluded from the sample in the statistical analysis, we did look for bound substellar companions around all of them. We found none and we present in the appendix the detection limits we achieved around these stars, along with the detection limits achieved around the stars we kept into the statistical sample. The median age of this statistical sample is 16.5Myr, while the median mass of the primary stellar hosts is 4.8\msun.

\begin{table}
\caption{Intermediate separation binaries in sample}
\label{tab:excluded_binaries}
\centering
\begin{tabular}{ccccc}
\hline\hline  
Name & Sep.(mas) & PA (°) & Epoch & Contrast \\
HIP50847   &   2212$\pm$5 &    352.4$\pm$0.4     & 2019-01-26     & $\Delta K$=5.6  \\  
HIP52742   &      1191 $\pm$2   &    9.3$\pm$0.1     &   2018-05-14     &  $\Delta J$=7.7\\
HIP59173   &      1270 $\pm$1   &    130.1$\pm$0.1     & 2020-03-02     &  $\Delta K$=5.3  \\
HIP60009$^1$   &   146     $\pm$4   &    171.4$\pm$0.5     &  2019-02-23     &  $\Delta K$=5.7 \\
HIP62434   &   4173    $\pm$ 2   &    121.8$\pm$0.1     &  2019-04-01       & $\Delta K$=8.2\\
HIP63005   &    262    $\pm$1.5   &    171.7$\pm$2.5     & 2019-03-05      & $\Delta K$=4.0  \\
HIP63945   &        1555$\pm$2    &   261.8$\pm$0.1      &2018-04-23      &  $\Delta K$=3.3  \\
HIP74100   &        549$\pm$1   &    277.4$\pm$0.1     &  2019-03-24     &  $\Delta K$=3.8 \\
\hline
\end{tabular}
\tablefoot{$^1$ HIP60009 was not excluded from the statistical sample.}
\end{table}

\begin{figure*}[t!]
    \centering
    \includegraphics[width=\textwidth]{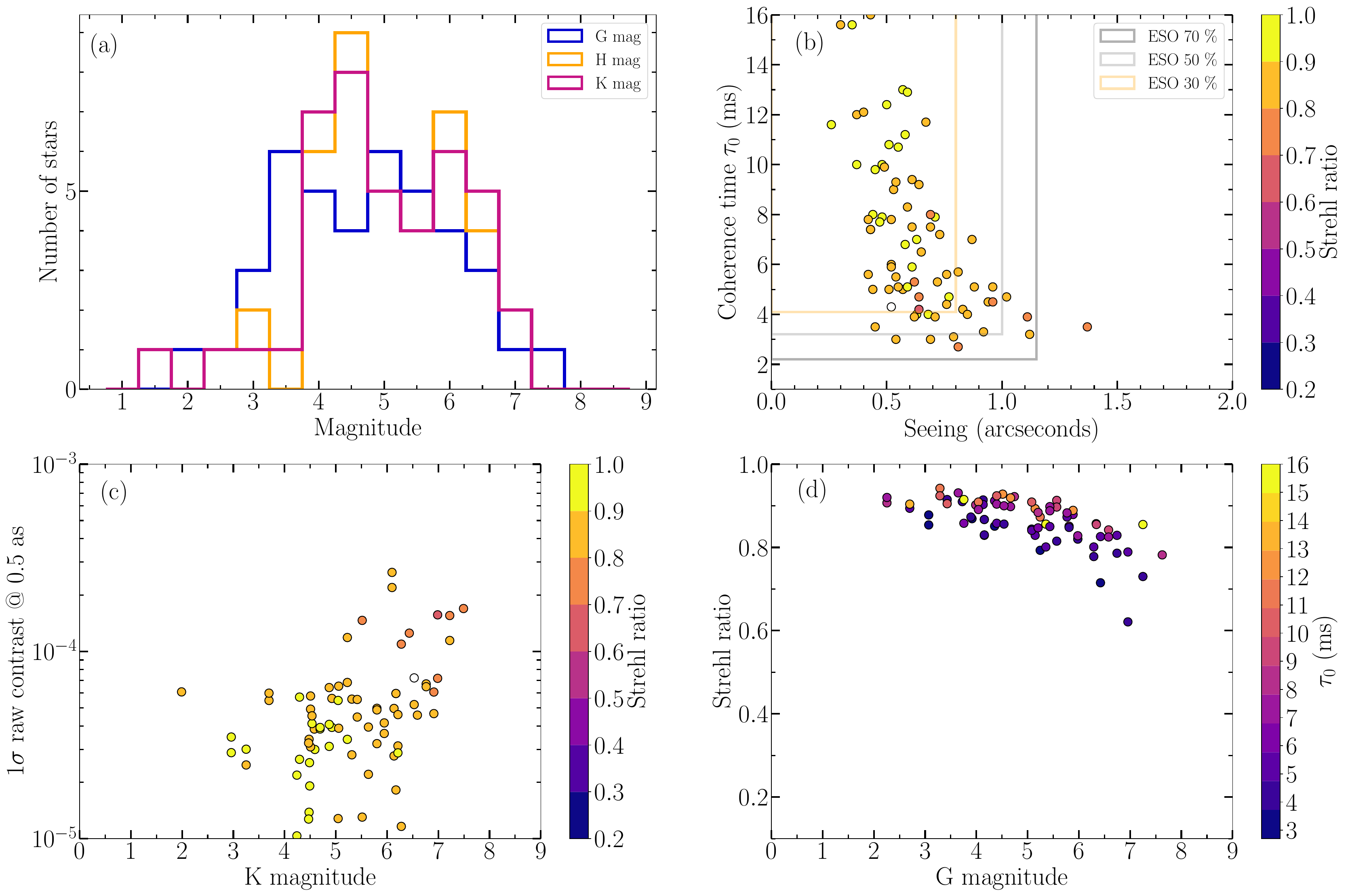}
    \caption{(a) Histogram of the star magnitudes (G, H, K) for the sample considered in this study. (b) Scatter plot showing the coherence time ($\tau_0$) against the seeing with the Strehl ratio color-coded. Data points without Strehl information are colored in white. The boxes delimits the ESO observing categories. (c) Scatter plot showing the $1\ sigma$ raw contrast measured at a separation of 0.5" (directly measured on the coronographic frame, before post-processing) with respect to the K-band star apparent magnitude. (d) Scatter plot displaying the Strehl ratio as a function of the star apparent magnitude in the G band.}
    \label{fig:combined_sample_properties}
\end{figure*}

\section{Observations and data processing}
\label{sec:analyze}

\subsection{Observation setup}
\label{subsec:obs_setup}

All of the observations were taken using SPHERE \citep{Beuzit.2019} in pupil stabilized mode. This mode introduces a field of view (FoV) rotation during the sequence, allowing the use of angular and spectral differential imaging \citep[ASDI,][]{Marois_adi} based post-processing techniques. Each observation uses the infrared dual-band imager and spectrograph \citep[IRDIS,][]{Dohlen_irdis} with the N-ALC-YJH-S coronagraph to observe in the K1/K2 bands (except dedicated followup observations) in combination with the YJH bands for the integral field spectrograph \citep[IFS,][]{Claudi_ifs}.

Each observation follows the same observational sequence: first, an unsaturated, noncoronagraphic image -- point spread function (PSF) -- of the primary is obtained for flux calibration purposes. Second, a coronagraphic exposure with a waffle pattern applied to the deformable mirror \citep{Cantalloube_2019}, for centering purposes. Then, the main coronagraphic exposures are recorded. Finally, another coronagraphic exposure with a waffle pattern is taken, followed by another PSF.

All of the observations were taken in service mode, ensuring the fulfilment of a set of atmospheric conditions for an observation to be accepted. All observations were scheduled to be taken within one hour of meridian crossing, ensuring at least around 20° of parallactic rotation during the coronagraphic sequence. Seeing was also required to be smaller than 0.9" and the coherence time greater than 4 ms (for more details, see \citetalias{Janson.2021_beast}). The observing conditions for each observation are summarized in Figure~\ref{fig:combined_sample_properties} as well as, in more detail, in Table~\ref{tab:obs_table}.

\subsection{Data processing}
\label{subsec:data_processing}

The whole data reduction was performed on the COBREX Data Center, a modified and improved server based on the High Contrast Data Center, \citep[HC-DC, formerly SPHERE Data Center;][]{Delorme_sphereDC}. The COBREX Data Center aims to improve the detection capabilities with existing SPHERE archival data by means of the patch covariance (\verb+PACO+; \citealp{Flasseur_paco}) algorithm. More specifically, we used \verb+PACO ASDI+ \citep{flasseur2020robustness, Flasseur_asdi} as our primary post-processing algorithm for both IRDIS and IFS, where the initial reduction was using the 
No-ADI (a simple temporal stack), cADI \citep{Marois_adi} and TLOCI-ADI \citep{Marois_TLOCI} algorithms embedded in the \verb+SPECAL+ software \citep{Galicher_specal} in the HC-DC for IRDIS. For IFS, the first reduction was also using three algorithms: cADI, TLOCI and PCA \citep{Soummer_PCA}. 

The pre-reduction pipeline (i.e., going from raw data to a calibrated 4D datacube) is identical to the one implemented in HC-DC, performing dark, flat, distortion, and bad pixel corrections. The full details on the improvements of the pre-reduction pipeline as well as the optimization regarding the ASDI mode of \verb+PACO+, and the obtained performances are described in \citet{Chomez.2023a}.

The algorithm used in this work, \verb+PACO+, models the noise using a multi-Gaussian model at a local scale on small patches, allowing for a better estimation of the spatial and spectral correlations of the noise. This modeling improves the contrast achieved between one and two magnitudes at all separations compared to more classical ADI approaches like TLOCI and PCA. \verb+PACO+ is also a data-driven algorithm, meaning that it requires no hyper-parameters tuning by its user. One of the other key advantage of \verb+PACO+ is its detection map, which is directly interpretable in terms of signal-to-noise (S/N) of detection. It thus follows the standard normal distribution $\mathcal{N}(0,\,1)$ by design, even close to the target star when noise is dominated by speckle residual noise. It allows two important features: the first one is a reliable automated detection threshold (fixed for this analysis at the classical $5\sigma$ threshold), which is particularly suitable for automated reductions of dozens of datasets. The second one is its statistical guarantee over a detection (i. e., the probability of it to be a statistical false positive), which is reliably linked to its S/N. 

Using \verb+PACO+ for both instruments allows to remove any systematics related to the use of different algorithms and easily compare the performances of both instruments to each other. Thanks also to the statistically grounded nature of the contrast limits provided by \verb+PACO+, it allows the derivation of more reliable constraints on the actual detection limits of the survey and the deeper sensitivity of \verb+PACO+ also enables us to look for new sources undetected by the previous analysis. 

\section{Detected companions within the sample}
\label{sec:comp}

The astrometry (separation and PA relative to the host star) and the photometry (IRDIS fluxes and IFS spectrum) of all sources detected within this analysis will be available via a VizieR catalog linked to this article. 


\subsection{Substellar companions}
Two substellar companions have been discovered within BEAST, \bcen b, by \citet{Janson.2021_bcen}, and \mds B by \citet{Squicciarini.2022}. \bcen b has a mass of $11\pm2$\mjup, and a mass ratio of 0.1--0.17\% with respect to its massive (6--10\msun) binary host star system, quite close to that of Jupiter in the Solar System. However its very large projected separation of 560~au and the very high mass of its stellar hosts make it a very distinct type of exoplanet. \mds B shares many of the peculiar properties of \bcen b, with a very large separation of 290~au from its very massive ($\sim$9\msun) host and a mass ratio below 0.2\%, and if a formation scenario could explain the existence of one of these companions it would certainly be compatible with the other. However their masses are slightly different, with \mds B being 14$\pm 1$\mjup and thus probably straddling the deuterium burning mass, that is being used \citep{IAUplanet2022} to draw a border between objects below that 13\mjup limit that are called exoplanets and those above that are called brown dwarfs. In the peculiar case of this study, this definition does not appear to bear any significant physical meaning and would only draw an artificial border between two objects that most likely share a very similar nature.

In the subsequent statistical analysis, we therefore decided to use 3 different mass cuts. (1) {\it Exoplanets}, for masses below the 13\mjup limits, one detection (\bcen b). (2) {\it Exoplanets and Low-mass Brown Dwarfs (ELBD)} for masses below 30\mjup limits, two detections (\bcen b and \mds B). (3) {\it Substellar objects} for masses below 73.3\mjup \citep[hydrogen burning minimum mass as defined by][ at solar metallicity]{Chabrier.2000}, two detections (\bcen b and \mds B). These mass cuts both affect the number of detections in each category but also the completeness of the survey when evaluating the significance of our non-detections.

The choice for the intermediate mass limit of 30\Mjup is of
course somewhat arbitrary, but motivated by the following argument. The most
massive stars systematically surveyed for the presence of giant planets at inner
orbits are probably G and K giant stars, with stellar masses
typically in the range between 1 and 3 solar massses 
\citep[see e.g.][]{Wolthoff.2022}. 
In these surveys, the most massive companions range from a few Jupiter masses
up to about 25 Jupiter masses continuously, with no obvious gap at certain masses.
However, beyond 25 Jupiter masses very few objects are found, if 
any\footnote{See e.g.\ the continuously updated table with planets discovered
around giant stars at
\href{https://www.lsw.uni-heidelberg.de/users/sreffert/giantplanets/giantplanets.php}{https://www.lsw.uni-heidelberg.de/users/sreffert/giantplanets/giantplanets.php}}. The next most massive companion after the 30 Jupiter mass limit has 
a minimum mass of 37 Jupiter masses, but there are indications from
{\it Gaia} astrometry that its real mass is in excess of 60 Jupiter masses when
accounted for inclination. Thus, we opted for a mass limit of 30 Jupiter masses
for the ELBD.

It is to be noted that other companions have been detected or suspected within the overall BEAST sample. First, \citet{Squicciarini.2022} identified a promising candidate companion \mds C, close to the coronographic mask, that - if bound - would have a mass of 18$\pm 2$\mjup. Since our statistical study consider the fraction of systems with at least one companion within a given mass range, it is not affected by the possible existence of another planetary-mass companion in the \mds system. However, if confirmed, \mds C, with is relatively short separation of $\sim20$~au, might bring constraints on the formation mechanisms that are quite distinct from the much more distant \bcen b and \mds B.
Second, \citet{Viswanath.2023} discovered a massive $\sim$67\mjup brown
dwarf (HIP81208B) and a very low mass star (HIP81208C) orbiting around HIP81208, and \citet{Chomez.2023b} discovered a $\sim$15\mjup planetary mass companion around the low-mass stellar companion HIP81208C. Though forming a fascinating hierarchical quadruple system with a chain of relative mass ratios that are difficult to explain by any planet or binary star formation model, the late first epoch observation of HIP81208 within the BEAST sample puts it outside of the BEAST intermediary sample that is considered by this article (see Sect.~\ref{sec:sample}).


\begin{figure}[t!]
    \centering
    \includegraphics[width=\linewidth]{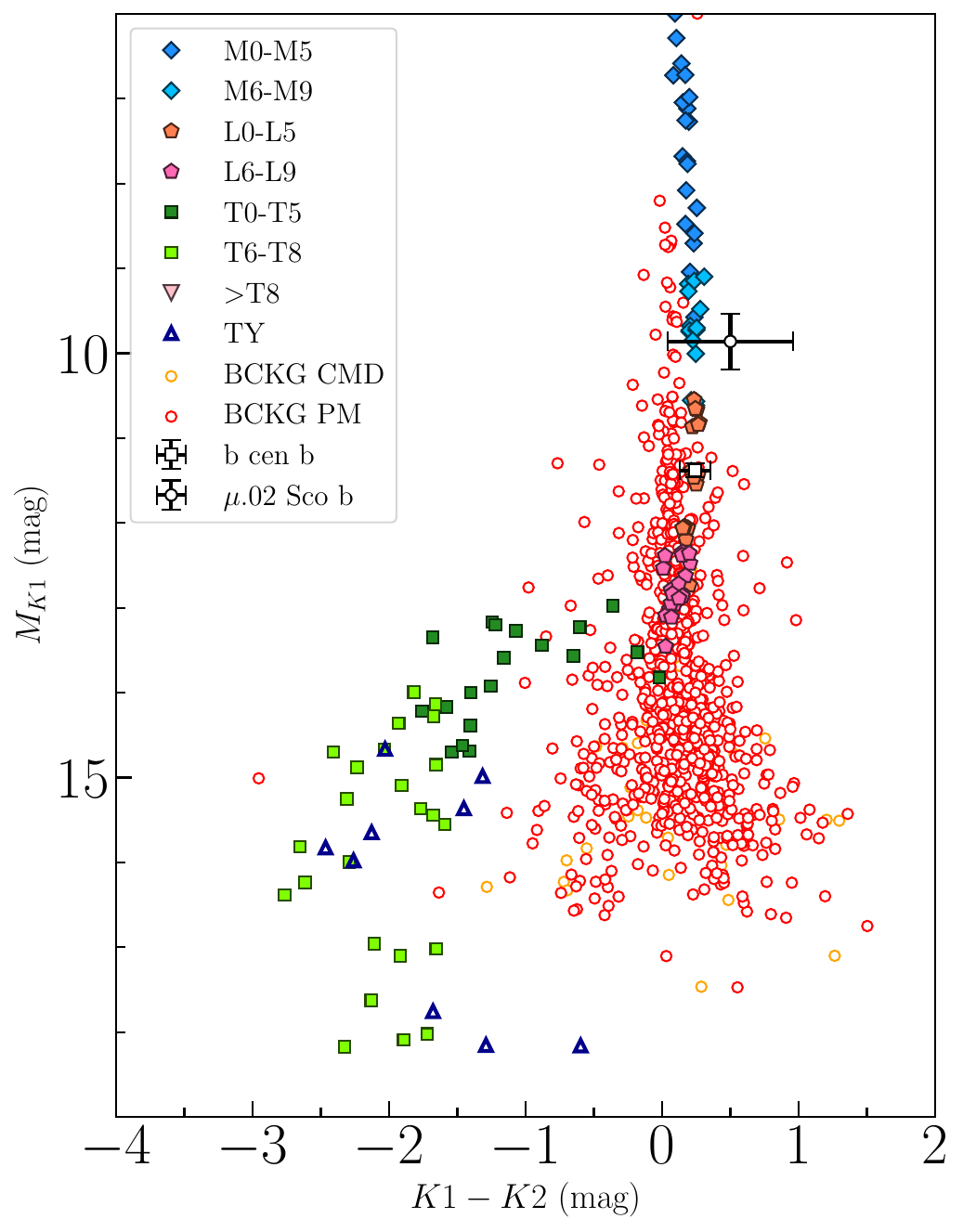}
    \caption{K1-K2 color-magnitude diagram of candidate companions found around the original, 42-strong, sample. Detected sources are represented by the hollow dots while the plain symbols represent the theoretical MLT track \citep[][Appendix C]{Bonnefoy_GJ504_2018}.}
    \label{fig:CMD}
\end{figure}

\subsection{Identified background sources}
\label{subsec:ident_bckg_pm}

Sco-Cen being located relatively close to the galactic center, it is an area of high stellar density and the large IRDIS field of view consequently yields many detections. Those detections are in vast majority field K and M stars \citep{Parravano.2011}, consistent with their position on the CMD (See Fig.~\ref{fig:CMD})  and located much further away than our target stars. The most effective method to identify those background stars (or in the other way around, find bound companions) is called the common proper motion test. Using (at least) two observations separated by a long enough time baseline it assesses if the measured motion of a given source between the two observations is consistent with a bound object sharing in that case the proper motion of the host star, or a background star with a different (and usually very small) proper motion. This method however has one major flaw: it requires the detection of the sources at both epochs, which can be challenging for faint sources. However, BEAST is conducted in service mode, ensuring consistent performances between all epochs and allowing us to systematically use this technique.

\subsection{Identification summary}

The sources are classified following three categories (visible in the color magnitude diagram (CMD) in Fig.~\ref{fig:CMD}): "CC" for confirm companion (i.e., planets, brown-dwarfs, and stellar companion), "BCKG\_PM" for background sources identified via the proper motion test (see Sect.~\ref{subsec:ident_bckg_pm}), and "BCKG\_CMD" for sources detected at one epoch (single epoch stars or unable to re-detect them at a second epoch) that are likely background sources given their positions on the CMD far from the exoplanet parameter range.

\section{Detection limits}
\label{sec:detlim}

\subsection{Dealing with single-epoch stars}
\label{subsec:contrast_SES}
Most of the stars present in this sample have been observed at least twice, allowing us to robustly determine the nature of all detected sources via proper motion. However some stars, as of the writing of this paper, have only been observed once. If no source has been detected, the detection limits of this single epoch dataset can be used in the following statistical analysis without introducing any bias. However, if a point source is detected, we cannot confirm its nature (bound or background) via proper motion, making the statistical interpretation of the data more challenging. One can use a CMD to try to distinguish between a background source and a bound one but, as shown in Figure~\ref{fig:CMD}, the CMD based on the K1-K2 color is not always discriminant, notably when the exoplanets and background stars track overlap. 

In our sample, only one star has such detection with no proper motion confirmation: HIP 76633, and we present here how we deal with its peculiar case in our statistical analysis. The previously undetected and very faint signal unveiled by \verb+PACO+ with a S/N = 5 is located at around 3 as from HIP 76633 and its nature is ambiguous. 
We modified the detection limits from this peculiar dataset so that they keep the same physical interpretation as detection limits for single epoch dataset with no companion detected:  it allows to rule out the existence of any companion brighter than this limit in contrast. We therefore want the detection limits $C$ at separation $s$ expressed in contrast unit and the contrast $ C_{\mathrm{amb}}$ of the ambiguous source located at the separation $s_{\mathrm{amb}}$ to verify the following relation :

\begin{eqnarray}
    \forall s \geq s_{\mathrm{amb}} ;
    C(s) = \max\left(C(s), C_{\mathrm{amb}}\right) 
\end{eqnarray}

\begin{figure}[t!]
    \centering
    \includegraphics[width=\linewidth]{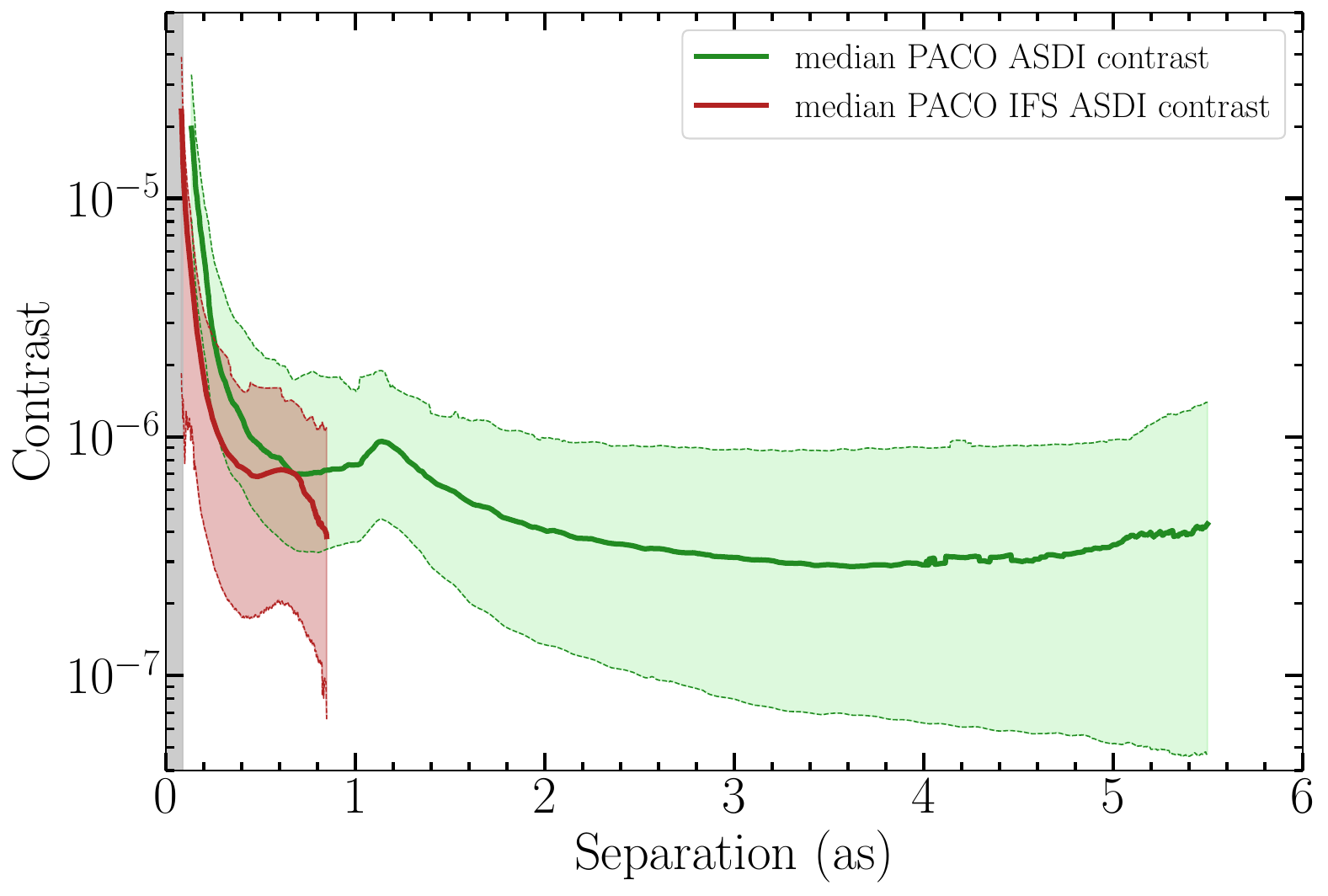}
    \caption{Contrast achieved over the initial, 42-strong sample. The median contrast for both instrument is highlighted with the bold line and the grey area represents the coronographic mask.}
    \label{fig:sample_contrast}
\end{figure}

This effectively removes from our detection map the ability to detect any companion brighter than the ambiguous source. The contrast achieved over the full sample after applying this correction is shown in Figure~\ref{fig:sample_contrast}.

\subsection{Achieved contrast and conversion to detection limits expressed in mass}
\label{subsec:detlim_mass}

To convert the 2D contrast map into mass maps expressed in $\Mjup$, we use a new feature implemented in \textsc{madys}
\citep{madys} that allow this conversion. We used the {\it Gaia} DR3 \citep{gaia_dr3} parallax and 2 MASS K magnitude for IRDIS and their associated uncertainties provided by Simbad\footnote{\href{http://simbad.u-strasbg.fr/simbad/}{http://simbad.u-strasbg.fr/simbad/}}. 
To convert magnitude into masses, we used several models: atmo2020 \citep{Phillips.2020}, ames-cond \citep{Baraffe.2003}, ames-dusty \citep{Chabrier.2000}, with the resulting variation from model to model being smaller than the mass uncertainties linked to the age uncertainties. In the following we chose to present results obtained with the ames-cond models. 

\begin{figure}[t!]
    \centering
    \includegraphics[width=\linewidth]{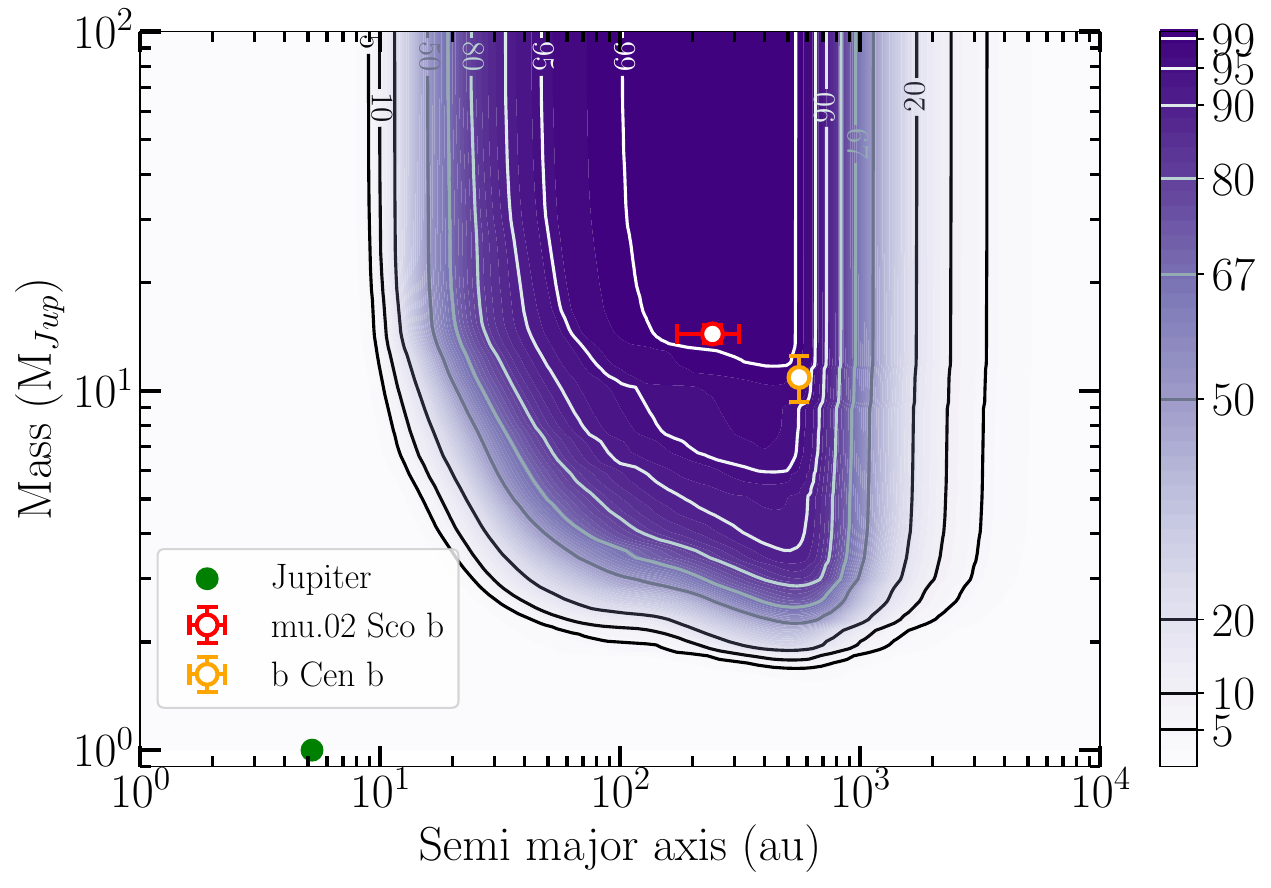}
    \caption{Average detection probability of a companion of given mass/sma in the BEAST intermediary survey using the cond atmospheric model. The two detected planets in this survey are also represented with the mass / sma extracted from \citet{Janson.2021_bcen} and \citet{squicciarini21}. Jupiter is also added to provide comparison. In the case of \bcen b, no semi-major axis estimation is available, so we represented its projected separation.}
    \label{fig:sample_detmaps}
\end{figure}

In order to produce probability detection maps for each target and for the sample, we used the \verb+EXO-DMC+ MCMC tool \citep{Bonavita.2020} with a semi major axis (sma) from 1 to 10\,000~au and all orbital parameters are uniformly distributed except for the eccentricity, which using a one-sided Gaussian prior with 0 mean and 0.3 as sigma (with a max of 1) for the distribution. The average sensitivity using the median age for each star for the whole sample can be seen in Figure~\ref{fig:sample_detmaps}, while detection maps using the min/max ages as well as the atmo2020 and ames-dusty models can be seen in Figure~\ref{fig:all_mean_detmaps}. Individual detection maps using the ames-cond model are displayed in Appendix~\ref{appendix:individual_det_maps}.

\section{The population of substellar companion around B stars}
\label{sec:pop}
\subsection{Companions around B-stars: Jupiter-like mass ratios}
\begin{figure}[t!]
    \centering
    \includegraphics[width=\linewidth]{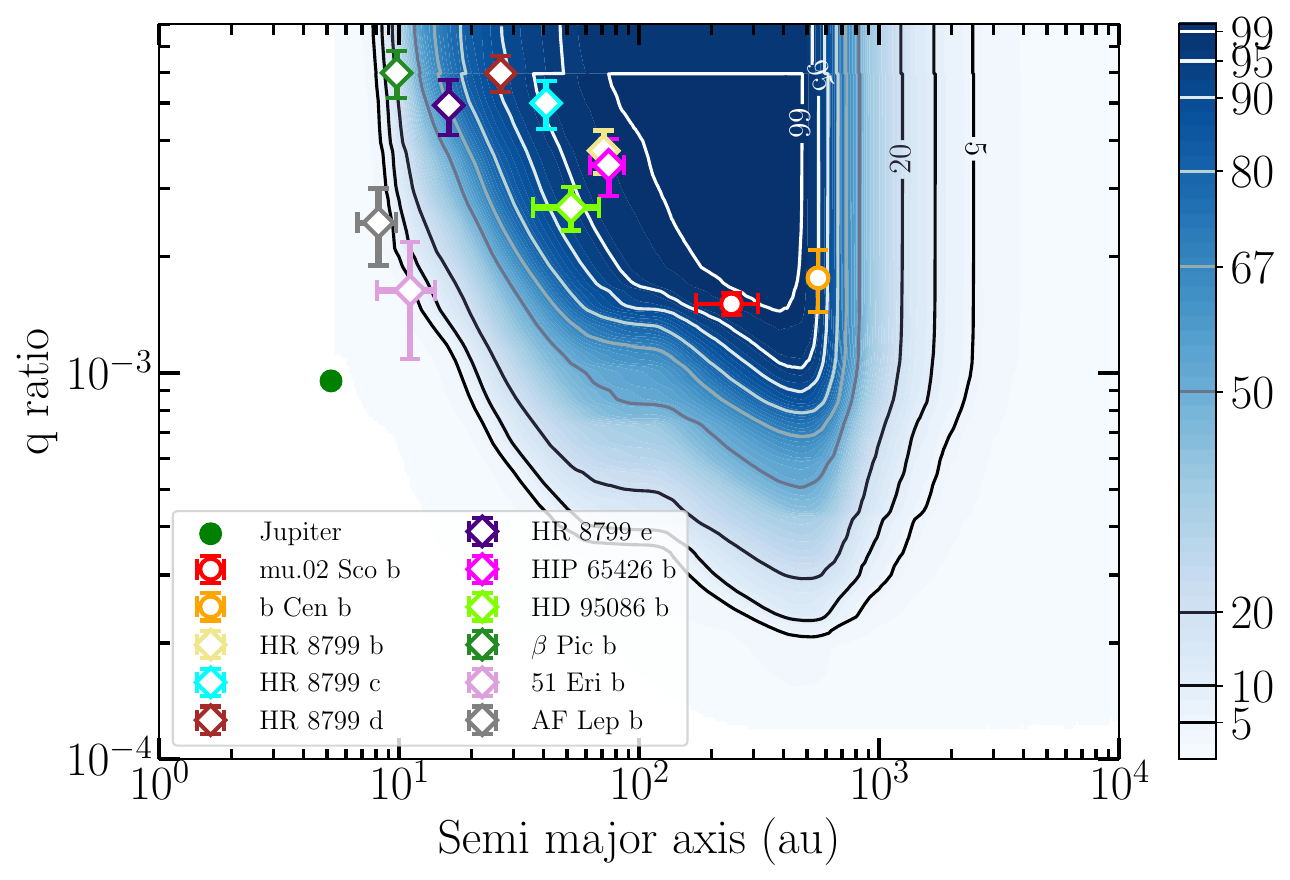}
    \caption{Same as Figure~\ref{fig:sample_detmaps} but expressed in companion to primary star mass-ratio. Several emblematic systems are added to the plot\textbf{, represented by the hollow diamonds}. Data are extracted from \citet{Zurlo_hr8799} for HR 8799 bcde, \citet{Desgranges_hd95086b} for HD 95086 b, \citet{Blunt_hip65426b} for HIP 65426 b, \citet{Lagrange_bpic_2020} for $\beta$ Pictoris b, \citet{51Erib_BS_2023} for 51 Eri b and \citet{Zhang_AF_Lep_b_2023} for AF Lep b.}
    \label{fig:sample_detmaps_q}
\end{figure}

Though straddling the planet/brown dwarf definition border, the substellar companions imaged by BEAST have mass-ratios relative to their host stars between 0.1 and 0.2\%, lower than those of other imaged exoplanets and close to the 0.1\% of Jupiter relative to the Sun. Since planetary formation takes place in discs around young stars, and since the properties of these protoplanetary discs are heavily related to the stellar mass, the mass ratio is quite informative to investigate the nature and possible formation mechanisms of substellar companions.
We therefore converted each detection probability map expressed in Jupiter masses into detection probability maps expressed in $q$-ratio, that is the planet to primary star mass ratio, using the nominal masses listed in Table~\ref{tab:star_table}. Once each individual $q$-ratio map has been computed, we interpolate them into a common $q$-ratio / sma grid and average them to produce the final average detection probability map presented in Figure~\ref{fig:sample_detmaps_q}. This figure clearly shows that \mds B and b Cen b populate a mass-ratio range that is close to that of Jupiter and in the same range as the lowest mass imaged exoplanets, 51 Eri b and  AF Lep b. In the case of \bcen b, the mass ratio can be as low as that of Jupiter if we follow \citet{Janson.2021_bcen}, and consider the total mass of the central binary (6--10\msun) instead of only the primary star (5.9$\pm 0.6$\msun)for the mass ratio calculation. Though the sma of the BEAST planets are much larger than Jupiter's, \citet{Squicciarini.2022} highlighted that the stellar flux at their observed separation is also quite similar to that of Jupiter by the Sun. However such a continuity in planetary formation from the Sun to supernovae progenitor such as \mds could come as surprising.  And there are some hints this similarity might be coincidental: \mds and \bcen are among the 20\% most massive stars of BEAST, and while we have the detection capability to detect planets with Jupiter-like mass ratios and insulation around lower mass B stars observed by BEAST, we found none, hinting that the population of exoplanets around B stars might differ depending on their mass. However we know that lower mass late B stars can host substellar companions, with HIP79098AB and HIP78530 which are two lower mass B stars outside of BEAST survey that are known to host substellar companions at large separation \citep{Janson.2019,Lafreniere.2011}. Our current sample is too small to be split into a higher-than-median-mass and lower-than-median-mass sub-samples to provide statistically meaningful constraints on possible variations of giant planet companions frequency with B star mass. The future analysis of the full BEAST sample should provide more informative results on this question.

\subsection{Frequency of companions around B-stars}
We want to estimate the frequency of planetary systems, so stars having at least one planetary mass companion. The average survey sensitivity (or completeness) for planetary companions below the IAU maximal mass, and detectable by our data (2--13 \mjup) with sma between 10 and 1000~au is 64.1\%. In this range, our sample of 34 stars therefore translates into an effective sample number of 21.8. With one exoplanet found in this range, a first order estimation of the frequency of such companions is 4.6\%. However the second companion, \mds B just straddles the 13\mjup limit. If we associate it with the population of Exoplanets and Low-mass Brown Dwarfs that we defined in section \ref{sec:comp}, the average survey sensitivity for such ELBD companions (2-30 \mjup) with sma between 10 and 1000~au is 70.5\%, resulting into an effective sample number of 24.0. Since we found two ELBD, a first order estimation of the frequency of such companions is 8.5\% .

We now refine these rough estimates using bayesian probability to be able to derive reliable confidence intervals for the frequency of planetary systems around B-stars. We followed the approach described in \citet{Carson.2012,Rameau.2013a,Lannier.2016} and derived the posterior distributions on the frequency of B-star systems with no stellar companion within 0.1" to 6" that host at least 1 substellar companion between 10 to 1000~au in various mass ranges. 

For a given target $j$ among the $N$ stars within our sample, the mean detection probability to find a companion of a given mass at a given semi-major axis is $p_j$. $p_j$ is derived from the 2D detection limit maps described above. We denote $f$ the fraction of stars around which there is at least one planet, with a mass included in the interval $[2,m_{max}]$\mjup ( with $m_{max}=13,30,73.3 $\mjup for exoplanets, ELBD and substellar companions respectively), and for separations inside the interval $[10,\,1000]$~au. Then, $fp_j$ is the probability to detect a companion around the star $j$ given its mean detection probability $p_j$ and the fraction of stars $f$ hosting at least one companion, and $1-fp_j$ is the probability not to find it. The detections and non-detections that are reported for a survey are denoted ${d_j}$: $d_j=1$ for stars around which we found a planet, and $0$ otherwise. The likelihood function, $L({d_j}|f)$, of the data represents the probability that an observation gives $d_j$ given the planet fraction $f$. We have: 
\\
\begin{equation}
L({d_j}|f)=\prod_{j=1}^N (1-fp_j)^{1-d_j} \times (fp_j)^{d_j}
\label{eq:likelyhood}
\end{equation}
\\
Bayes' theorem provides the probability density of $f$, the fraction of stars hosting at least one ELBD given the observed data $d_j$: this probability density is the posterior distribution that represents the distribution of $f$ given the observed data $d_j$.
\\
\begin{equation}
P(f|{d_j})=\frac{L({d_j}|f)P(f)}{\int_0^1 L({d_j}|f)P(f)df}
\end{equation}
\\
In the former equation, $P(f)$ is called the prior distribution. This prior is a distribution reporting any preexisting belief concerning the distribution of $f$, and can strongly impact the posterior result when scant observational evidence is available. To minimise the impact of the prior choice, we therefore chose a range in sma and companion masses where our survey sensitivity is good. For the sake of comparison with previous studies we used a uniform distribution in planetary frequency space as a prior $P(f)=1$ :

The median of the resulting posteriors, as well as the 68.27\% confidence intervals (equivalent to $\pm 1\sigma$) and 95.45\% confidence intervals (equivalent to $\pm 2\sigma$) are shown in Table~\ref{tab:frequency}.\\

\begin{table}
\caption{Derived occurrence rates for companions within various mass ranges around our sample of B stars. }
\label{tab:frequency}
\centering
\begin{tabular}{cccc}
\hline\hline
Occurrence :     &  Median &68\% confidence& 95\% confidence \\
\hline
Exoplanet$^1$  &   7.7\%        &  [3.3--14.6]\% &  [1.1--24.3]\%  \\
ELBD    $^2$     &    11.0\%       &   [5.8--18.4]\% & [2.6--28.0]\%   \\
Substellar$^3$  &    10.5\%       &[5.5--17.5]\%     & [2.5--26.7]\%   \\ \hline
\end{tabular}
\tablefoot{$^1$: $<$13\Mjup; $^2$: $<$30\Mjup; $^3$: $<$73.3 \Mjup}
\end{table}

\subsection{Comparison with the frequency of companions found around solar-type stars}
We use the results from \citet{Vigan.2021} SHINE-F150 survey as a comparison to explore whether the companion population is different around B-stars and around Sun-like stars. 

To account for the observational biases of the SHINE-F150 survey that attributed very high observational priority to some systems where the presence of planets was already known or suspected, we use the statistical weight from Table 1 of \citet{Vigan.2021} to each F150-detection. This attributes to each detection a weight ($\leq 1$) that corresponds to the probability that such stellar system would have been observed if the survey had been unbiased and observations carried out only on the basis of the survey-defined a priori merit function. Following this approach we attribute 3.4 detections of stellar systems with at least one {\it exoplanet} companion to SHINE F150 (accounting for HIP65426, \bpic, HR8799, HD95086, 51 Eri), 4.96 with at least one {\it ELBD} (adding HIP107412\,B, HIP\,78530\,B, GSC8047-0232\,B and AB Pic B) and  8.56 with at least one {\it substellar companion} (adding HIP64892\,B, $\eta$ Tel\,B, CD-35\,2722\,B, Pz Tel\,B ). We then used the same bayesian statistical framework described above to derive the posterior distribution on the fraction of systems hosting at least one companion from the SHINE-F150 survey. One difference is that we only had access to the average survey sensitivity and not the individual detection maps for the SHINE-F150 survey, and therefore set $fp_j$ to this average sensitivity in Equation~\ref{eq:likelyhood} for every star within this survey.

\begin{figure}[t!]
    \centering
    \includegraphics[width=\linewidth]{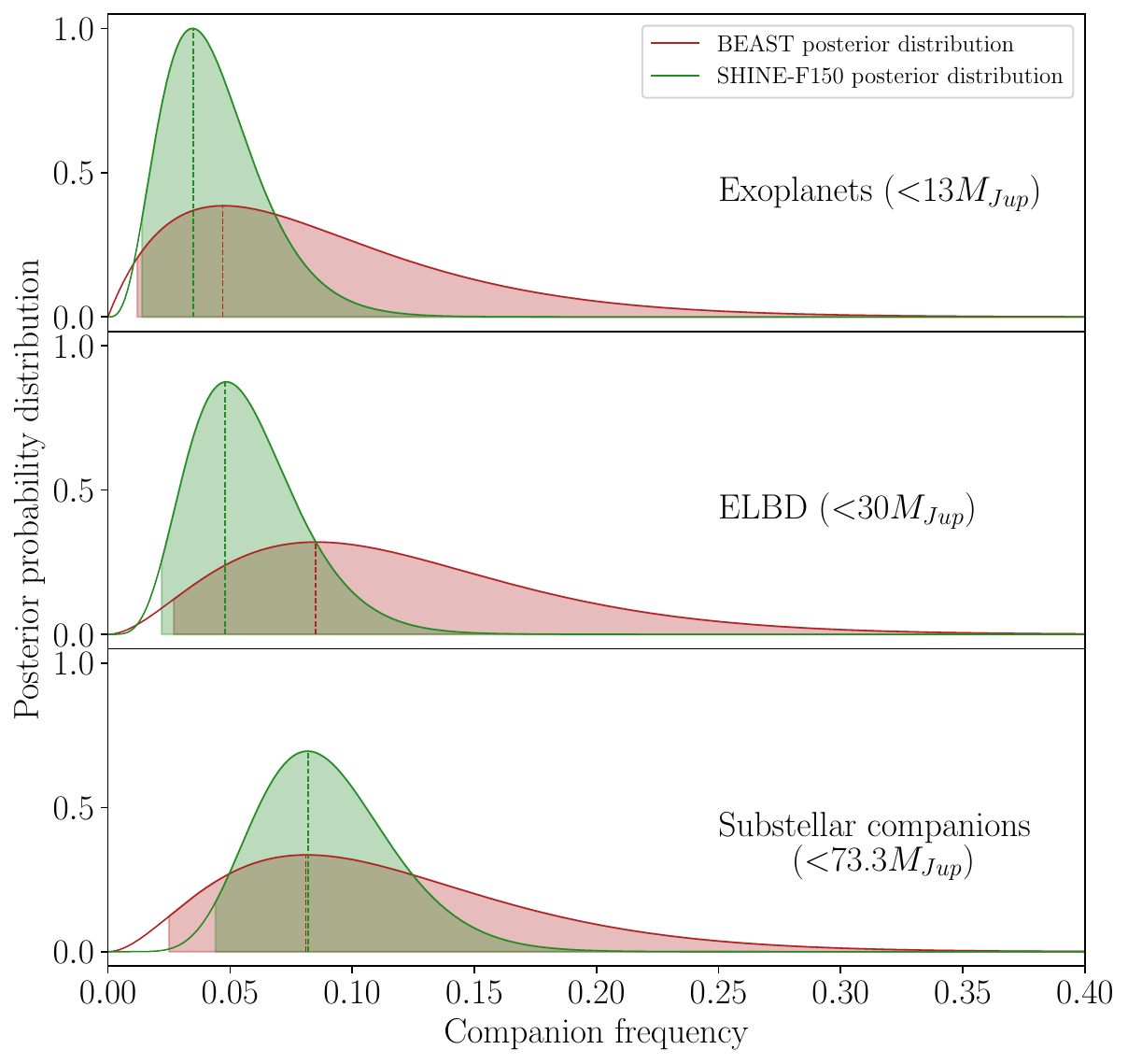}
    \caption{Comparison of the normalised posterior distribution for the frequency of systems with at least one companion below a given mass, between the SHINE-F150 survey targeting mostly solar-type stars and the BEAST survey targeting only B stars.The shaded areas highlight the 95\% confidence interval for each distribution and the vertical dashed line highlight the posterior maximum. }
    \label{fig:BEAST_F150_comp}
\end{figure}

As can be seen in Figure~\ref{fig:BEAST_F150_comp}, there is no statistically significant difference in the occurrence rate of either exoplanet, ELBD or brown dwarf companions. ELBDs might be more common around B stars, but the significance is below 2$\sigma$ and this possible trend needs to be confirmed by the study of larger samples. In fact the main difference in the companion properties observed in the BEAST sample is their very 
large separation from their host star which is replicated by none of the exoplanet or ELBD within the \citet{Vigan.2021}. However the YSES survey \citep{Bohn.2020a}, targeting solar-type stars located in Sco-Cen (as BEAST stars) shows that these lower mass stars also harbour exoplanets and ELBDs at very large separations. It is unclear if the absence of such very large separation companions in the SHINE-F150 is a random effect of small number statistics or is caused by some physical underlying processes. 

\subsection{Comparison to population synthesis models}
 
In order to discuss the formation of observed planets, it is helpful to compare them with synthetic populations that result from population synthesis calculations.
Such comparisons, however, are limited, because most population synthesis studies of the disk instability (DI) scenario are still lacking some physical processes and are typically focused on the planetary population around solar-type stars.

A recent development in population synthesis models in the DI model is the disc instability population synthesis project DIPSY (\citealt{2021A&A...645A..43S,2023A&A...669A..31S}, Schib~et~al.,2024a,2024b and in~prep).
The authors perform a population synthesis of objects formed through disc fragmentation, including the formation of the star-and-disc system by infall from the molecular cloud core.
They study a parameter space in final stellar mass from 0.05 to 5 \msun~ and find that about 10\% of the systems in the stellar mass range above 2.1 \msun~ have at least one surviving companion.
Among these systems, results show that a bit less than half would have been excluded from the BEAST survey because they have a stellar mass companion at intermediate separations (see Sec.~\ref{sec:sample}).
Based on preliminary results, the fraction of systems with surviving companions that would have been selected is $\sim6$\%.
This number  is in rather good agreement with the companion frequencies given above. Many of these companions also have semi-major axes of hundreds of astronomical units.

 However, the companions in systems compatible with the BEAST host star sample have masses from 1.5~\mjup to 320~\mjup, and most of them are intermediate or high mass brown dwarfs : 61~\% have masses between 30~\mjup and 73.3 ~\mjup.
The rest is divided into ELBD (34\%) and low mass stars (5\%).
Among the systems that do have a ELBD, 64\% contain an additional companion more massive than 30~\mjup.
Only 15~\% of the ELBD (corresponding to 5\% of the overall companion population) have no other companion in the system, so the probability to find two such systems and no other substellar companion as observed within the BEAST survey would be very small  according to these statistics from DIPSY.
It would be even more unlikely if instead considering the exoplanet subset (about one third of the ELBD). 

 At the moment it seems that the companions detected by BEAST are not the most typical outcomes of the DI model. However, it is possible that DIPSY over-estimates the final masses. This could be a result of various reasons such as: an overestimation of gas accretion rate, the assumption of perfect merging due to collisions and ideal merging, and the lack of magnetic fields in the models. 
Also, we note that DIPSY only studied a stellar mass range up to 5 \msun~and it is still unclear to what degree the conclusions are applicable to even more massive stars.

The same conclusions can be reached for studies of disc fragmentation around solar-type stars, such as the study by \citet{GI_paper6}. These authors perform a population synthesis of disc fragmentation, including tidal downsizing and gravitational interaction between fragments.
They ignore gas accretion but use a higher initial fragment mass than Schib~et~al.
Their resulting population is again dominated by objects with masses around 80~\mjup.

However, recent simulations 
using state-of-the-art grid-based as well as particle-based 3D MHD
simulations with self-gravity have uncovered a new regime
for disk fragmentation, in which a small seed magnetic field is grown
naturally by tapping the energy of the gravitational field \citep[e.g.,][]{Riols2019,Deng2020}. This results in a dynamo mechanism that is more powerful than  MRI which rapidly leads to a dynamical important field. In the latter regime, known as 
"GI dynamo", disk fragmentation has different characteristics, where the forming clumps are over an order of magnitude smaller in mass 
relative to conventional DI. In addition, in this case gas accretion is significantly less efficient 
due to the effect of magnetic pressure \citep{Deng2021,Kubli2023}. 
So far, this new fragmentation regime has only been explored 
for solar-type stars, where the characteristic masses were found to be 
in the regime of Super-Earth and mini-Neptunes. If the mass ratio  between star and clumps is constant in this scenario, one would expect that the inferred distribution of planetary masses in the case of B stars picks at about one Jupiter mass, so in this case lower masses than the observed 13-15\mjup BEAST companions. Even if this simple scaling does not quantitatively match the observed masses of the companions, it goes qualitatively in the right direction, by lowering the mass of the companions formed through disk fragmentation. This provides strong incentive to investigate GI dynamo in the intrinsically more massive disks expected around B stars in future research.

In core accretion, massive companions cannot form at large separation in-situ due to the long core growth time scale \citep{Pollack1996}.
 The accretion of pebbles may alleviate this problem \citep{ormelklahr2010a}, since pebbles can be accreted more efficiently due to the stronger gas drag.
However, when both accretion of planetesimals and pebbles are considered, the runaway gas accretion may still be delayed by late accretion of planetesimals \citep{2023A&A...674A.144K}.

It is, however, possible for massive objects to form closer to the star and be scattered to a large separation later.
This scenario should lead to the formation of wide companions only in a minority of systems.
For example the NG76 (100 embryos) population from the New Generation Planetary Population Synthesis \citep{emsenhuber21} features two companions above 1~\mjup outside of 100~au in 1000 systems, the 50 embryo population NG75 only has one.
However, these populations were conducted for 1~\msun\ stars.
A variant population with the same parameters as NG75, except with a stellar mass of 1.5~\msun\ exhibits five wide massive companions.
This hints towards a larger fraction of such objects around more massive stars. 
However, the trend of more giants with increasing stellar mass is unlikely to continue to such massive hosts.
Instead, the lifetimes of discs around stars more massive than $\sim3~\msun$ are expected to decline steeply with host mass because the internal photoevaporation rate has a strong dependence on the stellar mass \citep{2009ApJ...690.1539G,2021ApJ...910...51K}.
Lifetimes well below 1~Myr will be too short for core accretion to form giant planets.

\subsection{Gravitational capture scenario}
The serious challenges for giant planet formation around B stars led \citet[][hereafter \citetalias{parker22}]{parker22} to claim that the BEAST ELBD were so unlikely to have been formed around their current massive hosts that it was more likely they formed around less massive stars (or even as isolated objects) and were later gravitationally captured. 
 

In order to investigate this hypothesis, they set up $N_{\text{sim}}=20$ simulations where a circle of radius $\ell = 1$ pc is filled with 1000 stars following a box-fractal spatial distribution that mimics the initial state of new-born associations \citep[see][]{daffern22}. Given the random sampling of the stellar initial mass function, the number of B-type stars in the cubes, $N_B$, varies between 44 and 65. One half of non-B stars ($M_*<2.4 M_\odot$) comes provided with a 1~\mjup planet in a circular orbit at $a=30$ au. N-body integration allows following the dynamical evolution of each simulation for 10 million years.

A total of 18 planets is captured in the simulations; the frequency of captured planets per star is therefore $f_p \sim 18/(N_B \cdot N_\text{sim}) \approx 0.016$, where we assumed a constant $N_{B} = 55$ for simplicity. Therefore, the probability \pcapt of finding at least two captured planets in the 37-star sample considered in this work can be estimated via a binomial distribution:

\begin{equation}
    \pcapt= 1 - \sum_{k=0}^1{\binom{n}{k}f_p^k(1-f_p)^{n-k}} \approx 0.128.
\end{equation}

\noindent where $\binom{n}{k} = n! / [k! (n-k!)]$ is the binomial factor and $n=37$. Though this probability is relatively small it is probably overestimated. Indeed, the median stellar density, a crucial feature of these simulations, is of the order of $\sim 10^4~\text{pc}^{-3}$, a value that, in contrast with observations, should result in the presence of dense and gravitationally bound clusters at the age of Sco-Cen \citep{gieles10}. This value does not represent the mean density within the simulated volume ($n \sim 240~\text{pc}^{-3}$), but rather the local density of the fractal regions. However, we argue that such a high value of $n$ is not adequate to describe Scorpius-Centaurus, that has never been in such a compact configuration in the past \citep{galli18}. In addition to this, the assumed frequency of Jupiter-sized planets in wide orbits ($f_c = 0.5$) is unrealistically high if compared to observational evidence \citep{bowler16,nielsen19,Vigan.2021}. Therefore, we expect the value of \pcapt to be significantly lower.\\

We alternatively try to quantify an upper limit of the frequency of gravitational capture only based on empirical data. We assess it by summing the probability of two distinct such scenario: 1) the capture of a Sco-Cen object, either previously bound to a star or flee-floating; 2) as above, but with the object belonging to the population of old interlopers that happen to share the same physical volume with the association.

The timescale for a close encounter between two stars in a given stellar environment can be estimated as:
\begin{equation}\label{eq:encounter_timescale}
    \tau_{\mathrm{enc}} \approx 3.3 \times10^7 \text{ yr} \left (\frac{100~\text{pc}^{-3}}{n} \right ) \left ( \frac{v_\infty}{1~\kms} \right ) \left ( \frac{10^3~\text{au}}{\rmin} \right ) \left ( \frac{M_\odot}{m_t} \right ),
\end{equation}
where $n$ is the stellar density, $v_\infty$ the stellar mean relative speed at infinity, \rmin the encounter distance, and $m_t$ the total mass participating in the close encounter \citep{malmberg07}. The last factor $(M_\odot/m_t$) accounts for gravitational focusing -- i.e., the deflection of stellar trajectories caused by their mutual attraction --, a process largely extending the effective stellar cross section.

We adopt the stellar density of the relatively compact Lower Scorpius \citep{Squicciarini.2022} group ($n \approx 1~\text{pc}^{-3}$) as an  estimate for a the initial state of a typical star-forming subgroup within Sco-Cen; assuming the present-day velocity dispersion of the Upper Scorpius subgroup \citep[$v=3.0$ km s$^{-1}$,][]{squicciarini21}, a typical $m_l=5 M_\odot$, and $\rmin = 1000$ au, Eq.~\ref{eq:encounter_timescale} yields:
\begin{equation}
    \tau_{\mathrm{enc},\,\text{SC}} \approx 1980~\text{Myr}.
\end{equation}

An estimate for the frequency of substellar objects can be obtained based on the results from the SHINE survey \citet{Vigan.2021}: the fraction of $1-75$ \mjup companions with a semimajor axis $\in [5,300]$ au was found to be $5.8^{+4.7}_{2.8} \%$ and $12.6^{+12.9}_{7.1} \%$ around FGK and M hosts, respectively. We assume a constant fraction $f_{c} = 10\%$.\footnote{The value is a comfortable upper limit since the BEAST survey is virtually insensitive to $1 \text{\mjup} \lesssim M \lesssim 2$ \mjup objects and only weakly sensitive to the regime $2 \text{\mjup} \lesssim M \lesssim 3$ \mjup (Fig.~\ref{fig:sample_detmaps}).}
 As regards free-floating objects ($5 \text{\mjup} \lesssim M \lesssim 75$ \mjup), we estimate their fraction with respect to the total Sco-Cen population as $f_{\text{f}} \sim 20\%$ \citep{miret-roig22}. In order to set an upper limit on the likelihood of this scenario, we set the efficiency $\eta_{\text{capt}}$ of the capture process, that is the probability that a close encounter with a planet-bearing star or a free-floating object results in a capture event to 100\%. Then the expected number of captured objects per close encounter $N_{p, SC}$ after 20 Myr can be computed as:
\begin{equation}\label{eq:captured_objects_per_star}
N_{p, \text{SC}} \approx \frac{20~\text{Myr}}{460~\text{Myr}} \cdot (f_{c} \cdot (1-f_{\text{f}}) + f_{\text{f}}) = 0.0028,
\end{equation}
As $N_{p, \text{SC}} \ll 1$, we assume the capture probability for a single star to be $f_{p, SC} = N_{p, \text{SC}}$. Considered over the whole survey,
\begin{equation}\label{eq:captured_objects_survey}
    p_{capt, \text{SC}}= 1 - \sum_{k=0}^1{\binom{37}{k}f_{p,SC}^k(1-f_{p,SC})^{37-k}} \approx 0.005.
\end{equation}

In a similar fashion, we estimate the capture probability of a non-Sco-Cen object to be $p_{capt, \text{f}} = 3 \cdot 10^{-5}$ (Appendix~\ref{sec:capture_field}). The total upper limit on the probability of the capture scenario is therefore $\pcapt = p_{capt, SC} + p_{capt, f} \approx p_{capt, SC} = 0.005$, corresponding to a confidence level of 2.8$\sigma$ on the hypothesis that the two substellar companions are \emph{not} captured. We stress that the probability \pcapt\ depends linearly on the capture efficiency, and while we set \pcapt\ to 1 to assess an upper limit of the capture scenario, more realistic values of \pcapt, accounting for the possibility of ejection or simply that the companion stays within its birth system, would result in even lower capture probabilities. While these numbers are not sufficient to completely rule out the capture scenario, they indicate that this mechanism is unlikely to account for the population of B-star companions unveiled by BEAST.


\section{Conclusion}

 We discussed here several plausible scenarios as pathways to form ELBDs around very massive B stars, namely core fragmentation, gravitational instability in a disk, gravitational capture of free floating or ejected planets, or core accretion. Although all of them might lead to the formation of planetary mass objects, none of them provides a straightforward explanation for the population of ELBD we observe around the stars of the BEAST sample. In the case of gravitational instability, companions are usually more massive and ELBD make up only the tail of the distribution of companions. It is therefore unlikely to observe two ELBDs and none of the higher-mass brown dwarfs that would be more natural outcomes of gravitational instability. However, including magnetic field in gravitational instability models might improve the agreement with our observations. Indeed, when a magnetic field is present in the disk, and is dynamically important, the character of gravitational instability might be significantly altered, leading to lower mass companions.
 
 The same issue arises for core fragmentation, that typically forms stellar multiple systems and occasionally brown dwarfs/exoplanet systems \citep[eg. 2M1207B][]{Chauvin.2004}, and is complemented by the very unusual mass ratio of the observed ELBD, almost two orders of magnitude lower than the lowest observed in multiple systems \citep[see][]{Gratton.2023}. Core accretion is the typical planet formation scenario that is put forward to explain the formation of Solar System planets, and the mass ratio as well as the stellar insulation at the observed orbital separation for the BEAST ELBD are of the correct order of magnitude. However, the timescale for the formation of giant planets by core accretion are larger than the quite fast timescale of disk photo-evaporation around a star as hot and luminous as BEAST B stars \citep{2009ApJ...690.1539G,2021ApJ...910...51K}. 
 We also investigated gravitational capture as a distinct pathway to account for the presence of ELBD around the stars BEAST survey sample. Even when favourable values are used for the various parameters involved in this scenario, we find that gravitational capture is unlikely to account for the observed ELBD populations.
 
 It therefore turns out that none of the scenarios we explore readily explains the population of planetary mass companions around B stars of our sample. The significant uncertainty affecting each of these scenarios prevents us from identifying which ones could be the least unlikely. At the same time the physical complexity behind each of these processes means that current models and simulations are still limited in their predictive power, and further progress on the theoretical side must
 be ultimately guided by observations, such as those presented in this work, probing uncharted regions of parameter  space for exoplanets and other sub-stellar objects.

\label{sec:form}

\begin{acknowledgements}
This project has received funding from the European Research Council (ERC) under the European Union's Horizon 2020 research and innovation programme (COBREX; grant agreement n° 885593). 
SPHERE is an instrument designed and built by a consortium
consisting of IPAG (Grenoble, France), MPIA (Heidelberg, Germany),
LAM (Marseille, France), LESIA (Paris, France), Laboratoire Lagrange
(Nice, France), INAF - Osservatorio di Padova (Italy), Observatoire de
Genève (Switzerland), ETH Zürich (Switzerland), NOVA (Netherlands), ONERA
(France) and ASTRON (Netherlands) in collaboration with ESO. SPHERE
was funded by ESO, with additional contributions from CNRS (France),
MPIA (Germany), INAF (Italy), FINES (Switzerland) and NOVA (Netherlands).
SPHERE also received funding from the European Commission Sixth and Seventh
Framework Programmes as part of the Optical Infrared Coordination Network
for Astronomy (OPTICON) under grant number RII3-Ct-2004-001566 for
FP6 (2004-2008), grant number 226604 for FP7 (2009-2012) and grant number
312430 for FP7 (2013-2016). 
This work has made use of the High Contrast Data Centre, jointly operated by OSUG/IPAG (Grenoble), PYTHEAS/LAM/CeSAM (Marseille), OCA/Lagrange (Nice), Observatoire de Paris/LESIA (Paris), and Observatoire de Lyon/CRAL, and is supported by a grant from Labex OSUG@2020 (Investissements d'avenir - ANR10 LABX56).
Th.H.\ and G.-D.M.\ acknowledge support from the European Research Council under the Horizon 2020 Framework Program via the ERC Advanced Grant Origins 83 24 28.
G.-D.M.\ also acknowledges the support of the DFG priority program SPP 1992 ``Exploring the Diversity of Extrasolar Planets'' under grant MA~9185/1-1.
R.G.\ and S.D.\ acknowledge the support of PRIN-INAF 2019 Planetary Systems At Early Ages (PLATEA). O.S. acknowledges support from the Swiss National Science Foundation under grant 200021\_204847 ``PlanetsInTime''.
This research has made use of the SIMBAD database and VizieR catalogue access tool, operated at CDS, Strasbourg, France. This work is supported by the French National Research Agency in the framework of the Investissements d’Avenir program (ANR-15-IDEX-02), through the funding of the "Origin of Life" project of the Univ. Grenoble-Alpes.
This work was supported by the Action Spécifique Haute Résolution Angulaire (ASHRA) of CNRS/INSU co-funded by CNES.
This research has made use of data obtained from or tools provided by the portal \href{https://exoplanet.eu}{exoplanet.eu} of The Extrasolar Planets Encyclopaedia.
\end{acknowledgements}

\bibliographystyle{aa}
\bibliography{main.bib}

\begin{thebibliography}{94}
\expandafter\ifx\csname natexlab\endcsname\relax\def\natexlab#1{#1}\fi

\bibitem[{{Alecian} {et~al.}(2013){Alecian}, {Wade}, {Catala}, {Grunhut},
  {Landstreet}, {B{\"o}hm}, {Folsom}, \& {Marsden}}]{alecian13}
{Alecian}, E., {Wade}, G.~A., {Catala}, C., {et~al.} 2013, \mnras, 429, 1027

\bibitem[{{Baraffe} {et~al.}(2003){Baraffe}, {Chabrier}, {Barman}, {Allard}, \&
  {Hauschildt}}]{Baraffe.2003}
{Baraffe}, I., {Chabrier}, G., {Barman}, T.~S., {Allard}, F., \& {Hauschildt},
  P.~H. 2003, \aap, 402, 701

\bibitem[{{Beuzit} {et~al.}(2019){Beuzit}, {Vigan}, {Mouillet}, {Dohlen},
  {Gratton}, {Boccaletti}, {Sauvage}, {Schmid}, {Langlois}, {Petit},
  {Baruffolo}, {Feldt}, {Milli}, {Wahhaj}, {Abe}, {Anselmi}, {Antichi},
  {Barette}, {Baudrand}, {Baudoz}, {Bazzon}, {Bernardi}, {Blanchard}, {Brast},
  {Bruno}, {Buey}, {Carbillet}, {Carle}, {Cascone}, {Chapron}, {Charton},
  {Chauvin}, {Claudi}, {Costille}, {De Caprio}, {de Boer}, {Delboulb{\'e}},
  {Desidera}, {Dominik}, {Downing}, {Dupuis}, {Fabron}, {Fantinel}, {Farisato},
  {Feautrier}, {Fedrigo}, {Fusco}, {Gigan}, {Ginski}, {Girard}, {Giro},
  {Gisler}, {Gluck}, {Gry}, {Henning}, {Hubin}, {Hugot}, {Incorvaia}, {Jaquet},
  {Kasper}, {Lagadec}, {Lagrange}, {Le Coroller}, {Le Mignant}, {Le Ruyet},
  {Lessio}, {Lizon}, {Llored}, {Lundin}, {Madec}, {Magnard}, {Marteaud},
  {Martinez}, {Maurel}, {M{\'e}nard}, {Mesa}, {M{\"o}ller-Nilsson}, {Moulin},
  {Moutou}, {Orign{\'e}}, {Parisot}, {Pavlov}, {Perret}, {Pragt}, {Puget},
  {Rabou}, {Ramos}, {Reess}, {Rigal}, {Rochat}, {Roelfsema}, {Rousset}, {Roux},
  {Saisse}, {Salasnich}, {Santambrogio}, {Scuderi}, {Segransan}, {Sevin},
  {Siebenmorgen}, {Soenke}, {Stadler}, {Suarez}, {Tiph{\`e}ne}, {Turatto},
  {Udry}, {Vakili}, {Waters}, {Weber}, {Wildi}, {Zins}, \&
  {Zurlo}}]{Beuzit.2019}
{Beuzit}, J.~L., {Vigan}, A., {Mouillet}, D., {et~al.} 2019, \aap, 631, A155

\bibitem[{{Blunt} {et~al.}(2023){Blunt}, {Balmer}, {Wang}, {Lacour}, {Petrus},
  {Bourdarot}, {Kammerer}, {Pourr{\'e}}, {Rickman}, {Shangguan},
  {Winterhalder}, {Abuter}, {Amorim}, {Asensio-Torres}, {Benisty}, {Berger},
  {Beust}, {Boccaletti}, {Bohn}, {Bonnefoy}, {Bonnet}, {Brandner},
  {Cantalloube}, {Caselli}, {Charnay}, {Chauvin}, {Chavez}, {Choquet},
  {Christiaens}, {Cl{\'e}net}, {Du Foresto}, {Cridland}, {Dembet}, {Drescher},
  {Duvert}, {Eckart}, {Eisenhauer}, {Feuchtgruber}, {Garcia}, {Garcia Lopez},
  {Gendron}, {Genzel}, {Gillessen}, {Girard}, {Haubois}, {Hei{\ss}el},
  {Henning}, {Hinkley}, {Hippler}, {Horrobin}, {Houll{\'e}}, {Hubert}, {Jocou},
  {Keppler}, {Kervella}, {Kreidberg}, {Lagrange}, {Lapeyr{\`e}re}, {Le
  Bouquin}, {L{\'e}na}, {Lutz}, {Maire}, {Mang}, {Marleau}, {M{\'e}rand},
  {Molli{\`e}re}, {Monnier}, {Mordasini}, {Mouillet}, {Nasedkin}, {Nowak},
  {Ott}, {Otten}, {Paladini}, {Paumard}, {Perraut}, {Perrin}, {Pfuhl}, {Pueyo},
  {Rameau}, {Rodet}, {Rustamkulov}, {Shimizu}, {Sing}, {Stolker},
  {Straubmeier}, {Sturm}, {Tacconi}, {van Dishoeck}, {Vigan}, {Vincent},
  {Ward-Duong}, {Widmann}, {Wieprecht}, {Wiezorrek}, {Woillez}, {Yazici},
  {Young}, \& {Exogravity Collaboration}}]{Blunt_hip65426b}
{Blunt}, S., {Balmer}, W.~O., {Wang}, J.~J., {et~al.} 2023, \aj, 166, 257

\bibitem[{{Bohn} {et~al.}(2021){Bohn}, {Ginski}, {Kenworthy}, {Mamajek},
  {Pecaut}, {Mugrauer}, {Vogt}, {Adam}, {Meshkat}, {Reggiani}, \&
  {Snik}}]{Bohn.2021}
{Bohn}, A.~J., {Ginski}, C., {Kenworthy}, M.~A., {et~al.} 2021, \aap, 648, A73

\bibitem[{{Bohn} {et~al.}(2020){Bohn}, {Kenworthy}, {Ginski}, {Manara},
  {Pecaut}, {de Boer}, {Keller}, {Mamajek}, {Meshkat}, {Reggiani}, {Todorov},
  \& {Snik}}]{Bohn.2020a}
{Bohn}, A.~J., {Kenworthy}, M.~A., {Ginski}, C., {et~al.} 2020, \mnras, 492,
  431

\bibitem[{{Bonavita}(2020)}]{Bonavita.2020}
{Bonavita}, M. 2020, {Exo-DMC: Exoplanet Detection Map Calculator},
  Astrophysics Source Code Library, record ascl:2010.008

\bibitem[{{Bonnefoy} {et~al.}(2018){Bonnefoy}, {Perraut}, {Lagrange},
  {Delorme}, {Vigan}, {Line}, {Rodet}, {Ginski}, {Mourard}, {Marleau},
  {Samland}, {Tremblin}, {Ligi}, {Cantalloube}, {Molli{\`e}re}, {Charnay},
  {Kuzuhara}, {Janson}, {Morley}, {Homeier}, {D'Orazi}, {Klahr}, {Mordasini},
  {Lavie}, {Baudino}, {Beust}, {Peretti}, {Musso Bartucci}, {Mesa},
  {B{\'e}zard}, {Boccaletti}, {Galicher}, {Hagelberg}, {Desidera}, {Biller},
  {Maire}, {Allard}, {Borgniet}, {Lannier}, {Meunier}, {Desort}, {Alecian},
  {Chauvin}, {Langlois}, {Henning}, {Mugnier}, {Mouillet}, {Gratton}, {Brandt},
  {Mc Elwain}, {Beuzit}, {Tamura}, {Hori}, {Brandner}, {Buenzli}, {Cheetham},
  {Cudel}, {Feldt}, {Kasper}, {Keppler}, {Kopytova}, {Meyer}, {Perrot},
  {Rouan}, {Salter}, {Schmidt}, {Sissa}, {Zurlo}, {Wildi}, {Blanchard}, {De
  Caprio}, {Delboulb{\'e}}, {Maurel}, {Moulin}, {Pavlov}, {Rabou}, {Ramos},
  {Roelfsema}, {Rousset}, {Stadler}, {Rigal}, \& {Weber}}]{Bonnefoy_GJ504_2018}
{Bonnefoy}, M., {Perraut}, K., {Lagrange}, A.~M., {et~al.} 2018, \aap, 618, A63

\bibitem[{{Bowler}(2016)}]{bowler16}
{Bowler}, B.~P. 2016, \pasp, 128, 102001

\bibitem[{{Brown-Sevilla} {et~al.}(2023){Brown-Sevilla}, {Maire},
  {Molli{\`e}re}, {Samland}, {Feldt}, {Brandner}, {Henning}, {Gratton},
  {Janson}, {Stolker}, {Hagelberg}, {Zurlo}, {Cantalloube}, {Boccaletti},
  {Bonnefoy}, {Chauvin}, {Desidera}, {D'Orazi}, {Lagrange}, {Langlois},
  {Menard}, {Mesa}, {Meyer}, {Pavlov}, {Petit}, {Rochat}, {Rouan}, {Schmidt},
  {Vigan}, \& {Weber}}]{51Erib_BS_2023}
{Brown-Sevilla}, S.~B., {Maire}, A.~L., {Molli{\`e}re}, P., {et~al.} 2023,
  \aap, 673, A98

\bibitem[{{Cantalloube} {et~al.}(2019){Cantalloube}, {Dohlen}, {Milli},
  {Brandner}, \& {Vigan}}]{Cantalloube_2019}
{Cantalloube}, F., {Dohlen}, K., {Milli}, J., {Brandner}, W., \& {Vigan}, A.
  2019, The Messenger, 176, 25

\bibitem[{{Carson} {et~al.}(2012){Carson}, {Thalmann}, {Janson}, {Kozakis},
  {Wong}, {Goto}, {Henning}, {Brandner}, {Biller}, {Bonnefoy}, {Feldt},
  {McElwain}, {Kandori}, {Tamura}, \& {SEEDS Team}}]{Carson.2012}
{Carson}, J., {Thalmann}, C., {Janson}, M., {et~al.} 2012, in American
  Astronomical Society Meeting Abstracts, Vol. 219, American Astronomical
  Society Meeting Abstracts \#219, 432.02

\bibitem[{{Chabrier} {et~al.}(2000){Chabrier}, {Baraffe}, {Allard}, \&
  {Hauschildt}}]{Chabrier.2000}
{Chabrier}, G., {Baraffe}, I., {Allard}, F., \& {Hauschildt}, P. 2000, \apj,
  542, 464

\bibitem[{{Chauvin} {et~al.}(2004){Chauvin}, {Lagrange}, {Dumas}, {Zuckerman},
  {Mouillet}, {Song}, {Beuzit}, \& {Lowrance}}]{Chauvin.2004}
{Chauvin}, G., {Lagrange}, A.~M., {Dumas}, C., {et~al.} 2004, \aap, 425, L29

\bibitem[{{Chini} {et~al.}(2012){Chini}, {Hoffmeister}, {Nasseri}, {Stahl}, \&
  {Zinnecker}}]{chini12}
{Chini}, R., {Hoffmeister}, V.~H., {Nasseri}, A., {Stahl}, O., \& {Zinnecker},
  H. 2012, \mnras, 424, 1925

\bibitem[{{Choi} {et~al.}(2016){Choi}, {Dotter}, {Conroy}, {Cantiello},
  {Paxton}, \& {Johnson}}]{choi16}
{Choi}, J., {Dotter}, A., {Conroy}, C., {et~al.} 2016, \apj, 823, 102

\bibitem[{{Chomez} {et~al.}(2023{\natexlab{a}}){Chomez}, {Lagrange}, {Delorme},
  {Langlois}, {Chauvin}, {Flasseur}, {Dallant}, {Philipot}, {Bergeon},
  {Albert}, {Meunier}, \& {Rubini}}]{Chomez.2023a}
{Chomez}, A., {Lagrange}, A.~M., {Delorme}, P., {et~al.} 2023{\natexlab{a}},
  \aap, 675, A205

\bibitem[{{Chomez} {et~al.}(2023{\natexlab{b}}){Chomez}, {Squicciarini},
  {Lagrange}, {Delorme}, {Viswanath}, {Janson}, {Flasseur}, {Chauvin},
  {Langlois}, {Rubini}, {Bergeon}, {Albert}, {Bonnefoy}, {Desidera}, {Engler},
  {Gratton}, {Henning}, {Mamajek}, {Marleau}, {Meyer}, {Reffert}, {Ringqvist},
  \& {Samland}}]{Chomez.2023b}
{Chomez}, A., {Squicciarini}, V., {Lagrange}, A.~M., {et~al.}
  2023{\natexlab{b}}, \aap, 676, L10

\bibitem[{{Claudi} {et~al.}(2008){Claudi}, {Turatto}, {Gratton}, {Antichi},
  {Bonavita}, {Bruno}, {Cascone}, {De Caprio}, {Desidera}, {Giro}, {Mesa},
  {Scuderi}, {Dohlen}, {Beuzit}, \& {Puget}}]{Claudi_ifs}
{Claudi}, R.~U., {Turatto}, M., {Gratton}, R.~G., {et~al.} 2008, in Society of
  Photo-Optical Instrumentation Engineers (SPIE) Conference Series, Vol. 7014,
  Ground-based and Airborne Instrumentation for Astronomy II, ed. I.~S.
  {McLean} \& M.~M. {Casali}, 70143E

\bibitem[{{Cutri} {et~al.}(2003){Cutri}, {Skrutskie}, {van Dyk}, {Beichman},
  {Carpenter}, {Chester}, {Cambresy}, {Evans}, {Fowler}, {Gizis}, {Howard},
  {Huchra}, {Jarrett}, {Kopan}, {Kirkpatrick}, {Light}, {Marsh}, {McCallon},
  {Schneider}, {Stiening}, {Sykes}, {Weinberg}, {Wheaton}, {Wheelock}, \&
  {Zacarias}}]{2MASS_MAG}
{Cutri}, R.~M., {Skrutskie}, M.~F., {van Dyk}, S., {et~al.} 2003, VizieR Online
  Data Catalog, II/246

\bibitem[{{Daffern-Powell} {et~al.}(2022){Daffern-Powell}, {Parker}, \&
  {Quanz}}]{daffern22}
{Daffern-Powell}, E.~C., {Parker}, R.~J., \& {Quanz}, S.~P. 2022, \mnras, 514,
  920

\bibitem[{{Damiani} {et~al.}(2019){Damiani}, {Prisinzano}, {Pillitteri},
  {Micela}, \& {Sciortino}}]{damiani19}
{Damiani}, F., {Prisinzano}, L., {Pillitteri}, I., {Micela}, G., \&
  {Sciortino}, S. 2019, \aap, 623, A112

\bibitem[{{Delorme} {et~al.}(2017){Delorme}, {Meunier}, {Albert}, {Lagadec},
  {Le Coroller}, {Galicher}, {Mouillet}, {Boccaletti}, {Mesa}, {Meunier},
  {Beuzit}, {Lagrange}, {Chauvin}, {Sapone}, {Langlois}, {Maire},
  {Montarg{\`e}s}, {Gratton}, {Vigan}, \& {Surace}}]{Delorme_sphereDC}
{Delorme}, P., {Meunier}, N., {Albert}, D., {et~al.} 2017, in SF2A-2017:
  Proceedings of the Annual meeting of the French Society of Astronomy and
  Astrophysics, ed. C.~{Reyl{\'e}}, P.~{Di Matteo}, F.~{Herpin}, E.~{Lagadec},
  A.~{Lan{\c{c}}on}, Z.~{Meliani}, \& F.~{Royer}, Di

\bibitem[{{Deng} {et~al.}(2021){Deng}, {Mayer}, \& {Helled}}]{Deng2021}
{Deng}, H., {Mayer}, L., \& {Helled}, R. 2021, Nature Astronomy, 5, 967

\bibitem[{{Deng} {et~al.}(2020){Deng}, {Mayer}, \& {Latter}}]{Deng2020}
{Deng}, H., {Mayer}, L., \& {Latter}, H. 2020, \apj, 891, 154

\bibitem[{{Desgrange} {et~al.}(2022){Desgrange}, {Chauvin}, {Christiaens},
  {Cantalloube}, {Lefranc}, {Le Coroller}, {Rubini}, {Otten}, {Beust},
  {Bonavita}, {Delorme}, {Devinat}, {Gratton}, {Lagrange}, {Langlois}, {Mesa},
  {Milli}, {Szul{\'a}gyi}, {Nowak}, {Rodet}, {Rojo}, {Petrus}, {Janson},
  {Henning}, {Kral}, {van Holstein}, {M{\'e}nard}, {Beuzit}, {Biller},
  {Boccaletti}, {Bonnefoy}, {Brown}, {Costille}, {Delboulbe}, {Desidera},
  {D'Orazi}, {Feldt}, {Fusco}, {Galicher}, {Hagelberg}, {Lazzoni}, {Ligi},
  {Maire}, {Messina}, {Meyer}, {Potier}, {Ramos}, {Rouan}, {Schmidt}, {Vigan},
  \& {Zurlo}}]{Desgranges_hd95086b}
{Desgrange}, C., {Chauvin}, G., {Christiaens}, V., {et~al.} 2022, \aap, 664,
  A139

\bibitem[{{Dohlen} {et~al.}(2008){Dohlen}, {Langlois}, {Saisse}, {Hill},
  {Origne}, {Jacquet}, {Fabron}, {Blanc}, {Llored}, {Carle}, {Moutou}, {Vigan},
  {Boccaletti}, {Carbillet}, {Mouillet}, \& {Beuzit}}]{Dohlen_irdis}
{Dohlen}, K., {Langlois}, M., {Saisse}, M., {et~al.} 2008, in Society of
  Photo-Optical Instrumentation Engineers (SPIE) Conference Series, Vol. 7014,
  Ground-based and Airborne Instrumentation for Astronomy II, ed. I.~S.
  {McLean} \& M.~M. {Casali}, 70143L

\bibitem[{{Dotter}(2016)}]{dotter16}
{Dotter}, A. 2016, \apjs, 222, 8

\bibitem[{{Emsenhuber} {et~al.}(2021){Emsenhuber}, {Mordasini}, {Burn},
  {Alibert}, {Benz}, \& {Asphaug}}]{emsenhuber21}
{Emsenhuber}, A., {Mordasini}, C., {Burn}, R., {et~al.} 2021, \aap, 656, A70

\bibitem[{{Esplin} \& {Luhman}(2020)}]{esplin20}
{Esplin}, T.~L. \& {Luhman}, K.~L. 2020, \aj, 159, 282

\bibitem[{{Flasseur} {et~al.}(2018){Flasseur}, {Denis}, {Thi{\'e}baut}, \&
  {Langlois}}]{Flasseur_paco}
{Flasseur}, O., {Denis}, L., {Thi{\'e}baut}, {\'E}., \& {Langlois}, M. 2018,
  \aap, 618, A138

\bibitem[{{Flasseur} {et~al.}(2020){Flasseur}, {Denis}, {Thi{\'e}baut}, \&
  {Langlois}}]{Flasseur_asdi}
{Flasseur}, O., {Denis}, L., {Thi{\'e}baut}, {\'E}., \& {Langlois}, M. 2020,
  \aap, 637, A9

\bibitem[{Flasseur {et~al.}(2020)Flasseur, Denis, Thi{\'e}baut, \&
  Langlois}]{flasseur2020robustness}
Flasseur, O., Denis, L., Thi{\'e}baut, {\'E}., \& Langlois, M. 2020, Astronomy
  \& Astrophysics, 634, A2

\bibitem[{{Forgan} {et~al.}(2018){Forgan}, {Hall}, {Meru}, \&
  {Rice}}]{GI_paper6}
{Forgan}, D.~H., {Hall}, C., {Meru}, F., \& {Rice}, W.~K.~M. 2018, \mnras, 474,
  5036

\bibitem[{{Gagn{\'e}} {et~al.}(2018){Gagn{\'e}}, {Mamajek}, {Malo}, {Riedel},
  {Rodriguez}, {Lafreni{\`e}re}, {Faherty}, {Roy-Loubier}, {Pueyo}, {Robin}, \&
  {Doyon}}]{gagne18}
{Gagn{\'e}}, J., {Mamajek}, E.~E., {Malo}, L., {et~al.} 2018, \apj, 856, 23

\bibitem[{{Gaia Collaboration} {et~al.}(2018){Gaia Collaboration}, {Brown},
  {Vallenari}, {Prusti}, {de Bruijne}, {Babusiaux}, {Bailer-Jones}, {Biermann},
  {Evans}, {Eyer}, {Jansen}, {Jordi}, {Klioner}, {Lammers}, {Lindegren},
  {Luri}, {Mignard}, {Panem}, {Pourbaix}, {Randich}, {Sartoretti}, {Siddiqui},
  {Soubiran}, {van Leeuwen}, {Walton}, {Arenou}, {Bastian}, {Cropper},
  {Drimmel}, {Katz}, {Lattanzi}, {Bakker}, {Cacciari}, {Casta{\~n}eda},
  {Chaoul}, {Cheek}, {De Angeli}, {Fabricius}, {Guerra}, {Holl}, {Masana},
  {Messineo}, {Mowlavi}, {Nienartowicz}, {Panuzzo}, {Portell}, {Riello},
  {Seabroke}, {Tanga}, {Th{\'e}venin}, {Gracia-Abril}, {Comoretto},
  {Garcia-Reinaldos}, {Teyssier}, {Altmann}, {Andrae}, {Audard},
  {Bellas-Velidis}, {Benson}, {Berthier}, {Blomme}, {Burgess}, {Busso},
  {Carry}, {Cellino}, {Clementini}, {Clotet}, {Creevey}, {Davidson}, {De
  Ridder}, {Delchambre}, {Dell'Oro}, {Ducourant},
  {Fern{\'a}ndez-Hern{\'a}ndez}, {Fouesneau}, {Fr{\'e}mat}, {Galluccio},
  {Garc{\'\i}a-Torres}, {Gonz{\'a}lez-N{\'u}{\~n}ez}, {Gonz{\'a}lez-Vidal},
  {Gosset}, {Guy}, {Halbwachs}, {Hambly}, {Harrison}, {Hern{\'a}ndez},
  {Hestroffer}, {Hodgkin}, {Hutton}, {Jasniewicz}, {Jean-Antoine-Piccolo},
  {Jordan}, {Korn}, {Krone-Martins}, {Lanzafame}, {Lebzelter}, {L{\"o}ffler},
  {Manteiga}, {Marrese}, {Mart{\'\i}n-Fleitas}, {Moitinho}, {Mora}, {Muinonen},
  {Osinde}, {Pancino}, {Pauwels}, {Petit}, {Recio-Blanco}, {Richards},
  {Rimoldini}, {Robin}, {Sarro}, {Siopis}, {Smith}, {Sozzetti}, {S{\"u}veges},
  {Torra}, {van Reeven}, {Abbas}, {Abreu Aramburu}, {Accart}, {Aerts},
  {Altavilla}, {{\'A}lvarez}, {Alvarez}, {Alves}, {Anderson}, {Andrei},
  {Anglada Varela}, {Antiche}, {Antoja}, {Arcay}, {Astraatmadja}, {Bach},
  {Baker}, {Balaguer-N{\'u}{\~n}ez}, {Balm}, {Barache}, {Barata}, {Barbato},
  {Barblan}, {Barklem}, {Barrado}, {Barros}, {Barstow}, {Bartholom{\'e}
  Mu{\~n}oz}, {Bassilana}, {Becciani}, {Bellazzini}, {Berihuete}, {Bertone},
  {Bianchi}, {Bienaym{\'e}}, {Blanco-Cuaresma}, {Boch}, {Boeche}, {Bombrun},
  {Borrachero}, {Bossini}, {Bouquillon}, {Bourda}, {Bragaglia}, {Bramante},
  {Breddels}, {Bressan}, {Brouillet}, {Br{\"u}semeister}, {Brugaletta},
  {Bucciarelli}, {Burlacu}, {Busonero}, {Butkevich}, {Buzzi}, {Caffau},
  {Cancelliere}, {Cannizzaro}, {Cantat-Gaudin}, {Carballo}, {Carlucci},
  {Carrasco}, {Casamiquela}, {Castellani}, {Castro-Ginard}, {Charlot},
  {Chemin}, {Chiavassa}, {Cocozza}, {Costigan}, {Cowell}, {Crifo}, {Crosta},
  {Crowley}, {Cuypers}, {Dafonte}, {Damerdji}, {Dapergolas}, {David}, {David},
  {de Laverny}, {De Luise}, {De March}, {de Martino}, {de Souza}, {de Torres},
  {Debosscher}, {del Pozo}, {Delbo}, {Delgado}, {Delgado}, {Di Matteo},
  {Diakite}, {Diener}, {Distefano}, {Dolding}, {Drazinos}, {Dur{\'a}n},
  {Edvardsson}, {Enke}, {Eriksson}, {Esquej}, {Eynard Bontemps}, {Fabre},
  {Fabrizio}, {Faigler}, {Falc{\~a}o}, {Farr{\`a}s Casas}, {Federici},
  {Fedorets}, {Fernique}, {Figueras}, {Filippi}, {Findeisen}, {Fonti},
  {Fraile}, {Fraser}, {Fr{\'e}zouls}, {Gai}, {Galleti}, {Garabato},
  {Garc{\'\i}a-Sedano}, {Garofalo}, {Garralda}, {Gavel}, {Gavras}, {Gerssen},
  {Geyer}, {Giacobbe}, {Gilmore}, {Girona}, {Giuffrida}, {Glass}, {Gomes},
  {Granvik}, {Gueguen}, {Guerrier}, {Guiraud}, {Guti{\'e}rrez-S{\'a}nchez},
  {Haigron}, {Hatzidimitriou}, {Hauser}, {Haywood}, {Heiter}, {Helmi}, {Heu},
  {Hilger}, {Hobbs}, {Hofmann}, {Holland}, {Huckle}, {Hypki}, {Icardi},
  {Jan{\ss}en}, {Jevardat de Fombelle}, {Jonker}, {Juh{\'a}sz}, {Julbe},
  {Karampelas}, {Kewley}, {Klar}, {Kochoska}, {Kohley}, {Kolenberg},
  {Kontizas}, {Kontizas}, {Koposov}, {Kordopatis}, {Kostrzewa-Rutkowska},
  {Koubsky}, {Lambert}, {Lanza}, {Lasne}, {Lavigne}, {Le Fustec}, {Le
  Poncin-Lafitte}, {Lebreton}, {Leccia}, {Leclerc}, {Lecoeur-Taibi},
  {Lenhardt}, {Leroux}, {Liao}, {Licata}, {Lindstr{\o}m}, {Lister}, {Livanou},
  {Lobel}, {L{\'o}pez}, {Managau}, {Mann}, {Mantelet}, {Marchal}, {Marchant},
  {Marconi}, {Marinoni}, {Marschalk{\'o}}, {Marshall}, {Martino}, {Marton},
  {Mary}, {Massari}, {Matijevi{\v{c}}}, {Mazeh}, {McMillan}, {Messina},
  {Michalik}, {Millar}, {Molina}, {Molinaro}, {Moln{\'a}r}, {Montegriffo},
  {Mor}, {Morbidelli}, {Morel}, {Morris}, {Mulone}, {Muraveva}, {Musella},
  {Nelemans}, {Nicastro}, {Noval}, {O'Mullane}, {Ord{\'e}novic},
  {Ord{\'o}{\~n}ez-Blanco}, {Osborne}, {Pagani}, {Pagano}, {Pailler},
  {Palacin}, {Palaversa}, {Panahi}, {Pawlak}, {Piersimoni}, {Pineau}, {Plachy},
  {Plum}, {Poggio}, {Poujoulet}, {Pr{\v{s}}a}, {Pulone}, {Racero}, {Ragaini},
  {Rambaux}, {Ramos-Lerate}, {Regibo}, {Reyl{\'e}}, {Riclet}, {Ripepi}, {Riva},
  {Rivard}, {Rixon}, {Roegiers}, {Roelens}, {Romero-G{\'o}mez}, {Rowell},
  {Royer}, {Ruiz-Dern}, {Sadowski}, {Sagrist{\`a} Sell{\'e}s}, {Sahlmann},
  {Salgado}, {Salguero}, {Sanna}, {Santana-Ros}, {Sarasso}, {Savietto},
  {Schultheis}, {Sciacca}, {Segol}, {Segovia}, {S{\'e}gransan}, {Shih},
  {Siltala}, {Silva}, {Smart}, {Smith}, {Solano}, {Solitro}, {Sordo}, {Soria
  Nieto}, {Souchay}, {Spagna}, {Spoto}, {Stampa}, {Steele},
  {Steidelm{\"u}ller}, {Stephenson}, {Stoev}, {Suess}, {Surdej}, {Szabados},
  {Szegedi-Elek}, {Tapiador}, {Taris}, {Tauran}, {Taylor}, {Teixeira},
  {Terrett}, {Teyssandier}, {Thuillot}, {Titarenko}, {Torra Clotet}, {Turon},
  {Ulla}, {Utrilla}, {Uzzi}, {Vaillant}, {Valentini}, {Valette}, {van Elteren},
  {Van Hemelryck}, {van Leeuwen}, {Vaschetto}, {Vecchiato}, {Veljanoski},
  {Viala}, {Vicente}, {Vogt}, {von Essen}, {Voss}, {Votruba}, {Voutsinas},
  {Walmsley}, {Weiler}, {Wertz}, {Wevers}, {Wyrzykowski}, {Yoldas},
  {{\v{Z}}erjal}, {Ziaeepour}, {Zorec}, {Zschocke}, {Zucker}, {Zurbach}, \&
  {Zwitter}}]{gaia_dr2}
{Gaia Collaboration}, {Brown}, A.~G.~A., {Vallenari}, A., {et~al.} 2018, \aap,
  616, A1

\bibitem[{{Gaia Collaboration} {et~al.}(2023){Gaia Collaboration}, {Vallenari},
  {Brown}, {Prusti}, {de Bruijne}, {Arenou}, {Babusiaux}, {Biermann},
  {Creevey}, {Ducourant}, {Evans}, {Eyer}, {Guerra}, {Hutton}, {Jordi},
  {Klioner}, {Lammers}, {Lindegren}, {Luri}, {Mignard}, {Panem}, {Pourbaix},
  {Randich}, {Sartoretti}, {Soubiran}, {Tanga}, {Walton}, {Bailer-Jones},
  {Bastian}, {Drimmel}, {Jansen}, {Katz}, {Lattanzi}, {van Leeuwen}, {Bakker},
  {Cacciari}, {Casta{\~n}eda}, {De Angeli}, {Fabricius}, {Fouesneau},
  {Fr{\'e}mat}, {Galluccio}, {Guerrier}, {Heiter}, {Masana}, {Messineo},
  {Mowlavi}, {Nicolas}, {Nienartowicz}, {Pailler}, {Panuzzo}, {Riclet}, {Roux},
  {Seabroke}, {Sordo}, {Th{\'e}venin}, {Gracia-Abril}, {Portell}, {Teyssier},
  {Altmann}, {Andrae}, {Audard}, {Bellas-Velidis}, {Benson}, {Berthier},
  {Blomme}, {Burgess}, {Busonero}, {Busso}, {C{\'a}novas}, {Carry}, {Cellino},
  {Cheek}, {Clementini}, {Damerdji}, {Davidson}, {de Teodoro}, {Nu{\~n}ez
  Campos}, {Delchambre}, {Dell'Oro}, {Esquej}, {Fern{\'a}ndez-Hern{\'a}ndez},
  {Fraile}, {Garabato}, {Garc{\'\i}a-Lario}, {Gosset}, {Haigron}, {Halbwachs},
  {Hambly}, {Harrison}, {Hern{\'a}ndez}, {Hestroffer}, {Hodgkin}, {Holl},
  {Jan{\ss}en}, {Jevardat de Fombelle}, {Jordan}, {Krone-Martins}, {Lanzafame},
  {L{\"o}ffler}, {Marchal}, {Marrese}, {Moitinho}, {Muinonen}, {Osborne},
  {Pancino}, {Pauwels}, {Recio-Blanco}, {Reyl{\'e}}, {Riello}, {Rimoldini},
  {Roegiers}, {Rybizki}, {Sarro}, {Siopis}, {Smith}, {Sozzetti}, {Utrilla},
  {van Leeuwen}, {Abbas}, {{\'A}brah{\'a}m}, {Abreu Aramburu}, {Aerts},
  {Aguado}, {Ajaj}, {Aldea-Montero}, {Altavilla}, {{\'A}lvarez}, {Alves},
  {Anders}, {Anderson}, {Anglada Varela}, {Antoja}, {Baines}, {Baker},
  {Balaguer-N{\'u}{\~n}ez}, {Balbinot}, {Balog}, {Barache}, {Barbato},
  {Barros}, {Barstow}, {Bartolom{\'e}}, {Bassilana}, {Bauchet}, {Becciani},
  {Bellazzini}, {Berihuete}, {Bernet}, {Bertone}, {Bianchi}, {Binnenfeld},
  {Blanco-Cuaresma}, {Blazere}, {Boch}, {Bombrun}, {Bossini}, {Bouquillon},
  {Bragaglia}, {Bramante}, {Breedt}, {Bressan}, {Brouillet}, {Brugaletta},
  {Bucciarelli}, {Burlacu}, {Butkevich}, {Buzzi}, {Caffau}, {Cancelliere},
  {Cantat-Gaudin}, {Carballo}, {Carlucci}, {Carnerero}, {Carrasco},
  {Casamiquela}, {Castellani}, {Castro-Ginard}, {Chaoul}, {Charlot}, {Chemin},
  {Chiaramida}, {Chiavassa}, {Chornay}, {Comoretto}, {Contursi}, {Cooper},
  {Cornez}, {Cowell}, {Crifo}, {Cropper}, {Crosta}, {Crowley}, {Dafonte},
  {Dapergolas}, {David}, {David}, {de Laverny}, {De Luise}, {De March}, {De
  Ridder}, {de Souza}, {de Torres}, {del Peloso}, {del Pozo}, {Delbo},
  {Delgado}, {Delisle}, {Demouchy}, {Dharmawardena}, {Di Matteo}, {Diakite},
  {Diener}, {Distefano}, {Dolding}, {Edvardsson}, {Enke}, {Fabre}, {Fabrizio},
  {Faigler}, {Fedorets}, {Fernique}, {Fienga}, {Figueras}, {Fournier},
  {Fouron}, {Fragkoudi}, {Gai}, {Garcia-Gutierrez}, {Garcia-Reinaldos},
  {Garc{\'\i}a-Torres}, {Garofalo}, {Gavel}, {Gavras}, {Gerlach}, {Geyer},
  {Giacobbe}, {Gilmore}, {Girona}, {Giuffrida}, {Gomel}, {Gomez},
  {Gonz{\'a}lez-N{\'u}{\~n}ez}, {Gonz{\'a}lez-Santamar{\'\i}a},
  {Gonz{\'a}lez-Vidal}, {Granvik}, {Guillout}, {Guiraud},
  {Guti{\'e}rrez-S{\'a}nchez}, {Guy}, {Hatzidimitriou}, {Hauser}, {Haywood},
  {Helmer}, {Helmi}, {Sarmiento}, {Hidalgo}, {Hilger}, {H{\l}adczuk}, {Hobbs},
  {Holland}, {Huckle}, {Jardine}, {Jasniewicz}, {Jean-Antoine Piccolo},
  {Jim{\'e}nez-Arranz}, {Jorissen}, {Juaristi Campillo}, {Julbe}, {Karbevska},
  {Kervella}, {Khanna}, {Kontizas}, {Kordopatis}, {Korn}, {K{\'o}sp{\'a}l},
  {Kostrzewa-Rutkowska}, {Kruszy{\'n}ska}, {Kun}, {Laizeau}, {Lambert},
  {Lanza}, {Lasne}, {Le Campion}, {Lebreton}, {Lebzelter}, {Leccia}, {Leclerc},
  {Lecoeur-Taibi}, {Liao}, {Licata}, {Lindstr{\o}m}, {Lister}, {Livanou},
  {Lobel}, {Lorca}, {Loup}, {Madrero Pardo}, {Magdaleno Romeo}, {Managau},
  {Mann}, {Manteiga}, {Marchant}, {Marconi}, {Marcos}, {Marcos Santos},
  {Mar{\'\i}n Pina}, {Marinoni}, {Marocco}, {Marshall}, {Martin Polo},
  {Mart{\'\i}n-Fleitas}, {Marton}, {Mary}, {Masip}, {Massari},
  {Mastrobuono-Battisti}, {Mazeh}, {McMillan}, {Messina}, {Michalik}, {Millar},
  {Mints}, {Molina}, {Molinaro}, {Moln{\'a}r}, {Monari}, {Mongui{\'o}},
  {Montegriffo}, {Montero}, {Mor}, {Mora}, {Morbidelli}, {Morel}, {Morris},
  {Muraveva}, {Murphy}, {Musella}, {Nagy}, {Noval}, {Oca{\~n}a}, {Ogden},
  {Ordenovic}, {Osinde}, {Pagani}, {Pagano}, {Palaversa}, {Palicio},
  {Pallas-Quintela}, {Panahi}, {Payne-Wardenaar}, {Pe{\~n}alosa Esteller},
  {Penttil{\"a}}, {Pichon}, {Piersimoni}, {Pineau}, {Plachy}, {Plum}, {Poggio},
  {Pr{\v{s}}a}, {Pulone}, {Racero}, {Ragaini}, {Rainer}, {Raiteri}, {Rambaux},
  {Ramos}, {Ramos-Lerate}, {Re Fiorentin}, {Regibo}, {Richards}, {Rios Diaz},
  {Ripepi}, {Riva}, {Rix}, {Rixon}, {Robichon}, {Robin}, {Robin}, {Roelens},
  {Rogues}, {Rohrbasser}, {Romero-G{\'o}mez}, {Rowell}, {Royer}, {Ruz Mieres},
  {Rybicki}, {Sadowski}, {S{\'a}ez N{\'u}{\~n}ez}, {Sagrist{\`a} Sell{\'e}s},
  {Sahlmann}, {Salguero}, {Samaras}, {Sanchez Gimenez}, {Sanna},
  {Santove{\~n}a}, {Sarasso}, {Schultheis}, {Sciacca}, {Segol}, {Segovia},
  {S{\'e}gransan}, {Semeux}, {Shahaf}, {Siddiqui}, {Siebert}, {Siltala},
  {Silvelo}, {Slezak}, {Slezak}, {Smart}, {Snaith}, {Solano}, {Solitro},
  {Souami}, {Souchay}, {Spagna}, {Spina}, {Spoto}, {Steele},
  {Steidelm{\"u}ller}, {Stephenson}, {S{\"u}veges}, {Surdej}, {Szabados},
  {Szegedi-Elek}, {Taris}, {Taylor}, {Teixeira}, {Tolomei}, {Tonello}, {Torra},
  {Torra}, {Torralba Elipe}, {Trabucchi}, {Tsounis}, {Turon}, {Ulla}, {Unger},
  {Vaillant}, {van Dillen}, {van Reeven}, {Vanel}, {Vecchiato}, {Viala},
  {Vicente}, {Voutsinas}, {Weiler}, {Wevers}, {Wyrzykowski}, {Yoldas}, {Yvard},
  {Zhao}, {Zorec}, {Zucker}, \& {Zwitter}}]{gaia_dr3}
{Gaia Collaboration}, {Vallenari}, A., {Brown}, A.~G.~A., {et~al.} 2023, \aap,
  674, A1

\bibitem[{{Galicher} {et~al.}(2018){Galicher}, {Boccaletti}, {Mesa}, {Delorme},
  {Gratton}, {Langlois}, {Lagrange}, {Maire}, {Le Coroller}, {Chauvin},
  {Biller}, {Cantalloube}, {Janson}, {Lagadec}, {Meunier}, {Vigan},
  {Hagelberg}, {Bonnefoy}, {Zurlo}, {Rocha}, {Maurel}, {Jaquet}, {Buey}, \&
  {Weber}}]{Galicher_specal}
{Galicher}, R., {Boccaletti}, A., {Mesa}, D., {et~al.} 2018, \aap, 615, A92

\bibitem[{{Galli} {et~al.}(2018){Galli}, {Joncour}, \& {Moraux}}]{galli18}
{Galli}, P. A.~B., {Joncour}, I., \& {Moraux}, E. 2018, \mnras, 477, L50

\bibitem[{{Gieles}(2010)}]{gieles10}
{Gieles}, M. 2010, in IAU Symposium, Vol. 266, Star Clusters: Basic Galactic
  Building Blocks Throughout Time and Space, ed. R.~{de Grijs} \& J.~R.~D.
  {L{\'e}pine}, 69--80

\bibitem[{{Gillon} {et~al.}(2016){Gillon}, {Jehin}, {Lederer}, {Delrez}, {de
  Wit}, {Burdanov}, {Van Grootel}, {Burgasser}, {Triaud}, {Opitom}, {Demory},
  {Sahu}, {Bardalez Gagliuffi}, {Magain}, \& {Queloz}}]{Gillon.2016}
{Gillon}, M., {Jehin}, E., {Lederer}, S.~M., {et~al.} 2016, \nat, 533, 221

\bibitem[{{Gorti} \& {Hollenbach}(2009)}]{2009ApJ...690.1539G}
{Gorti}, U. \& {Hollenbach}, D. 2009, \apj, 690, 1539

\bibitem[{{Gratton} {et~al.}(2023){Gratton}, {Squicciarini}, {Nascimbeni},
  {Janson}, {Reffert}, {Meyer}, {Delorme}, {Mamajek}, {Bonavita}, {Desidera},
  {Mesa}, {Rigliaco}, {D'Orazi}, {Vigan}, {Lazzoni}, {Chauvin}, \&
  {Langlois}}]{Gratton.2023}
{Gratton}, R., {Squicciarini}, V., {Nascimbeni}, V., {et~al.} 2023, \aap, 678,
  A93

\bibitem[{{Haemmerl{\'e}} {et~al.}(2019){Haemmerl{\'e}}, {Eggenberger},
  {Ekstr{\"o}m}, {Georgy}, {Meynet}, {Postel}, {Audard}, {S{\o}rensen}, \&
  {Fragos}}]{haemmerle19}
{Haemmerl{\'e}}, L., {Eggenberger}, P., {Ekstr{\"o}m}, S., {et~al.} 2019, \aap,
  624, A137

\bibitem[{{Holman} \& {Wiegert}(1999)}]{Holman.1999}
{Holman}, M.~J. \& {Wiegert}, P.~A. 1999, \aj, 117, 621

\bibitem[{{Janson} {et~al.}(2019){Janson}, {Asensio-Torres}, {Andr{\'e}},
  {Bonnefoy}, {Delorme}, {Reffert}, {Desidera}, {Langlois}, {Chauvin},
  {Gratton}, {Bohn}, {Eriksson}, {Marleau}, {Mamajek}, {Vigan}, \&
  {Carson}}]{Janson.2019}
{Janson}, M., {Asensio-Torres}, R., {Andr{\'e}}, D., {et~al.} 2019, \aap, 626,
  A99

\bibitem[{{Janson} {et~al.}(2011){Janson}, {Bonavita}, {Klahr},
  {Lafreni{\`e}re}, {Jayawardhana}, \& {Zinnecker}}]{Janson.2011}
{Janson}, M., {Bonavita}, M., {Klahr}, H., {et~al.} 2011, \apj, 736, 89

\bibitem[{{Janson} {et~al.}(2021{\natexlab{a}}){Janson}, {Gratton}, {Rodet},
  {Vigan}, {Bonnefoy}, {Delorme}, {Mamajek}, {Reffert}, {Stock}, {Marleau},
  {Langlois}, {Chauvin}, {Desidera}, {Ringqvist}, {Mayer}, {Viswanath},
  {Squicciarini}, {Meyer}, {Samland}, {Petrus}, {Helled}, {Kenworthy}, {Quanz},
  {Biller}, {Henning}, {Mesa}, {Engler}, \& {Carson}}]{Janson.2021_bcen}
{Janson}, M., {Gratton}, R., {Rodet}, L., {et~al.} 2021{\natexlab{a}}, \nat,
  600, 231

\bibitem[{{Janson} {et~al.}(2021{\natexlab{b}}){Janson}, {Squicciarini},
  {Delorme}, {Gratton}, {Bonnefoy}, {Reffert}, {Mamajek}, {Eriksson}, {Vigan},
  {Langlois}, {Engler}, {Chauvin}, {Desidera}, {Mayer}, {Marleau}, {Bohn},
  {Samland}, {Meyer}, {d'Orazi}, {Henning}, {Quanz}, {Kenworthy}, \&
  {Carson}}]{Janson.2021_beast}
{Janson}, M., {Squicciarini}, V., {Delorme}, P., {et~al.} 2021{\natexlab{b}},
  \aap, 646, A164

\bibitem[{{Kervella} {et~al.}(2022){Kervella}, {Arenou}, \&
  {Th{\'e}venin}}]{kervella22}
{Kervella}, P., {Arenou}, F., \& {Th{\'e}venin}, F. 2022, \aap, 657, A7

\bibitem[{Kessler \& Alibert(2023)}]{2023A&A...674A.144K}
Kessler, A. \& Alibert, Y. 2023, Astronomy and Astrophysics, 674, A144

\bibitem[{{Komaki} {et~al.}(2021){Komaki}, {Nakatani}, \&
  {Yoshida}}]{2021ApJ...910...51K}
{Komaki}, A., {Nakatani}, R., \& {Yoshida}, N. 2021, \apj, 910, 51

\bibitem[{{Kubli} {et~al.}(2023){Kubli}, {Mayer}, \& {Deng}}]{Kubli2023}
{Kubli}, N., {Mayer}, L., \& {Deng}, H. 2023, \mnras, 525, 2731

\bibitem[{{Lafreni{\`e}re} {et~al.}(2011){Lafreni{\`e}re}, {Jayawardhana},
  {Janson}, {Helling}, {Witte}, \& {Hauschildt}}]{Lafreniere.2011}
{Lafreni{\`e}re}, D., {Jayawardhana}, R., {Janson}, M., {et~al.} 2011, \apj,
  730, 42

\bibitem[{{Lagrange} {et~al.}(2020){Lagrange}, {Rubini}, {Nowak}, {Lacour},
  {Grandjean}, {Boccaletti}, {Langlois}, {Delorme}, {Gratton}, {Wang},
  {Flasseur}, {Galicher}, {Kral}, {Meunier}, {Beust}, {Babusiaux}, {Le
  Coroller}, {Thebault}, {Kervella}, {Zurlo}, {Maire}, {Wahhaj}, {Amorim},
  {Asensio-Torres}, {Benisty}, {Berger}, {Bonnefoy}, {Brandner}, {Cantalloube},
  {Charnay}, {Chauvin}, {Choquet}, {Cl{\'e}net}, {Christiaens}, {Coud{\'e} Du
  Foresto}, {de Zeeuw}, {Desidera}, {Duvert}, {Eckart}, {Eisenhauer},
  {Galland}, {Gao}, {Garcia}, {Garcia Lopez}, {Gendron}, {Genzel}, {Gillessen},
  {Girard}, {Hagelberg}, {Haubois}, {Henning}, {Heissel}, {Hippler},
  {Horrobin}, {Janson}, {Kammerer}, {Kenworthy}, {Keppler}, {Kreidberg},
  {Lapeyr{\`e}re}, {Le Bouquin}, {L{\'e}na}, {M{\'e}rand}, {Messina},
  {Molli{\`e}re}, {Monnier}, {Ott}, {Otten}, {Paumard}, {Paladini}, {Perraut},
  {Perrin}, {Pueyo}, {Pfuhl}, {Rodet}, {Rodriguez-Coira}, {Rousset}, {Samland},
  {Shangguan}, {Schmidt}, {Straub}, {Straubmeier}, {Stolker}, {Vigan},
  {Vincent}, {Widmann}, {Woillez}, \& {GRAVITY
  Collaboration}}]{Lagrange_bpic_2020}
{Lagrange}, A.~M., {Rubini}, P., {Nowak}, M., {et~al.} 2020, \aap, 642, A18

\bibitem[{{Lannier} {et~al.}(2016){Lannier}, {Delorme}, {Lagrange}, {Borgniet},
  {Rameau}, {Schlieder}, {Gagn{\'e}}, {Bonavita}, {Malo}, {Chauvin},
  {Bonnefoy}, \& {Girard}}]{Lannier.2016}
{Lannier}, J., {Delorme}, P., {Lagrange}, A.~M., {et~al.} 2016, \aap, 596, A83

\bibitem[{{Lecavelier des Etangs} \& {Lissauer}(2022)}]{IAUplanet2022}
{Lecavelier des Etangs}, A. \& {Lissauer}, J.~J. 2022, \nar, 94, 101641

\bibitem[{{Leike} {et~al.}(2020){Leike}, {Glatzle}, \& {En{\ss}lin}}]{leike20}
{Leike}, R.~H., {Glatzle}, M., \& {En{\ss}lin}, T.~A. 2020, \aap, 639, A138

\bibitem[{{Luhman}(2022)}]{luhman22}
{Luhman}, K.~L. 2022, \aj, 163, 24

\bibitem[{{Luhman} \& {Esplin}(2020)}]{luhman20}
{Luhman}, K.~L. \& {Esplin}, T.~L. 2020, \aj, 160, 44

\bibitem[{{Malmberg} {et~al.}(2007){Malmberg}, {de Angeli}, {Davies}, {Church},
  {Mackey}, \& {Wilkinson}}]{malmberg07}
{Malmberg}, D., {de Angeli}, F., {Davies}, M.~B., {et~al.} 2007, \mnras, 378,
  1207

\bibitem[{{Marois} {et~al.}(2014){Marois}, {Correia}, {V{\'e}ran}, \&
  {Currie}}]{Marois_TLOCI}
{Marois}, C., {Correia}, C., {V{\'e}ran}, J.-P., \& {Currie}, T. 2014, in
  Exploring the Formation and Evolution of Planetary Systems, ed. M.~{Booth},
  B.~C. {Matthews}, \& J.~R. {Graham}, Vol. 299, 48--49

\bibitem[{{Marois} {et~al.}(2006){Marois}, {Lafreni{\`e}re}, {Doyon},
  {Macintosh}, \& {Nadeau}}]{Marois_adi}
{Marois}, C., {Lafreni{\`e}re}, D., {Doyon}, R., {Macintosh}, B., \& {Nadeau},
  D. 2006, \apj, 641, 556

\bibitem[{{Miret-Roig} {et~al.}(2022){Miret-Roig}, {Bouy}, {Raymond}, {Tamura},
  {Bertin}, {Barrado}, {Olivares}, {Galli}, {Cuillandre}, {Sarro}, {Berihuete},
  \& {Hu{\'e}lamo}}]{miret-roig22}
{Miret-Roig}, N., {Bouy}, H., {Raymond}, S.~N., {et~al.} 2022, Nature
  Astronomy, 6, 89

\bibitem[{{Moe} \& {Di Stefano}(2017)}]{moe17}
{Moe}, M. \& {Di Stefano}, R. 2017, \apjs, 230, 15

\bibitem[{{Nguyen} {et~al.}(2022){Nguyen}, {Costa}, {Girardi}, {Volpato},
  {Bressan}, {Chen}, {Marigo}, {Fu}, \& {Goudfrooij}}]{nguyen22}
{Nguyen}, C.~T., {Costa}, G., {Girardi}, L., {et~al.} 2022, \aap, 665, A126

\bibitem[{{Nielsen} {et~al.}(2019{\natexlab{a}}){Nielsen}, {De Rosa},
  {Macintosh}, {Wang}, {Ruffio}, {Chiang}, {Marley}, {Saumon}, {Savransky},
  {Ammons}, {Bailey}, {Barman}, {Blain}, {Bulger}, {Burrows}, {Chilcote},
  {Cotten}, {Czekala}, {Doyon}, {Duch{\^e}ne}, {Esposito}, {Fabrycky},
  {Fitzgerald}, {Follette}, {Fortney}, {Gerard}, {Goodsell}, {Graham},
  {Greenbaum}, {Hibon}, {Hinkley}, {Hirsch}, {Hom}, {Hung}, {Dawson},
  {Ingraham}, {Kalas}, {Konopacky}, {Larkin}, {Lee}, {Lin}, {Maire}, {Marchis},
  {Marois}, {Metchev}, {Millar-Blanchaer}, {Morzinski}, {Oppenheimer},
  {Palmer}, {Patience}, {Perrin}, {Poyneer}, {Pueyo}, {Rafikov}, {Rajan},
  {Rameau}, {Rantakyr{\"o}}, {Ren}, {Schneider}, {Sivaramakrishnan}, {Song},
  {Soummer}, {Tallis}, {Thomas}, {Ward-Duong}, \& {Wolff}}]{Nielsen.2019}
{Nielsen}, E.~L., {De Rosa}, R.~J., {Macintosh}, B., {et~al.}
  2019{\natexlab{a}}, \aj, 158, 13

\bibitem[{{Nielsen} {et~al.}(2019{\natexlab{b}}){Nielsen}, {De Rosa},
  {Macintosh}, {Wang}, {Ruffio}, {Chiang}, {Marley}, {Saumon}, {Savransky},
  {Ammons}, {Bailey}, {Barman}, {Blain}, {Bulger}, {Burrows}, {Chilcote},
  {Cotten}, {Czekala}, {Doyon}, {Duch{\^e}ne}, {Esposito}, {Fabrycky},
  {Fitzgerald}, {Follette}, {Fortney}, {Gerard}, {Goodsell}, {Graham},
  {Greenbaum}, {Hibon}, {Hinkley}, {Hirsch}, {Hom}, {Hung}, {Dawson},
  {Ingraham}, {Kalas}, {Konopacky}, {Larkin}, {Lee}, {Lin}, {Maire}, {Marchis},
  {Marois}, {Metchev}, {Millar-Blanchaer}, {Morzinski}, {Oppenheimer},
  {Palmer}, {Patience}, {Perrin}, {Poyneer}, {Pueyo}, {Rafikov}, {Rajan},
  {Rameau}, {Rantakyr{\"o}}, {Ren}, {Schneider}, {Sivaramakrishnan}, {Song},
  {Soummer}, {Tallis}, {Thomas}, {Ward-Duong}, \& {Wolff}}]{nielsen19}
{Nielsen}, E.~L., {De Rosa}, R.~J., {Macintosh}, B., {et~al.}
  2019{\natexlab{b}}, \aj, 158, 13

\bibitem[{Ormel \& Klahr(2010)}]{ormelklahr2010a}
Ormel, C. \& Klahr, H. 2010, eprint arXiv, 1007, 916

\bibitem[{{Parker} \& {Daffern-Powell}(2022)}]{parker22}
{Parker}, R.~J. \& {Daffern-Powell}, E.~C. 2022, \mnras, 516, L91

\bibitem[{{Parravano} {et~al.}(2011){Parravano}, {McKee}, \&
  {Hollenbach}}]{Parravano.2011}
{Parravano}, A., {McKee}, C.~F., \& {Hollenbach}, D.~J. 2011, \apj, 726, 27

\bibitem[{{Pecaut} \& {Mamajek}(2013)}]{pecaut13}
{Pecaut}, M.~J. \& {Mamajek}, E.~E. 2013, \apjs, 208, 9

\bibitem[{{Pecaut} \& {Mamajek}(2016)}]{pecaut16}
{Pecaut}, M.~J. \& {Mamajek}, E.~E. 2016, \mnras, 461, 794

\bibitem[{{Phillips} {et~al.}(2020){Phillips}, {Tremblin}, {Baraffe},
  {Chabrier}, {Allard}, {Spiegelman}, {Goyal}, {Drummond}, \&
  {H{\'e}brard}}]{Phillips.2020}
{Phillips}, M.~W., {Tremblin}, P., {Baraffe}, I., {et~al.} 2020, \aap, 637, A38

\bibitem[{Pollack {et~al.}(1996)Pollack, Hubickyj, Bodenheimer, Lissauer,
  Podolak, \& Greenzweig}]{Pollack1996}
Pollack, J.~B., Hubickyj, O., Bodenheimer, P., {et~al.} 1996, Icarus, 124, 62

\bibitem[{{Rameau} {et~al.}(2013){Rameau}, {Chauvin}, {Lagrange}, {Klahr},
  {Bonnefoy}, {Mordasini}, {Bonavita}, {Desidera}, {Dumas}, \&
  {Girard}}]{Rameau.2013a}
{Rameau}, J., {Chauvin}, G., {Lagrange}, A.~M., {et~al.} 2013, \aap, 553, A60

\bibitem[{{Reffert} {et~al.}(2015){Reffert}, {Bergmann}, {Quirrenbach},
  {Trifonov}, \& {K{\"u}nstler}}]{Reffert.2015}
{Reffert}, S., {Bergmann}, C., {Quirrenbach}, A., {Trifonov}, T., \&
  {K{\"u}nstler}, A. 2015, \aap, 574, A116

\bibitem[{{Riols} \& {Latter}(2019)}]{Riols2019}
{Riols}, A. \& {Latter}, H. 2019, \mnras, 482, 3989

\bibitem[{{Rizzuto} {et~al.}(2015){Rizzuto}, {Ireland}, \& {Kraus}}]{rizzuto15}
{Rizzuto}, A.~C., {Ireland}, M.~J., \& {Kraus}, A.~L. 2015, \mnras, 448, 2737

\bibitem[{{Rizzuto} {et~al.}(2011){Rizzuto}, {Ireland}, \&
  {Robertson}}]{rizzuto11}
{Rizzuto}, A.~C., {Ireland}, M.~J., \& {Robertson}, J.~G. 2011, \mnras, 416,
  3108

\bibitem[{{Rizzuto} {et~al.}(2013){Rizzuto}, {Ireland}, {Robertson}, {Kok},
  {Tuthill}, {Warrington}, {Haubois}, {Tango}, {Norris}, {ten Brummelaar},
  {Kraus}, {Jacob}, \& {Laliberte-Houdeville}}]{Rizzuto.2013}
{Rizzuto}, A.~C., {Ireland}, M.~J., {Robertson}, J.~G., {et~al.} 2013, \mnras,
  436, 1694

\bibitem[{Schib {et~al.}(2023)Schib, Mordasini, \&
  Helled}]{2023A&A...669A..31S}
Schib, O., Mordasini, C., \& Helled, R. 2023, Astronomy and Astrophysics, 669,
  A31

\bibitem[{Schib {et~al.}(2021)Schib, Mordasini, Wenger, Marleau, \&
  Helled}]{2021A&A...645A..43S}
Schib, O., Mordasini, C., Wenger, N., Marleau, G.~D., \& Helled, R. 2021,
  Astronomy and Astrophysics, 645, A43

\bibitem[{{Soummer} {et~al.}(2012){Soummer}, {Pueyo}, \&
  {Larkin}}]{Soummer_PCA}
{Soummer}, R., {Pueyo}, L., \& {Larkin}, J. 2012, \apjl, 755, L28

\bibitem[{{Squicciarini} \& {Bonavita}(2022)}]{madys}
{Squicciarini}, V. \& {Bonavita}, M. 2022, \aap, 666, A15

\bibitem[{{Squicciarini} {et~al.}(2021){Squicciarini}, {Gratton}, {Bonavita},
  \& {Mesa}}]{squicciarini21}
{Squicciarini}, V., {Gratton}, R., {Bonavita}, M., \& {Mesa}, D. 2021, \mnras,
  507, 1381

\bibitem[{{Squicciarini} {et~al.}(2022){Squicciarini}, {Gratton}, {Janson},
  {Mamajek}, {Chauvin}, {Delorme}, {Langlois}, {Vigan}, {Ringqvist}, {Meeus},
  {Reffert}, {Kenworthy}, {Meyer}, {Bonnefoy}, {Bonavita}, {Mesa}, {Samland},
  {Desidera}, {D'Orazi}, {Engler}, {Alecian}, {Miglio}, {Henning}, {Quanz},
  {Mayer}, {Flasseur}, \& {Marleau}}]{Squicciarini.2022}
{Squicciarini}, V., {Gratton}, R., {Janson}, M., {et~al.} 2022, \aap, 664, A9

\bibitem[{{van Leeuwen}(2007)}]{hipparcos2}
{van Leeuwen}, F. 2007, \aap, 474, 653

\bibitem[{{Vigan} {et~al.}(2021){Vigan}, {Fontanive}, {Meyer}, {Biller},
  {Bonavita}, {Feldt}, {Desidera}, {Marleau}, {Emsenhuber}, {Galicher}, {Rice},
  {Forgan}, {Mordasini}, {Gratton}, {Le Coroller}, {Maire}, {Cantalloube},
  {Chauvin}, {Cheetham}, {Hagelberg}, {Lagrange}, {Langlois}, {Bonnefoy},
  {Beuzit}, {Boccaletti}, {D'Orazi}, {Delorme}, {Dominik}, {Henning}, {Janson},
  {Lagadec}, {Lazzoni}, {Ligi}, {Menard}, {Mesa}, {Messina}, {Moutou},
  {M{\"u}ller}, {Perrot}, {Samland}, {Schmid}, {Schmidt}, {Sissa}, {Turatto},
  {Udry}, {Zurlo}, {Abe}, {Antichi}, {Asensio-Torres}, {Baruffolo}, {Baudoz},
  {Baudrand}, {Bazzon}, {Blanchard}, {Bohn}, {Brown Sevilla}, {Carbillet},
  {Carle}, {Cascone}, {Charton}, {Claudi}, {Costille}, {De Caprio},
  {Delboulb{\'e}}, {Dohlen}, {Engler}, {Fantinel}, {Feautrier}, {Fusco},
  {Gigan}, {Girard}, {Giro}, {Gisler}, {Gluck}, {Gry}, {Hubin}, {Hugot},
  {Jaquet}, {Kasper}, {Le Mignant}, {Llored}, {Madec}, {Magnard}, {Martinez},
  {Maurel}, {M{\"o}ller-Nilsson}, {Mouillet}, {Moulin}, {Orign{\'e}}, {Pavlov},
  {Perret}, {Petit}, {Pragt}, {Puget}, {Rabou}, {Ramos}, {Rickman}, {Rigal},
  {Rochat}, {Roelfsema}, {Rousset}, {Roux}, {Salasnich}, {Sauvage}, {Sevin},
  {Soenke}, {Stadler}, {Suarez}, {Wahhaj}, {Weber}, \& {Wildi}}]{Vigan.2021}
{Vigan}, A., {Fontanive}, C., {Meyer}, M., {et~al.} 2021, \aap, 651, A72

\bibitem[{{Viswanath} {et~al.}(2023){Viswanath}, {Janson}, {Gratton},
  {Squicciarini}, {Rodet}, {Ringqvist}, {Mamajek}, {Reffert}, {Chauvin},
  {Delorme}, {Vigan}, {Bonnefoy}, {Engler}, {Desidera}, {Henning}, {Hagelberg},
  {Langlois}, \& {Meyer}}]{Viswanath.2023}
{Viswanath}, G., {Janson}, M., {Gratton}, R., {et~al.} 2023, \aap, 675, A54

\bibitem[{{Wenger} {et~al.}(2000){Wenger}, {Ochsenbein}, {Egret}, {Dubois},
  {Bonnarel}, {Borde}, {Genova}, {Jasniewicz}, {Lalo{\"e}}, {Lesteven}, \&
  {Monier}}]{simbad}
{Wenger}, M., {Ochsenbein}, F., {Egret}, D., {et~al.} 2000, \aaps, 143, 9

\bibitem[{{Wolthoff} {et~al.}(2022){Wolthoff}, {Reffert}, {Quirrenbach},
  {Jones}, {Wittenmyer}, \& {Jenkins}}]{Wolthoff.2022}
{Wolthoff}, V., {Reffert}, S., {Quirrenbach}, A., {et~al.} 2022, \aap, 661, A63

\bibitem[{{Zhang} {et~al.}(2023){Zhang}, {Molli{\`e}re}, {Hawkins}, {Manea},
  {Fortney}, {Morley}, {Skemer}, {Marley}, {Bowler}, {Carter}, {Franson},
  {Maas}, \& {Sneden}}]{Zhang_AF_Lep_b_2023}
{Zhang}, Z., {Molli{\`e}re}, P., {Hawkins}, K., {et~al.} 2023, \aj, 166, 198

\bibitem[{{Zurlo} {et~al.}(2022){Zurlo}, {Go{\'z}dziewski}, {Lazzoni}, {Mesa},
  {Nogueira}, {Desidera}, {Gratton}, {Marzari}, {Langlois}, {Pinna}, {Chauvin},
  {Delorme}, {Girard}, {Hagelberg}, {Henning}, {Janson}, {Rickman}, {Kervella},
  {Avenhaus}, {Bhowmik}, {Biller}, {Boccaletti}, {Bonaglia}, {Bonavita},
  {Bonnefoy}, {Cantalloube}, {Cheetham}, {Claudi}, {D'Orazi}, {Feldt},
  {Galicher}, {Ghose}, {Lagrange}, {le Coroller}, {Ligi}, {Kasper}, {Maire},
  {Medard}, {Meyer}, {Peretti}, {Perrot}, {Puglisi}, {Rossi}, {Rothberg},
  {Schmidt}, {Sissa}, {Vigan}, \& {Wahhaj}}]{Zurlo_hr8799}
{Zurlo}, A., {Go{\'z}dziewski}, K., {Lazzoni}, C., {et~al.} 2022, \aap, 666,
  A133

\end{thebibliography}

\begin{appendix}

\section{Uncertainties on stellar mass estimates}
\label{appendix:mass_error}

\begin{figure}
    \centering
    \includegraphics[width=\linewidth]{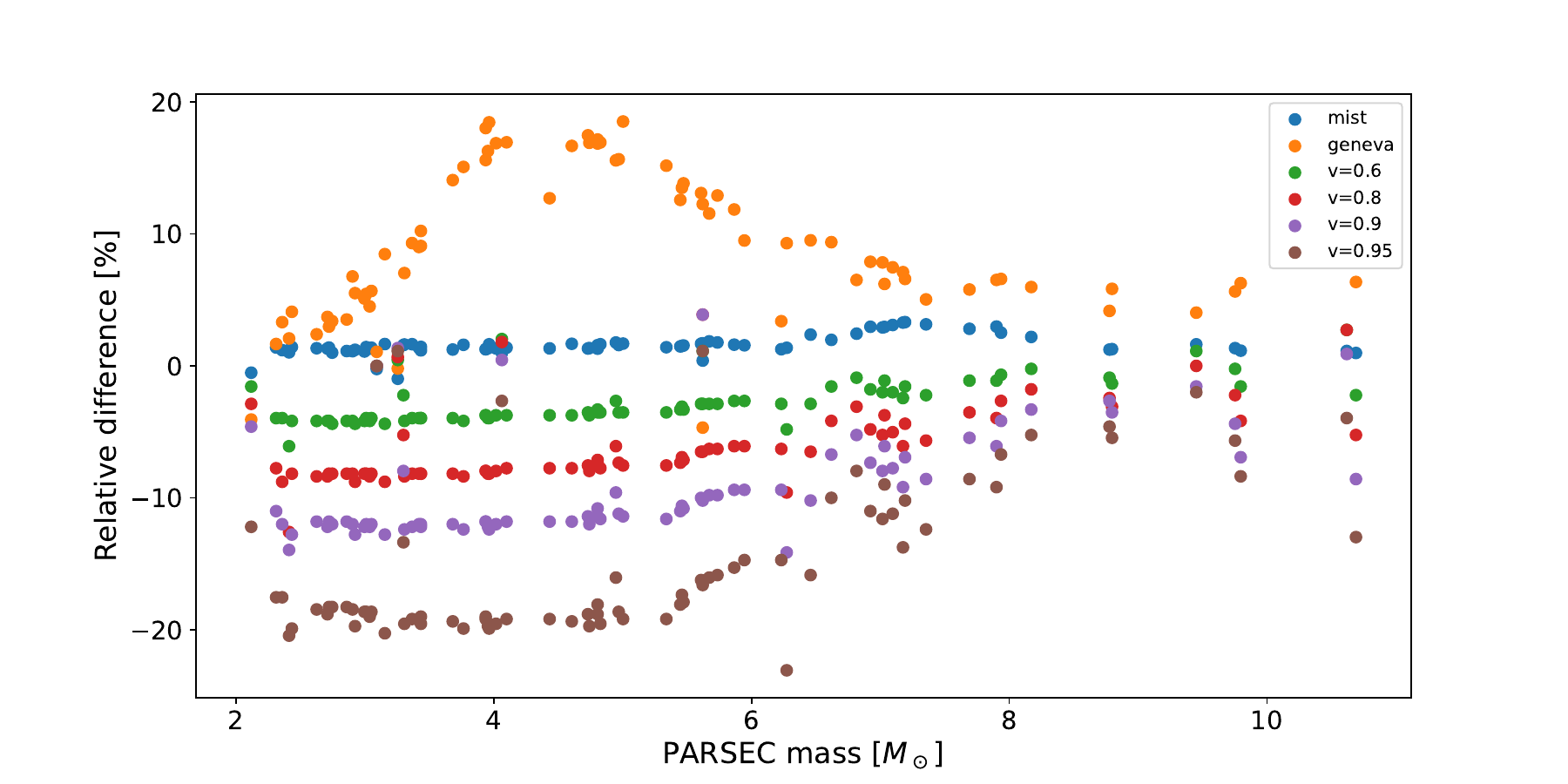}
    \caption{Systematic difference related to model selection and stellar rotation. Rotation velocities in the legend are expressed in units of the break-up velocity $v_{\text{break}}$.}
    \label{fig:systematic_mass_unc}
\end{figure}

 As described in Section~\ref{subsec:revised_prop}, we derived stellar masses using PARSEC isochrones, so as to be consistent with the underlying age analysis. The typical random uncertainty $\sigma_r$ of these, arising from the propagation of the uncertainties on parallax, apparent photometry and age, is of the order $0.01 < \sigma_r \lesssim 0.1$ \msun.

However, in addition to unresolved multiplicity and random uncertainties, additional factors can bias the photometric estimate of stellar mass for BEAST stars: 1) the choice of the input evolutionary model, and 2) stellar rotation, which can induce non-negligible luminosity variations in the B-star mass range.

In order to explore the dependency on these factors, we recomputed all stellar masses employing: 1) two different models -- MIST \citep{dotter16,choi16} and Geneva \citep{haemmerle19}; 2) PARSEC models with different values of stellar rotation \citep{nguyen22}.

A comparison of the results of these various models with the nominal values from the PARSEC non rotating analysis that we use in this article is shown in Figure~\ref{fig:systematic_mass_unc}.

In order to correctly interpret the results of this comparison, a couple of considerations have to be accounted for. Firstly, if one compares isochronal masses and spectral-type-based masses, a much lower compatibility is observed when using Geneva models, especially in the [4-6] \Msun range: in other words, systematically larger values are obtained using this model. Secondly, we expect the typical rotation velocity of our sample to be $<< 0.95 v_{\text{break}}$. \citet[][hereafter \citetalias{alecian13}]{alecian13} studied the rotation rate of a large sample of young (10 kyr to 10 Myr) Be stars. Even in the worst scenario of non-magnetic Be stars (see Fig. 5, bottom panel, blue bars of \citetalias{alecian13}), we expect about <5\% (15\%) of stars to have $v \sin{i} > 0.8 v_{\text{break}}$ for $2.3 \msun < M <3.5 \msun$ ($M > 3.5 \msun$). Moreover, only 7 stars of our sample are reported as Be stars by Simbad; standard B stars with no prominent emission lines are expected to possess lower rotation rates. Finally, our stars are generally older than the sample by \citetalias{alecian13}, which shows a clear trend of decreasing angular momentum over time (at least for stars with M>5 \Msun). For all these reasons, we consider it reasonable to (conservatively) assume a minimum fractional uncertainty of 10\% on all stellar masses. In the case of $\mu^2$ Sco, which is in a mass range where models differ weakly and for which a dedicated analysis, showing evidence for a very low rotation rate, was undertaken in a previous publication \citep{Squicciarini.2022}, we retain the nominal PARSEC uncertainty. 

\subsection{Notes on individual objects}

{\bf HIP 63210}: the classification of this star as B2 ($\sim$ 20000 K) is suspicious, as \teff measurements from the literature ($\sim$ 11000 K) are consistent with a late B. Also, the star is indicated as a SB2 by \citet{chini12}. An equal-mass B8V binary (with $M_A=M_B \approx 3.4$ \msun) would be in agreement with the age, temperature and luminosity indicators.

{\bf HIP 78384}: an equal-mass binary system with $M_A=M_B \approx 7$ \msun might explain the discrepancy between the two mass estimates; however, the pair separation would have to be < 1 au to be consistent with the non-detection by \cite{Rizzuto.2013}.

\onecolumn
\section{Properties of target stars}
\label{appendix:star_table}

\small 
{\setlength\tabcolsep{2pt} 
\begin{longtable}{cccccccccccc}\hline 
HIP ID & OTHER ID & RA$^a$ & DEC$^a$ & G & K & Dist. & Age & SpT & $M_{SpT}$ & $M_{iso}$ & $m_f$ \\  
 & &  \tiny{hms} & \tiny{dms} & \tiny{mag} & \tiny{mag} & \tiny{pc} & \tiny{Myr} & & \tiny{\msun} & \tiny{\msun} & \\ 
\hline 
\object{HIP50847} &    * 191 Car & 10 22 58.1465 & -66 54 05.385 &        4.93 & 5.31 &   $128.9 \pm 1.6$ &          $27 \pm 4$ &      B8V &  $3.4_{-0.6}^{+0.5}$ &        $4.9 \pm 0.5$ &    1 \\
\object{HIP52742} &     HD 93563 & 10 46 57.4734 &  -56 45 25.89 &        5.12 & 5.22 &   $178.5 \pm 3.6$ &        $150 \pm 50$ &      B5V &  $4.7_{-0.4}^{+0.4}$ &  $4.1_{-0.4}^{+0.6}$ &    1 \\
\object{HIP54767} &     HD 97583 & 11 12 45.2069 & -64 10 11.171 &         5.2 & 5.42 &    $97.7 \pm 0.6$ &   $100_{-40}^{+50}$ &      B8V &  $3.4_{-0.6}^{+0.5}$ &        $3.3 \pm 0.3$ &    1 \\
\object{HIP58452} &    HD 104080 & 11 59 10.6803 & -45 49 55.998 &        6.34 & 6.53 &   $134.0 \pm 0.9$ &          $20 \pm 5$ &    B8.5V &  $3.1_{-0.4}^{+0.6}$ &        $3.0 \pm 0.3$ &    0 \\
\object{HIP58901} &    HD 104900 & 12 04 45.2597 & -59 15 11.703 &        6.29 & 6.28 &   $117.0 \pm 1.4$ &          $13 \pm 3$ &      B9V &  $2.8_{-0.6}^{+0.6}$ &        $2.9 \pm 0.3$ &    1 \\
\object{HIP59173} &  V* V863 Cen & 12 08 05.2222 & -50 39 40.575 &         4.4 & 4.87 &   $108.9 \pm 2.3$ &          $15 \pm 3$ &      B5V &  $4.7_{-0.4}^{+0.4}$ &        $5.4 \pm 0.5$ &    1 \\
\object{HIP59747} &    * del Cru & 12 15 08.7184 & -58 44 56.126 &        2.77 & 3.53 &  $139.5 \pm 13.9$ &          $16 \pm 3$ &      B2V &  $7.0_{-1.0}^{+3.0}$ &       $11.0 \pm 1.0$ &    1 \\
\object{HIP60009} &    * zet Cru &  12 18 26.242 & -64 00 11.107 &        4.04 & 4.53 &   $108.2 \pm 2.3$ &      $12_{-2}^{+4}$ &    B2.5V &  $6.0_{-0.6}^{+1.3}$ &        $6.5 \pm 0.6$ &    1 \\
\object{HIP60379} &    HD 107696 & 12 22 49.4317 & -57 40 34.068 &        5.36 & 5.64 &   $104.1 \pm 0.8$ &          $13 \pm 3$ &      B7V &  $3.9_{-0.5}^{+0.4}$ &        $3.8 \pm 0.4$ &    0 \\
\object{HIP60710} &      * G Cen &   12 26 31.76 & -51 27 02.288 &         4.8 & 5.23 &   $136.1 \pm 3.4$ &      $18_{-3}^{+2}$ &      B3V &  $5.4_{-0.3}^{+0.6}$ &        $5.6 \pm 0.6$ &    1 \\
\object{HIP60823} &    * sig Cen & 12 28 02.3821 & -50 13 50.296 &        3.89 & 4.48 &   $126.2 \pm 3.7$ &          $17 \pm 3$ &      B2V &  $7.0_{-1.0}^{+3.0}$ &        $7.2 \pm 0.7$ &    1 \\
\object{HIP60855} &      * u Cen & 12 28 22.4671 &  -39 02 28.19 &        5.43 & 5.54 &   $150.4 \pm 3.7$ &          $16 \pm 3$ &    B8.5V &  $3.1_{-0.4}^{+0.6}$ &        $5.0 \pm 0.5$ &    1 \\
\object{HIP61257} &    HD 109195 & 12 33 12.1871 & -52 04 58.236 &        6.55 & 6.61 &   $125.8 \pm 0.6$ &          $17 \pm 3$ &    B9.5V &  $2.7_{-0.6}^{+0.4}$ &        $2.7 \pm 0.3$ &    1 \\
\object{HIP61585} &    * alf Mus & 12 37 11.0178 & -69 08 08.033 &         2.7 & 3.25 &  $116.0 \pm 11.2$ &          $15 \pm 3$ &      B2V &  $7.0_{-1.0}^{+3.0}$ &  $9.8_{-1.0}^{+1.1}$ &    1 \\
\object{HIP62058} &    HD 110506 & 12 43 09.1786 & -56 10 34.418 &        5.98 & 6.17 &   $123.5 \pm 0.6$ &          $15 \pm 3$ &    B7.5V &  $3.6_{-0.6}^{+0.5}$ &        $3.4 \pm 0.3$ &    0 \\
\object{HIP62327} &    HD 110956 & 12 46 22.7144 & -56 29 19.735 &        4.61 & 5.06 & $117.9 \pm 3.1^b$ &          $17 \pm 3$ &    B2.5V &  $6.0_{-0.6}^{+1.3}$ &        $5.4 \pm 0.5$ &    1 \\
\object{HIP62434} &    * bet Cru & 12 47 43.2687 & -59 41 19.579 & \textemdash & 1.99 &  $85.4 \pm 7.1^b$ &          $16 \pm 3$ &      B1V & $11.0_{-1.0}^{+4.0}$ &       $11.0 \pm 1.0$ &    1 \\
\object{HIP62786} &    HD 111774 &  12 51 56.931 &  -39 40 49.55 &        5.97 & 6.22 &   $140.3 \pm 1.1$ &          $17 \pm 3$ &    B7.5V &  $3.6_{-0.6}^{+0.5}$ &        $3.7 \pm 0.4$ &    0 \\
\object{HIP63003} &  * mu.01 Cru & 12 54 35.6242 & -57 10 40.528 &         4.0 & 4.58 &   $124.8 \pm 3.3$ &          $16 \pm 3$ &    B2V-V &  $7.0_{-1.0}^{+3.0}$ &        $7.1 \pm 0.7$ &    1 \\
\object{HIP63005} &  * mu.02 Cru & 12 54 36.8841 & -57 10 07.193 &        5.15 & 5.31 &   $121.0 \pm 1.7$ &          $16 \pm 3$ &      B5V &  $4.7_{-0.4}^{+0.4}$ &        $4.7 \pm 0.5$ &    1 \\
\object{HIP63210} &      * H Cen & 12 57 04.3514 & -51 11 55.512 &        5.14 & 5.34 &   $128.8 \pm 1.7$ &          $19 \pm 3$ &      B2V &  $7.0_{-1.0}^{+3.0}$ &        $4.8 \pm 0.5$ &    1 \\
\object{HIP63945} &      * f Cen & 13 06 16.7036 & -48 27 47.846 &        4.67 & 5.05 &   $122.6 \pm 2.3$ &          $15 \pm 3$ &      B4V &  $5.1_{-0.4}^{+0.3}$ &        $5.5 \pm 0.5$ &    1 \\
\object{HIP64004} &  * ksi02 Cen & 13 06 54.6394 & -49 54 22.487 &        4.22 & 4.84 &   $147.1 \pm 3.9$ &          $20 \pm 4$ &      B3V &  $5.4_{-0.3}^{+0.6}$ &        $7.0 \pm 0.7$ &    1 \\
\object{HIP65021} &    HD 115583 & 13 19 43.4213 & -67 21 51.567 &        7.25 & 7.22 &   $171.4 \pm 0.7$ &          $16 \pm 3$ &      B9V &  $2.8_{-0.6}^{+0.6}$ &        $2.9 \pm 0.3$ &    0 \\
\object{HIP65112} &  V* V964 Cen & 13 20 37.8253 & -52 44 52.169 &        5.43 &  5.8 &   $119.2 \pm 1.4$ &          $15 \pm 3$ &      B5V &  $4.7_{-0.4}^{+0.4}$ &        $4.0 \pm 0.4$ &    0 \\
\object{HIP66454} &    HD 118354 &  13 37 23.476 & -46 25 40.432 &        5.89 & 6.14 &   $143.0 \pm 2.7$ &          $17 \pm 3$ &      B8V &  $3.4_{-0.6}^{+0.5}$ &        $4.0 \pm 0.4$ &    1 \\
\object{HIP67464} &    * nu. Cen & 13 49 30.2765 &  -41 41 15.75 &        3.38 & 4.24 &   $124.3 \pm 5.4$ &      $18_{-5}^{+3}$ &      B2V &  $7.0_{-1.0}^{+3.0}$ &        $8.2 \pm 0.8$ &    1 \\
\object{HIP67669} &    * 3 Cen A & 13 51 49.6013 & -32 59 38.706 &        4.51 & 4.97 & $105.4 \pm 9.9^b$ &          $17 \pm 3$ &  B5V+B9V &  $4.7_{-0.4}^{+0.4}$ &        $4.8 \pm 0.5$ &    1 \\
\object{HIP67703} &      * N Cen & 13 52 04.8623 & -52 48 41.506 &        5.25 & 5.51 &    $93.1 \pm 0.8$ &          $16 \pm 3$ &      B8V &  $3.4_{-0.6}^{+0.5}$ &        $3.4 \pm 0.3$ &    1 \\
\object{HIP68245} &    * phi Cen & 13 58 16.2661 & -42 06 02.724 &        3.81 & 4.49 &   $140.8 \pm 5.4$ &          $17 \pm 3$ &      B2V &  $7.0_{-1.0}^{+3.0}$ &        $7.9 \pm 0.8$ &    0 \\
\object{HIP68282} &  * ups01 Cen & 13 58 40.7486 & -44 48 12.908 &        3.85 & 4.47 & $131.1 \pm 2.7^b$ &      $21_{-2}^{+4}$ &    B2V-V &  $7.0_{-1.0}^{+3.0}$ &        $6.6 \pm 0.7$ &    1 \\
\object{HIP68862} &    * chi Cen &  14 06 02.768 & -41 10 46.678 &        4.31 & 4.93 &   $154.1 \pm 5.1$ &          $25 \pm 5$ &      B2V &  $7.0_{-1.0}^{+3.0}$ &        $6.8 \pm 0.7$ &    0 \\
\object{HIP69011} &    HD 123247 &  14 07 40.808 & -48 42 14.498 &        6.42 & 6.43 &    $97.6 \pm 0.5$ &          $17 \pm 3$ &    B9.5V &  $2.7_{-0.6}^{+0.4}$ &        $2.3 \pm 0.2$ &    0 \\
\object{HIP69618} &  V* V795 Cen & 14 14 57.1383 & -57 05 10.049 &        5.01 & 4.77 &   $131.5 \pm 2.3$ &           $2 \pm 1$ &      B4V &  $5.1_{-0.4}^{+0.3}$ &  $6.3_{-3.0}^{+0.6}$ &    0 \\
\object{HIP70300} &      * a Cen & 14 23 02.2387 & -39 30 42.544 &        4.36 & 4.92 &   $120.1 \pm 2.6$ &      $17_{-5}^{+3}$ &      B2V &  $7.0_{-1.0}^{+3.0}$ &        $5.9 \pm 0.6$ &    2 \\
\object{HIP70626} &    HD 126475 & 14 26 49.8732 & -39 52 26.344 &        6.34 & 6.59 &   $141.0 \pm 1.1$ &          $15 \pm 3$ &      B9V &  $2.8_{-0.6}^{+0.6}$ &        $3.3 \pm 0.3$ &    0 \\
\object{HIP71352} &    * eta Cen & 14 35 30.4241 &  -42 09 28.17 &        2.25 & 2.96 &  $93.7 \pm 1.8^b$ &          $15 \pm 3$ &      B2V &  $7.0_{-1.0}^{+3.0}$ &        $9.7 \pm 1.0$ &    3 \\
\object{HIP71353} &    HD 127971 &  14 35 31.479 & -41 31 02.792 &        5.86 & 6.03 &   $108.3 \pm 2.6$ &          $15 \pm 3$ &      B7V &  $3.9_{-0.5}^{+0.4}$ &        $3.2 \pm 0.3$ &    1 \\
\object{HIP71453} &    HD 128207 & 14 36 44.1322 & -40 12 41.696 &        5.72 & 6.01 &   $136.4 \pm 1.5$ &          $16 \pm 3$ &      B8V &  $3.4_{-0.6}^{+0.5}$ &        $4.0 \pm 0.4$ &    0 \\
\object{HIP71536} &    * rho Lup & 14 37 53.2261 & -49 25 32.981 &         4.0 & 4.51 &   $107.0 \pm 2.3$ &          $14 \pm 4$ &    B3/4V &  $5.2_{-0.3}^{+0.8}$ &        $6.2 \pm 0.6$ &    1 \\
\object{HIP71860} &    * alf Lup & 14 41 55.7557 & -47 23 17.515 &        2.32 & 2.67 & $142.5 \pm 3.4^b$ &          $19 \pm 3$ &    B1.5V & $10.0_{-3.0}^{+1.0}$ &       $11.0 \pm 1.0$ &    1 \\
\object{HIP71865} &      * b Cen & 14 41 57.5905 &  -37 47 36.58 &        3.99 & 4.49 &    $99.7 \pm 3.1$ &          $15 \pm 2$ &    B2.5V &  $6.0_{-0.6}^{+1.3}$ &        $5.9 \pm 0.6$ &    1 \\
\object{HIP73266} &    HD 132094 & 14 58 24.2646 & -37 21 44.902 &        7.26 & 7.28 &   $170.9 \pm 1.2$ &          $15 \pm 3$ &      B9V &  $2.8_{-0.6}^{+0.6}$ &        $2.7 \pm 0.3$ &    1 \\
\object{HIP73624} &    HD 132955 & 15 02 59.2764 & -32 38 35.849 &        5.42 & 5.73 &   $146.7 \pm 2.8$ &          $17 \pm 3$ &      B3V &  $5.4_{-0.3}^{+0.6}$ &        $5.0 \pm 0.5$ &    1 \\
\object{HIP74100} &    HD 133937 &  15 08 39.198 & -42 52 04.522 &        5.81 &  6.1 &   $144.2 \pm 1.6$ &          $19 \pm 3$ &    B5/7V &  $4.3_{-0.4}^{+0.4}$ &        $4.1 \pm 0.4$ &    1 \\
\object{HIP74449} &      * e Lup & 15 12 49.5875 &  -44 30 01.49 &        4.79 & 5.28 &   $138.2 \pm 3.9$ &          $18 \pm 3$ &      B3V &  $5.4_{-0.3}^{+0.6}$ &        $5.7 \pm 0.6$ &    1 \\
\object{HIP74657} &    HD 135174 & 15 15 19.6424 &  -44 08 58.18 &        6.73 & 6.95 &   $154.9 \pm 1.2$ &          $18 \pm 3$ &      B9V &  $2.8_{-0.6}^{+0.6}$ &        $3.0 \pm 0.3$ &    0 \\
\object{HIP74752} &    HD 135454 & 15 16 37.1496 & -42 22 12.565 &        6.75 & 6.83 &   $130.9 \pm 0.7$ &          $14 \pm 3$ &    B9.5V &  $2.7_{-0.6}^{+0.4}$ &        $2.6 \pm 0.3$ &    1 \\
\object{HIP74950} &    V* GG Lup & 15 18 56.3746 & -40 47 17.596 &        5.57 & 6.21 &   $150.1 \pm 2.0$ &          $18 \pm 3$ &      B7V &  $3.9_{-0.5}^{+0.4}$ &        $4.4 \pm 0.4$ &    1 \\
\object{HIP75141} &    * del Lup & 15 21 22.3138 & -40 38 51.109 &        3.21 & 3.96 &  $148.6 \pm 10.6$ &          $16 \pm 4$ &    B1.5V & $10.0_{-3.0}^{+1.0}$ &  $9.8_{-1.0}^{+1.1}$ &    1 \\
\object{HIP75304} &  * phi02 Lup & 15 23 09.3506 & -36 51 30.557 &        4.51 & 4.94 & $159.2 \pm 5.1^b$ &          $15 \pm 3$ &      B4V &  $5.1_{-0.4}^{+0.3}$ &        $7.1 \pm 0.7$ &    1 \\
\object{HIP75647} &    HD 137432 & 15 27 18.1304 & -36 46 03.214 &        5.43 & 5.84 &   $142.3 \pm 2.4$ &          $17 \pm 3$ &      B5V &  $4.7_{-0.4}^{+0.4}$ &        $4.6 \pm 0.5$ &    1 \\
\object{HIP76048} &    HD 138221 & 15 31 50.2295 & -32 52 52.015 &        6.45 & 6.25 &   $163.7 \pm 1.0$ &           $2 \pm 1$ &      B7V &  $3.9_{-0.5}^{+0.4}$ &  $3.3_{-0.3}^{+1.6}$ &    0 \\
\object{HIP76126} &    * zet Lib & 15 32 55.2212 & -16 51 10.241 &        5.48 & 5.89 &   $230.8 \pm 8.4$ &          $11 \pm 3$ &      B3V &  $5.4_{-0.3}^{+0.6}$ &        $7.2 \pm 0.7$ &    1 \\
\object{HIP76591} &    HD 139233 & 15 38 32.6385 &  -39 09 38.52 &        6.58 & 6.76 &   $145.7 \pm 1.0$ &          $17 \pm 3$ &      B9V &  $2.8_{-0.6}^{+0.6}$ &        $3.0 \pm 0.3$ &    0 \\
\object{HIP76600} &    * tau Lib & 15 38 39.3694 & -29 46 39.895 &        3.65 & 4.12 & $112.5 \pm 2.5^b$ &      $13_{-3}^{+4}$ &    B2.5V &  $6.0_{-0.6}^{+1.3}$ &        $7.5 \pm 0.7$ &    1 \\
\object{HIP76633} &    HD 139486 & 15 39 00.0577 & -19 43 57.199 &        7.63 & 7.49 &   $164.7 \pm 1.0$ &       $7_{-2}^{+3}$ &      B9V &  $2.8_{-0.6}^{+0.6}$ &        $2.4 \pm 0.2$ &    0 \\
\object{HIP77562} &    HD 141168 & 15 50 07.0836 & -53 12 35.174 &        5.77 & 5.95 &    $96.5 \pm 0.5$ &          $17 \pm 3$ &      B9V &  $2.8_{-0.6}^{+0.6}$ &        $2.9 \pm 0.3$ &    0 \\
\object{HIP77968} &    HD 142256 & 15 55 22.8856 & -44 31 33.651 &        6.96 & 6.99 &   $181.9 \pm 1.0$ &      $17_{-2}^{+5}$ &      B8V &  $3.4_{-0.6}^{+0.5}$ &        $3.4 \pm 0.3$ &    0 \\
\object{HIP78104} &    * rho Sco & 15 56 53.0772 & -29 12 50.667 &        3.87 & 4.46 &   $136.6 \pm 5.5$ &      $14_{-3}^{+2}$ &    B2V-V &  $7.0_{-1.0}^{+3.0}$ &        $7.9 \pm 0.8$ &    1 \\
\object{HIP78168} &    HD 142883 & 15 57 40.4636 & -20 58 59.082 &        5.81 & 5.73 &   $157.4 \pm 1.3$ &          $10 \pm 3$ &      B3V &  $5.4_{-0.3}^{+0.6}$ &        $5.3 \pm 0.5$ &    1 \\
\object{HIP78207} &     * 48 Lib & 15 58 11.3682 & -14 16 45.681 &        4.74 & 4.59 &   $139.8 \pm 2.4$ & $0.5_{-0.3}^{+0.5}$ & B5Vp &  $4.7_{-0.4}^{+0.4}$ &        $4.7 \pm 0.5$ &    0 \\
\object{HIP78324} &    HD 143022 & 15 59 30.8817 & -40 51 54.598 &        8.12 & 7.49 &   $167.2 \pm 1.0$ &           $7 \pm 2$ &    B9.5V &  $2.7_{-0.6}^{+0.4}$ &        $2.1 \pm 0.2$ &    0 \\
\object{HIP78384} &    * eta Lup & 16 00 07.3279 & -38 23 48.138 &        3.41 &  4.1 & $135.5 \pm 3.3^b$ &          $14 \pm 3$ &    B2.5V &  $6.0_{-0.6}^{+1.3}$ &        $9.1 \pm 0.9$ &    1 \\
\object{HIP78655} &    HD 143699 & 16 03 24.1895 & -38 36 09.141 &        4.87 & 5.27 & $122.5 \pm 4.5^b$ &          $14 \pm 3$ &    B5/7V &  $4.3_{-0.4}^{+0.4}$ &        $5.1 \pm 0.5$ &    0 \\
\object{HIP78702} &    HD 143956 & 16 04 00.2383 & -19 46 02.926 &        7.76 & 7.24 &   $148.4 \pm 0.8$ &      $10_{-5}^{+2}$ &      B9V &  $2.8_{-0.6}^{+0.6}$ &        $2.4 \pm 0.2$ &    1 \\
\object{HIP78918} &    * tet Lup & 16 06 35.5439 & -36 48 08.173 &        4.16 &  4.7 &   $135.7 \pm 4.7$ &          $17 \pm 3$ &    B2.5V &  $6.0_{-0.6}^{+1.3}$ &        $6.9 \pm 0.7$ &    1 \\
\object{HIP78933} &    * ome Sco & 16 06 48.4249 & -20 40 09.079 &        3.92 & 4.01 &   $140.9 \pm 5.4$ &          $11 \pm 3$ &      B1V & $11.0_{-1.0}^{+4.0}$ &        $9.8 \pm 1.0$ &    0 \\
\object{HIP78968} &    HD 144586 & 16 07 14.9283 & -17 56 09.731 &        7.76 &  7.4 &   $149.1 \pm 0.6$ &       $9_{-2}^{+3}$ &      B9V &  $2.8_{-0.6}^{+0.6}$ &        $2.4 \pm 0.2$ &    0 \\
\object{HIP79044} &    HD 144591 & 16 08 04.3798 & -36 13 54.598 &        6.74 & 6.91 &   $138.5 \pm 0.9$ &          $17 \pm 3$ &      B9V &  $2.8_{-0.6}^{+0.6}$ &        $2.7 \pm 0.3$ &    1 \\
\object{HIP79404} &     * 13 Sco & 16 12 18.2046 & -27 55 34.953 &        4.53 & 4.98 &   $147.4 \pm 3.1$ &          $11 \pm 3$ &      B2V &  $7.0_{-1.0}^{+3.0}$ &        $7.4 \pm 0.7$ &    1 \\
\object{HIP80142} &    HD 147001 & 16 21 27.0327 & -48 11 19.041 &         6.5 & 6.62 &   $176.3 \pm 1.2$ &          $17 \pm 3$ &      B7V &  $3.9_{-0.5}^{+0.4}$ &        $3.9 \pm 0.4$ &    1 \\
\object{HIP80208} &    HD 147152 & 16 22 27.9955 & -49 34 20.468 &        5.07 &  5.4 & $141.8 \pm 6.4^b$ &          $17 \pm 3$ &      B3V &  $5.4_{-0.3}^{+0.6}$ &        $5.4 \pm 0.5$ &    0 \\
\object{HIP80569} &    * chi Oph & 16 27 01.4355 & -18 27 22.499 &         4.2 & 3.02 &   $152.9 \pm 4.6$ &       $7_{-4}^{+3}$ &      B2V &  $7.0_{-1.0}^{+3.0}$ &       $11.0 \pm 1.0$ &    1 \\
\object{HIP80911} &      * N Sco & 16 31 22.9344 & -34 42 15.725 &        4.18 & 4.69 &   $182.4 \pm 6.8$ &          $16 \pm 4$ &    B2V-V &  $7.0_{-1.0}^{+3.0}$ &        $8.8 \pm 0.9$ &    0 \\
\object{HIP81208} &    HD 149274 & 16 35 13.8392 & -35 43 28.725 &        6.63 & 6.77 &   $146.1 \pm 1.0$ &      $17_{-4}^{+3}$ &      B9V &  $2.8_{-0.6}^{+0.6}$ &        $2.6 \pm 0.3$ &    1 \\
\object{HIP81266} &    * tau Sco & 16 35 52.9528 & -28 12 57.661 &         2.8 & 3.57 &  $195.2 \pm 42.3$ &      $12_{-4}^{+2}$ &    B0.2V & $17.7_{-2.7}^{+0.8}$ & $12.0_{-1.0}^{+2.0}$ &    1 \\
\object{HIP81316} &    HD 149425 &  16 36 28.673 & -40 18 10.916 &        7.07 & 6.71 &   $184.4 \pm 1.0$ &          $18 \pm 4$ &      B9V &  $2.8_{-0.6}^{+0.6}$ &        $3.9 \pm 0.4$ &    1 \\
\object{HIP81472} & V* V1003 Sco & 16 38 26.2919 & -43 23 54.328 &        5.81 & 5.93 &   $190.9 \pm 1.9$ &          $17 \pm 3$ &      B2V &  $7.0_{-1.0}^{+3.0}$ &        $5.7 \pm 0.6$ &    1 \\
\object{HIP81474} &    HD 149914 & 16 38 28.6527 & -18 13 13.713 &        6.64 & 5.69 &   $154.4 \pm 0.6$ & $2.0_{-0.2}^{+0.5}$ &    B9.5V &  $2.7_{-0.6}^{+0.4}$ &  $3.1_{-0.3}^{+1.6}$ &    1 \\
\object{HIP81891} &    HD 150638 & 16 43 38.7239 & -32 06 21.406 &        6.45 & 6.63 &   $153.0 \pm 1.0$ &          $14 \pm 3$ &      B8V &  $3.4_{-0.6}^{+0.5}$ &        $3.4 \pm 0.3$ &    0 \\
\object{HIP81914} &    HD 150591 &  16 43 54.082 & -41 06 48.038 &        6.12 & 6.29 &   $174.6 \pm 1.6$ &          $17 \pm 3$ &    B6/7V &  $4.1_{-0.5}^{+0.4}$ &        $4.7 \pm 0.5$ &    1 \\
\object{HIP81972} &    HD 150742 & 16 44 42.5926 &  -40 50 22.83 &        5.61 & 5.84 &   $173.8 \pm 1.9$ &          $18 \pm 3$ &      B3V &  $5.4_{-0.3}^{+0.6}$ &        $5.6 \pm 0.6$ &    1 \\
\object{HIP82514} &  * mu.01 Sco & 16 51 52.2283 & -38 02 50.638 &        3.07 &  3.7 & $169.5 \pm 8.6^c$ &          $20 \pm 4$ &   B1V+BV & $11.0_{-1.0}^{+4.0}$ &       $10.0 \pm 1.0$ &    1 \\
\object{HIP82545} &  * mu.02 Sco & 16 52 20.1453 & -38 01 03.125 &        3.54 & 4.29 & $169.5 \pm 8.6^c$ &          $20 \pm 4$ &      B2V &  $7.0_{-1.0}^{+3.0}$ &        $9.1 \pm 0.3$ &    0 \\
\hline \hline 
\caption{Revised stellar parameters for all the targets of the BEAST survey. The multiplicity flag $m_f$ indicates whether the star is known to be a multiple system (1: \citealt{Gratton.2023}, 2: this work, 3: \citealt{chini12}) or not (0).}
\label{tab:star_table}
\end{longtable}} \par 
\tablefoot{$^a$: coordinates are given in J2000 IRCS. $^b$: using Hipparcos parallax. $^c$: see \citet{Squicciarini.2022}.}

\newpage
\section{Observation table}
\label{appendix:obs_table} 
\begin{center}
\small

{\setlength\tabcolsep{2pt} 
\begin{longtable}{cccccccccc}
\multicolumn{1}{c}{\textbf{HIP ID}} & \multicolumn{1}{c}{\textbf{OTHER ID}} & \multicolumn{1}{c}{\textbf{DATE OBS}} & \multicolumn{1}{c}{\textbf{FILTER}} & \multicolumn{1}{c}{\textbf{DIT(s)$\times$NDIT$\times$NEXP}} & \multicolumn{1}{c}{\textbf{$\Delta$ PA (\degree)$^a$}} & \multicolumn{1}{c}{\textbf{Seeing (")$^b$}} & \multicolumn{1}{c}{\textbf{Airmass$^b$}} & \multicolumn{1}{c}{\textbf{$\tau_0$ (ms)$^{a,b}$}} & \multicolumn{1}{c}{\textbf{Program ID}}\\ \hline 
\endfirsthead

\multicolumn{10}{c}%
{{\bfseries \tablename\ \thetable{} -- continued from previous page}} \\
\multicolumn{1}{c}{\textbf{HIP ID}} & \multicolumn{1}{c}{\textbf{OTHER ID}} & \multicolumn{1}{c}{\textbf{DATE OBS}} & \multicolumn{1}{c}{\textbf{FILTER}} & \multicolumn{1}{c}{\textbf{DIT(s)$\times$NDIT$\times$NEXP}} & \multicolumn{1}{c}{\textbf{$\Delta$ PA (\degree)$^a$}} & \multicolumn{1}{c}{\textbf{Seeing (")$^b$}} & \multicolumn{1}{c}{\textbf{Airmass$^b$}} & \multicolumn{1}{c}{\textbf{$\tau_0$ (ms)$^{a,b}$}} & \multicolumn{1}{c}{\textbf{Program ID}}\\ \hline 
\endhead

\hline \multicolumn{10}{r}{{\textit{Continued on next page}}} \\ 
\endfoot

\endlastfoot

\hline 
\hline 
HIP  67669 & *   3 Cen A & 2019-02-22 & DB\_K12 & 32x6x46 & 72.5 & 0.5 & 1.01 & 12.4 & 1101.C-0258(B) \\ 
HIP  67669 & *   3 Cen A & 2019-02-22 & OBS\_H & 64x1x46 & 71.9 & 0.5 & 1.01 & 12.4 & 1101.C-0258(B) \\ 
\hline 
HIP  50847 & * L Car & 2019-01-26 & DB\_K12 & 64x3x46 & 15.2 & 0.41 & 1.49 & 17.0 & 1101.C-0258(A) \\ 
\hline 
HIP  52742 & HD  93563 & 2018-05-14 & DB\_K12 & 64x3x46 & 22.0 & 0.78 & 1.18 & 3.1 & 1101.C-0258(A) \\ 
HIP  52742 & HD  93563 & 2018-05-14 & OBS\_H & 64x1x46 & 22.1 & 0.79 & 1.18 & 3.1 & 1101.C-0258(A) \\ 
HIP  52742 & HD  93563 & 2019-01-26 & DB\_K12 & 64x3x46 & 22.1 & 0.51 & 1.18 & 10.7 & 1101.C-0258(A) \\ 
HIP  52742 & HD  93563 & 2019-01-26 & OBS\_H & 64x1x46 & 22.2 & 0.51 & 1.18 & 10.8 & 1101.C-0258(A) \\ 
HIP  52742 & HD  93563 & 2020-03-20 & DB\_K12 & 64x3x46 & 22.3 & 0.95 & 1.18 & 5.0 & 1101.C-0258(D) \\ 
HIP  52742 & HD  93563 & 2020-03-20 & OBS\_H & 64x1x46 & 22.4 & 0.96 & 1.18 & 5.1 & 1101.C-0258(D) \\ 
\hline 
HIP  54767 & HD  97583 & 2019-02-24 & DB\_K12 & 32x6x42 & 11.7 & 0.55 & 1.3 & 19.5 & 1101.C-0258(B) \\ 
HIP  54767 & HD  97583 & 2019-02-24 & OBS\_H & 64x1x42 & 16.0 & 0.56 & 1.3 & 18.4 & 1101.C-0258(B) \\ 
HIP  54767 & HD  97583 & 2019-03-12 & DB\_K12 & 64x3x46 & 18.5 & 0.59 & 1.3 & 8.3 & 1101.C-0258(B) \\ 
HIP  54767 & HD  97583 & 2019-03-12 & OBS\_H & 64x1x46 & 18.6 & 0.59 & 1.3 & 8.3 & 1101.C-0258(B) \\ 
HIP  54767 & HD  97583 & 2020-02-21 & DB\_K12 & 64x3x46 & 18.7 & 0.64 & 1.3 & 6.5 & 1101.C-0258(D) \\ 
HIP  54767 & HD  97583 & 2020-02-21 & OBS\_H & 64x1x46 & 18.7 & 0.65 & 1.3 & 6.5 & 1101.C-0258(D) \\ 
\hline 
HIP  58452 & HD 104080 & 2018-04-09 & DB\_K12 & 32x6x46 & 31.9 & 0.5 & 1.08 & 9.5 & 1101.C-0258(A) \\ 
HIP  58452 & HD 104080 & 2018-04-09 & OBS\_H & 32x1x46 & 32.5 & 0.49 & 1.08 & 9.9 & 1101.C-0258(A) \\ 
HIP  58452 & HD 104080 & 2020-02-01 & DB\_K12 & 64x3x46 & 32.4 & 0.52 & 1.07 & 4.4 & 1101.C-0258(C) \\ 
HIP  58452 & HD 104080 & 2020-02-01 & OBS\_H & 64x1x46 & 32.4 & 0.52 & 1.07 & 4.3 & 1101.C-0258(C) \\ 
\hline 
HIP  58901 & HD 104900 & 2019-02-20 & DB\_K12 & 64x3x46 & 20.7 & 1.11 & 1.22 & 3.8 & 1101.C-0258(B) \\ 
HIP  58901 & HD 104900 & 2019-02-20 & OBS\_H & 64x1x46 & 20.8 & 1.11 & 1.22 & 3.9 & 1101.C-0258(B) \\ 
HIP  58901 & HD 104900 & 2021-03-17 & DB\_K12 & 64x3x46 & 20.8 & 0.54 & 1.22 & 5.4 & 1101.C-0258(D) \\ 
HIP  58901 & HD 104900 & 2021-03-17 & OBS\_H & 64x1x46 & 20.9 & 0.54 & 1.22 & 5.5 & 1101.C-0258(D) \\ 
\hline 
HIP  59173 & HD 105382 & 2019-03-12 & DB\_K12 & 32x6x46 & 26.9 & 0.58 & 1.12 & 6.9 & 1101.C-0258(B) \\ 
HIP  59173 & HD 105382 & 2019-03-12 & OBS\_H & 64x1x46 & 26.6 & 0.58 & 1.12 & 6.8 & 1101.C-0258(B) \\ 
HIP  59173 & HD 105382 & 2020-02-21 & DB\_K12 & 32x6x46 & 27.2 & 0.73 & 1.12 & 5.2 & 1101.C-0258(D) \\ 
HIP  59173 & HD 105382 & 2020-02-21 & OBS\_H & 64x1x46 & 26.7 & 0.72 & 1.12 & 5.3 & 1101.C-0258(D) \\ 
HIP  59173 & HD 105382 & 2020-03-02 & DB\_K12 & 32x6x46 & 27.2 & 0.55 & 1.12 & 10.8 & 1101.C-0258(D) \\ 
HIP  59173 & HD 105382 & 2020-03-02 & OBS\_H & 64x1x46 & 26.8 & 0.55 & 1.12 & 10.7 & 1101.C-0258(D) \\ 
\hline 
HIP  60009 & * zet Cru & 2019-02-23 & DB\_K12 & 32x6x46 & 19.0 & 0.57 & 1.3 & 11.5 & 1101.C-0258(B) \\ 
HIP  60009 & * zet Cru & 2019-02-23 & OBS\_H & 32x1x46 & 19.3 & 0.58 & 1.3 & 11.2 & 1101.C-0258(B) \\ 
HIP  60009 & * zet Cru & 2020-02-08 & DB\_K12 & 32x6x46 & 19.1 & 0.52 & 1.3 & 7.8 & 1101.C-0258(D) \\ 
HIP  60009 & * zet Cru & 2020-02-08 & OBS\_H & 32x1x46 & 19.3 & 0.52 & 1.3 & 7.8 & 1101.C-0258(D) \\ 
\hline 
HIP  60379 & HD 107696 & 2019-02-24 & DB\_K12 & 32x6x44 & 22.0 & 0.43 & 1.2 & 16.0 & 1101.C-0258(B) \\ 
HIP  60379 & HD 107696 & 2019-02-24 & OBS\_H & 64x1x44 & 21.7 & 0.43 & 1.2 & 16.0 & 1101.C-0258(B) \\ 
HIP  60379 & HD 107696 & 2021-03-19 & DB\_K12 & 64x3x46 & 21.5 & 0.44 & 1.2 & 5.0 & 1101.C-0258(D) \\ 
HIP  60379 & HD 107696 & 2021-03-19 & OBS\_H & 64x1x46 & 21.5 & 0.44 & 1.2 & 5.0 & 1101.C-0258(D) \\ 
\hline 
HIP  60823 & * sig Cen & 2019-02-25 & DB\_K12 & 32x6x46 & 27.6 & 0.91 & 1.11 & 4.7 & 1101.C-0258(B) \\ 
HIP  60823 & * sig Cen & 2019-02-25 & OBS\_H & 32x1x46 & 28.1 & 0.94 & 1.11 & 4.5 & 1101.C-0258(B) \\ 
HIP  60823 & * sig Cen & 2021-03-07 & DB\_K12 & 32x6x46 & 27.6 & 0.44 & 1.11 & 16.9 & 1101.C-0258(D) \\ 
HIP  60823 & * sig Cen & 2021-03-07 & OBS\_H & 32x1x46 & 28.0 & 0.44 & 1.11 & 16.7 & 1101.C-0258(D) \\ 
\hline 
HIP  61585 & * alf Mus & 2019-03-15 & DB\_K12 & 8x23x45 & 18.0 & 0.59 & 1.4 & 5.0 & 1101.C-0258(B) \\ 
HIP  61585 & * alf Mus & 2019-03-15 & OBS\_H & 8x1x45 & 18.5 & 0.57 & 1.4 & 5.2 & 1101.C-0258(B) \\ 
HIP  61585 & * alf Mus & 2021-04-06 & DB\_K12 & 8x23x45 & 18.0 & 0.57 & 1.41 & 12.2 & 1101.C-0258(D) \\ 
HIP  61585 & * alf Mus & 2021-04-06 & OBS\_H & 8x1x45 & 18.4 & 0.57 & 1.41 & 12.5 & 1101.C-0258(D) \\ 
\hline 
HIP  62058 & HD 110506 & 2019-03-16 & DB\_K12 & 64x3x46 & 22.2 & 0.85 & 1.18 & 4.0 & 1101.C-0258(B) \\ 
HIP  62058 & HD 110506 & 2019-03-16 & OBS\_H & 64x1x46 & 20.8 & 0.85 & 1.18 & 4.0 & 1101.C-0258(B) \\ 
HIP  62058 & HD 110506 & 2021-04-06 & DB\_K12 & 64x3x46 & 18.8 & 0.87 & 1.24 & 6.9 & 1101.C-0258(D) \\ 
HIP  62058 & HD 110506 & 2021-04-06 & OBS\_H & 64x1x46 & 18.9 & 0.87 & 1.24 & 7.0 & 1101.C-0258(D) \\ 
\hline 
HIP  62327 & HD 110956 & 2019-03-29 & DB\_K12 & 32x6x46 & 22.7 & 0.72 & 1.18 & 7.2 & 1101.C-0258(B) \\ 
HIP  62327 & HD 110956 & 2019-03-29 & OBS\_H & 64x1x46 & 22.5 & 0.73 & 1.18 & 7.2 & 1101.C-0258(B) \\ 
HIP  62327 & HD 110956 & 2020-03-17 & DB\_K12 & 32x6x46 & 22.8 & 0.76 & 1.18 & 4.4 & 1101.C-0258(D) \\ 
HIP  62327 & HD 110956 & 2020-03-17 & OBS\_H & 64x1x46 & 22.4 & 0.76 & 1.18 & 4.4 & 1101.C-0258(D) \\ 
\hline 
HIP  62434 & * bet Cru & 2019-04-01 & DB\_K12 & 2x72x42 & 22.6 & 0.87 & 1.22 & 3.2 & 1101.C-0258(B) \\ 
HIP  62434 & * bet Cru & 2022-04-05 & DB\_K12 & 2x72x42 & 22.6 & 0.49 & 1.23 & 6.4 & 1101.C-0258(D) \\ 
\hline 
HIP  63003 & * mu.01 Cru & 2018-06-04 & DB\_K12 & 32x6x46 & 18.1 & 0.44 & 1.19 & 3.5 & 1101.C-0258(A) \\ 
HIP  63003 & * mu.01 Cru & 2018-06-04 & OBS\_H & 32x1x46 & 17.0 & 0.45 & 1.19 & 3.5 & 1101.C-0258(A) \\ 
HIP  63003 & * mu.01 Cru & 2019-04-09 & DB\_K12 & 16x12x46 & 22.9 & 0.55 & 1.19 & 9.0 & 1101.C-0258(C) \\ 
HIP  63003 & * mu.01 Cru & 2019-04-09 & OBS\_H & 16x1x46 & 23.9 & 0.56 & 1.19 & 8.8 & 1101.C-0258(C) \\ 
\hline 
HIP  63005 & * mu.02 Cru & 2019-03-05 & DB\_K12 & 64x3x46 & 21.8 & 0.51 & 1.19 & 5.0 & 1101.C-0258(B) \\ 
HIP  63005 & * mu.02 Cru & 2019-03-05 & OBS\_H & 64x1x46 & 21.9 & 0.51 & 1.19 & 5.0 & 1101.C-0258(B) \\ 
HIP  63005 & * mu.02 Cru & 2022-02-27 & DB\_K12 & 64x3x46 & 22.0 & 0.37 & 1.19 & 12.0 & 1101.C-0258(D) \\ 
HIP  63005 & * mu.02 Cru & 2022-02-27 & OBS\_H & 64x1x46 & 22.0 & 0.37 & 1.19 & 12.1 & 1101.C-0258(D) \\ 
\hline 
HIP  63945 & * f Cen & 2018-04-23 & DB\_K12 & 32x6x46 & 29.3 & 0.69 & 1.1 & 7.3 & 1101.C-0258(A) \\ 
HIP  63945 & * f Cen & 2018-04-23 & OBS\_H & 32x1x46 & 29.8 & 0.69 & 1.1 & 7.5 & 1101.C-0258(A) \\ 
HIP  63945 & * f Cen & 2019-04-10 & DB\_K12 & 32x3x42 & 14.9 & 0.56 & 1.1 & 13.2 & 1101.C-0258(C) \\ 
HIP  63945 & * f Cen & 2019-04-10 & OBS\_H & 32x1x42 & 20.0 & 0.59 & 1.1 & 12.9 & 1101.C-0258(C) \\ 
\hline 
HIP  65021 & HD 115583 & 2018-05-17 & DB\_K12 & 64x3x46 & 17.3 & 0.3 & 1.37 & 15.6 & 1101.C-0258(A) \\ 
HIP  65021 & HD 115583 & 2018-05-17 & OBS\_H & 64x1x46 & 17.4 & 0.3 & 1.37 & 15.6 & 1101.C-0258(A) \\ 
HIP  65021 & HD 115583 & 2020-02-21 & DB\_K12 & 64x3x46 & 17.5 & 0.96 & 1.37 & 4.4 & 1101.C-0258(C) \\ 
HIP  65021 & HD 115583 & 2020-02-21 & OBS\_H & 64x1x46 & 17.6 & 0.96 & 1.37 & 4.5 & 1101.C-0258(C) \\ 
\hline 
HIP  65112 & V* V964 Cen & 2019-03-21 & DB\_K12 & 64x3x46 & 24.8 & 0.43 & 1.14 & 7.4 & 1101.C-0258(B) \\ 
HIP  65112 & V* V964 Cen & 2019-03-21 & OBS\_H & 64x1x46 & 25.0 & 0.43 & 1.14 & 7.4 & 1101.C-0258(B) \\ 
HIP  65112 & V* V964 Cen & 2020-03-02 & DB\_K12 & 64x3x46 & 25.1 & 0.61 & 1.14 & 7.5 & 1101.C-0258(D) \\ 
HIP  65112 & V* V964 Cen & 2020-03-02 & OBS\_H & 64x1x46 & 25.1 & 0.61 & 1.14 & 7.5 & 1101.C-0258(D) \\ 
HIP  65112 & V* V964 Cen & 2021-04-16 & DB\_H23 & 64x3x46 & 25.0 & 0.55 & 1.14 & 5.2 & 1101.C-0258(C) \\ 
HIP  65112 & V* V964 Cen & 2021-04-16 & OBS\_YJ & 64x1x46 & 24.8 & 0.55 & 1.14 & 5.1 & 1101.C-0258(C) \\ 
\hline 
HIP  66454 & HD 118354 & 2018-05-16 & DB\_K12 & 64x3x46 & 30.7 & 0.52 & 1.08 & 5.9 & 1101.C-0258(A) \\ 
HIP  66454 & HD 118354 & 2018-05-16 & OBS\_H & 64x1x46 & 30.8 & 0.52 & 1.08 & 5.9 & 1101.C-0258(A) \\ 
HIP  66454 & HD 118354 & 2020-02-24 & DB\_K12 & 64x3x46 & 31.6 & 0.4 & 1.08 & 12.4 & 1101.C-0258(C) \\ 
HIP  66454 & HD 118354 & 2020-02-24 & OBS\_H & 64x1x46 & 31.3 & 0.4 & 1.08 & 12.1 & 1101.C-0258(C) \\ 
\hline 
HIP  67464 & * nu. Cen & 2018-05-05 & DB\_K12 & 2x96x42 & 10.7 & 0.28 & 1.05 & 11.7 & 1101.C-0258(A) \\ 
HIP  67464 & * nu. Cen & 2018-05-05 & OBS\_H & 16x1x42 & 32.4 & 0.26 & 1.05 & 11.6 & 1101.C-0258(A) \\ 
HIP  67464 & * nu. Cen & 2019-02-23 & DB\_K12 & 16x12x46 & 41.0 & 0.38 & 1.05 & 10.0 & 1101.C-0258(A) \\ 
HIP  67464 & * nu. Cen & 2019-02-23 & OBS\_H & 16x1x46 & 42.9 & 0.37 & 1.05 & 10.2 & 1101.C-0258(A) \\ 
\hline 
HIP  67703 & * N Cen & 2019-02-24 & DB\_K12 & 32x6x42 & 15.8 & 0.66 & 1.14 & 12.0 & 1101.C-0258(B) \\ 
HIP  67703 & * N Cen & 2019-02-24 & OBS\_H & 64x1x42 & 17.1 & 0.67 & 1.14 & 11.7 & 1101.C-0258(B) \\ 
HIP  67703 & * N Cen & 2019-02-25 & DB\_K12 & 64x3x46 & 24.7 & 1.37 & 1.14 & 3.5 & 1101.C-0258(B) \\ 
HIP  67703 & * N Cen & 2019-02-25 & OBS\_H & 64x1x46 & 24.8 & 1.37 & 1.14 & 3.5 & 1101.C-0258(B) \\ 
HIP  67703 & * N Cen & 2019-03-24 & DB\_K12 & 64x3x46 & 24.9 & 0.61 & 1.14 & 8.6 & 1101.C-0258(B) \\ 
HIP  67703 & * N Cen & 2019-03-24 & OBS\_H & 64x1x46 & 24.95 & 0.61 & 1.14 & 8.6 & 1101.C-0258(B) \\ 
HIP  67703 & * N Cen & 2020-03-20 & DB\_K12 & 64x3x46 & 25.0 & 0.87 & 1.14 & 4.6 & 1101.C-0258(D) \\ 
HIP  67703 & * N Cen & 2020-03-20 & OBS\_H & 64x1x46 & 26.39 & 1.15 & 1.14 & 3.8 & 1101.C-0258(B) \\ 
\hline 
HIP  68245 & * phi Cen & 2018-05-28 & DB\_K12 & 32x6x46 & 39.1 & 0.68 & 1.05 & 3.9 & 1101.C-0258(A) \\ 
HIP  68245 & * phi Cen & 2018-05-28 & OBS\_H & 32x1x46 & 39.7 & 0.68 & 1.05 & 3.9 & 1101.C-0258(A) \\ 
\hline 
HIP  68282 & * ups01 Cen & 2019-03-27 & DB\_K12 & 16x12x42 & 34.9 & 0.52 & 1.07 & 6.2 & 1101.C-0258(B) \\ 
HIP  68282 & * ups01 Cen & 2019-03-27 & OBS\_H & 16x1x42 & 36.4 & 0.52 & 1.07 & 6.4 & 1101.C-0258(B) \\ 
HIP  68282 & * ups01 Cen & 2021-03-07 & DB\_K12 & 32x6x46 & 34.3 & 0.35 & 1.07 & 15.3 & 1101.C-0258(D) \\ 
HIP  68282 & * ups01 Cen & 2021-03-07 & OBS\_H & 32x1x46 & 34.8 & 0.35 & 1.07 & 15.6 & 1101.C-0258(D) \\ 
\hline 
HIP  69011 & HD 123247 & 2019-05-31 & DB\_K12 & 64x3x46 & 28.8 & 0.81 & 1.1 & 2.7 & 1101.C-0258(B) \\ 
HIP  69011 & HD 123247 & 2019-05-31 & OBS\_H & 64x1x46 & 28.9 & 0.81 & 1.1 & 2.7 & 1101.C-0258(B) \\ 
\hline 
HIP  70300 & * a Cen & 2019-03-16 & DB\_K12 & 32x6x46 & 44.9 & 1.12 & 1.04 & 3.2 & 1101.C-0258(A) \\ 
HIP  70300 & * a Cen & 2019-03-16 & OBS\_H & 64x1x46 & 44.5 & 1.12 & 1.04 & 3.2 & 1101.C-0258(A) \\ 
HIP  70300 & * a Cen & 2019-04-09 & DB\_K12 & 32x6x46 & 44.9 & 0.6 & 1.04 & 5.9 & 1101.C-0258(A) \\ 
HIP  70300 & * a Cen & 2019-04-09 & OBS\_H & 64x1x46 & 44.5 & 0.61 & 1.04 & 5.9 & 1101.C-0258(A) \\ 
\hline 
HIP  70626 & HD 126475 & 2018-04-23 & DB\_K12 & 64x3x46 & 43.4 & 0.53 & 1.04 & 9.2 & 1101.C-0258(A) \\ 
HIP  70626 & HD 126475 & 2018-04-23 & OBS\_H & 64x1x46 & 43.4 & 0.53 & 1.04 & 9.1 & 1101.C-0258(A) \\ 
\hline 
HIP  71352 & * eta Cen & 2019-03-12 & DB\_K12 & 4x42x44 & 37.5 & 0.72 & 1.06 & 8.0 & 1101.C-0258(B) \\ 
HIP  71352 & * eta Cen & 2019-03-12 & OBS\_H & 4x1x44 & 38.1 & 0.71 & 1.06 & 7.9 & 1101.C-0258(B) \\ 
HIP  71352 & * eta Cen & 2020-03-02 & DB\_K12 & 4x42x44 & 41.2 & 0.63 & 1.05 & 6.8 & 1101.C-0258(D) \\ 
HIP  71352 & * eta Cen & 2020-03-02 & OBS\_H & 4x1x44 & 42.1 & 0.63 & 1.05 & 6.9 & 1101.C-0258(D) \\ 
\hline 
HIP  71536 & * rho Lup & 2019-03-15 & DB\_K12 & 32x6x46 & 28.2 & 0.89 & 1.11 & 3.3 & 1101.C-0258(B) \\ 
HIP  71536 & * rho Lup & 2019-03-15 & OBS\_H & 32x1x46 & 28.7 & 0.92 & 1.11 & 3.3 & 1101.C-0258(B) \\ 
HIP  71536 & * rho Lup & 2020-03-11 & OBS\_H & 32x1x42 & 33.8 & 0.76 & 1.1 & 5.6 & 1101.C-0258(D) \\ 
HIP  71536 & * rho Lup & 2021-04-16 & DB\_K12 & 32x6x46 & 28.1 & 0.82 & 1.11 & 4.2 & 1101.C-0258(D) \\ 
HIP  71536 & * rho Lup & 2021-04-16 & OBS\_H & 32x1x46 & 28.5 & 0.83 & 1.11 & 4.2 & 1101.C-0258(D) \\ 
HIP  71536 & * rho Lup & 2021-07-17 & DB\_K12 & 32x6x46 & 28.3 & 0.61 & 1.11 & 4.1 & 1101.C-0258(D) \\ 
HIP  71536 & * rho Lup & 2021-07-17 & OBS\_H & 32x1x46 & 28.7 & 0.62 & 1.11 & 3.9 & 1101.C-0258(D) \\ 
\hline 
HIP  71865 & * b Cen & 2019-03-21 & DB\_K12 & 16x12x42 & 51.5 & 0.44 & 1.03 & 7.3 & 1101.C-0258(B) \\ 
HIP  71865 & * b Cen & 2019-03-21 & OBS\_H & 16x1x42 & 53.8 & 0.44 & 1.03 & 7.6 & 1101.C-0258(B) \\ 
HIP  71865 & * b Cen & 2021-04-11 & DB\_J23 & 16x12x23 & 49.1 & 0.7 & 1.03 & 4.4 & 1101.C-0258(D) \\ 
\hline 
HIP  74100 & HD 133937 & 2019-03-24 & DB\_K12 & 64x3x46 & 36.6 & 0.52 & 1.06 & 9.4 & 1101.C-0258(A) \\ 
HIP  74100 & HD 133937 & 2019-03-24 & OBS\_H & 64x1x46 & 36.7 & 0.54 & 1.06 & 9.3 & 1101.C-0258(A) \\ 
HIP  74100 & HD 133937 & 2020-03-07 & DB\_K12 & 64x3x46 & 37.0 & 1.02 & 1.06 & 4.7 & 1101.C-0258(D) \\ 
HIP  74100 & HD 133937 & 2020-03-07 & OBS\_H & 64x1x46 & 37.0 & 1.02 & 1.06 & 4.7 & 1101.C-0258(D) \\ 
\hline 
HIP  74950 & V* GG Lup & 2019-03-27 & DB\_K12 & 64x3x46 & 25.9 & 0.71 & 1.04 & 3.9 & 1101.C-0258(A) \\ 
HIP  74950 & V* GG Lup & 2019-03-27 & OBS\_H & 64x1x46 & 25.9 & 0.71 & 1.04 & 3.9 & 1101.C-0258(A) \\ 
HIP  74950 & V* GG Lup & 2019-04-29 & DB\_K12 & 64x3x46 & 41.4 & 0.45 & 1.04 & 9.8 & 1101.C-0258(A) \\ 
HIP  74950 & V* GG Lup & 2019-04-29 & OBS\_H & 64x1x46 & 41.4 & 0.45 & 1.04 & 9.8 & 1101.C-0258(A) \\ 
HIP  74950 & V* GG Lup & 2020-03-23 & DB\_K12 & 64x3x46 & 41.4 & 0.61 & 1.04 & 9.4 & 1101.C-0258(D) \\ 
HIP  74950 & V* GG Lup & 2020-03-23 & OBS\_H & 64x1x46 & 41.5 & 0.61 & 1.04 & 9.4 & 1101.C-0258(D) \\ 
\hline 
HIP  76591 & HD 139233 & 2018-04-23 & DB\_K12 & 64x3x46 & 45.2 & 0.63 & 1.03 & 9.3 & 1101.C-0258(A) \\ 
HIP  76591 & HD 139233 & 2018-04-23 & OBS\_H & 64x1x46 & 45.3 & 0.64 & 1.03 & 9.2 & 1101.C-0258(A) \\ 
HIP  76591 & HD 139233 & 2021-07-21 & DB\_K12 & 64x3x46 & 45.3 & 0.42 & 1.04 & 7.8 & 1101.C-0258(C) \\ 
HIP  76591 & HD 139233 & 2021-07-21 & OBS\_H & 64x1x46 & 45.4 & 0.42 & 1.04 & 7.8 & 1101.C-0258(C) \\ 
\hline 
HIP  76600 & * tau Lib & 2019-04-09 & OBS\_H & 16x1x46 & 3.6 & 0.47 & 1.01 & 7.7 & 1101.C-0258(A) \\ 
HIP  76600 & * tau Lib & 2019-07-27 & DB\_K12 & 16x12x46 & 90.9 & 0.54 & 1.01 & 4.6 & 1101.C-0258(A) \\ 
\hline 
HIP  76633 & HD 139486 & 2018-05-16 & DB\_K12 & 64x3x46 & 100.2 & 0.69 & 1.01 & 7.9 & 1101.C-0258(A) \\ 
HIP  76633 & HD 139486 & 2018-05-16 & OBS\_H & 64x1x46 & 100.4 & 0.69 & 1.01 & 8.0 & 1101.C-0258(A) \\ 
\hline 
HIP  77562 & HD 141168 & 2019-03-21 & DB\_K12 & 64x3x46 & 24.5 & 0.88 & 1.14 & 5.3 & 1101.C-0258(B) \\ 
HIP  77562 & HD 141168 & 2019-03-21 & OBS\_H & 64x1x46 & 24.6 & 0.88 & 1.14 & 5.1 & 1101.C-0258(B) \\ 
HIP  77562 & HD 141168 & 2021-05-17 & DB\_K12 & 64x3x46 & 24.8 & 0.43 & 1.14 & 7.5 & 1101.C-0258(D) \\ 
HIP  77562 & HD 141168 & 2021-05-17 & OBS\_H & 64x1x46 & 24.8 & 0.43 & 1.14 & 7.4 & 1101.C-0258(D) \\ 
\hline 
HIP  77968 & HD 142256 & 2018-06-23 & DB\_K12 & 64x3x46 & 34.1 & 0.62 & 1.07 & 5.3 & 1101.C-0258(A) \\ 
HIP  77968 & HD 142256 & 2018-06-23 & OBS\_H & 64x1x46 & 34.2 & 0.62 & 1.07 & 5.3 & 1101.C-0258(A) \\ 
HIP  77968 & HD 142256 & 2021-06-27 & DB\_K12 & 64x3x46 & 34.3 & 0.64 & 1.07 & 4.2 & 1101.C-0258(C) \\ 
HIP  77968 & HD 142256 & 2021-06-27 & OBS\_H & 64x1x46 & 34.4 & 0.64 & 1.07 & 4.2 & 1101.C-0258(C) \\ 
\hline 
HIP  78207 & *  48 Lib & 2018-04-09 & DB\_K12 & 16x12x46 & 46.3 & 0.48 & 1.03 & 7.8 & 1101.C-0258(A) \\ 
HIP  78207 & *  48 Lib & 2018-04-09 & OBS\_H & 32x1x46 & 46.2 & 0.48 & 1.03 & 7.9 & 1101.C-0258(A) \\ 
\hline 
HIP  79044 & HD 144591 & 2019-03-22 & DB\_K12 & 64x3x46 & 53.9 & 0.81 & 1.02 & 5.7 & 1101.C-0258(B) \\ 
HIP  79044 & HD 144591 & 2019-03-22 & OBS\_H & 64x1x46 & 53.9 & 0.81 & 1.02 & 5.7 & 1101.C-0258(B) \\ 
HIP  79044 & HD 144591 & 2021-04-11 & DB\_K12 & 64x3x46 & 55.0 & 0.64 & 1.02 & 4.7 & 1101.C-0258(D) \\ 
HIP  79044 & HD 144591 & 2021-04-11 & OBS\_H & 64x1x46 & 55.2 & 0.64 & 1.02 & 4.7 & 1101.C-0258(D) \\ 
\hline 
HIP  80911 & * N Sco & 2018-08-13 & DB\_K12 & 32x6x46 & 63.0 & 0.78 & 1.02 & 4.7 & 1101.C-0258(A) \\ 
HIP  80911 & * N Sco & 2018-08-13 & OBS\_H & 32x1x46 & 63.9 & 0.77 & 1.02 & 4.7 & 1101.C-0258(A) \\ 
HIP  80911 & * N Sco & 2019-07-06 & DB\_K12 & 32x6x46 & 63.1 & 0.6 & 1.02 & 5.1 & 1101.C-0258(C) \\ 
HIP  80911 & * N Sco & 2019-07-06 & OBS\_H & 32x1x46 & 63.9 & 0.59 & 1.02 & 5.1 & 1101.C-0258(C) \\ 
\hline 
HIP  82514 & * mu.01 Sco & 2018-05-15 & DB\_K12 & 8x24x46 & 48.8 & 0.7 & 1.03 & 3.0 & 1101.C-0258(A) \\ 
HIP  82514 & * mu.01 Sco & 2018-05-15 & OBS\_H & 8x1x46 & 52.1 & 0.69 & 1.04 & 3.0 & 1101.C-0258(A) \\ 
HIP  82514 & * mu.01 Sco & 2019-08-04 & DB\_K12 & 8x24x46 & 53.4 & 0.52 & 1.03 & 3.7 & 1101.C-0258(C) \\ 
HIP  82514 & * mu.01 Sco & 2019-08-04 & OBS\_H & 8x1x46 & 57.3 & 0.54 & 1.03 & 3.4 & 1101.C-0258(C) \\ 
\hline 
HIP  82545 & * mu.02 Sco & 2018-04-24 & DB\_K12 & 16x12x46 & 51.1 & 0.64 & 1.03 & 4.0 & 1101.C-0258(A) \\ 
HIP  82545 & * mu.02 Sco & 2018-04-24 & OBS\_H & 16x1x46 & 53.3 & 0.63 & 1.03 & 4.0 & 1101.C-0258(A) \\ 
HIP  82545 & * mu.02 Sco & 2021-06-04 & DB\_K12 & 16x12x46 & 51.3 & 0.47 & 1.03 & 10.4 & 1101.C-0258(C) \\ 
HIP  82545 & * mu.02 Sco & 2021-06-04 & OBS\_H & 16x1x46 & 53.2 & 0.48 & 1.03 & 10.1 & 1101.C-0258(C) \\ 
\hline
\hline
\caption{Observing logs for all stars within the sample}
\label{tab:obs_table}
\end{longtable}} \par 
\end{center}
\tablefoot{ $^a$: DIT correspond to the Detector Integration Time per frame. $\Delta$ PA is the amplitude of the parallactic rotation. $\tau_0$ correspond to the coherence time. $^b$: values extracted from the updated DIMM info, averaged over the sequence.}
\twocolumn

\section{Assessing the capture scenario for a field object}\label{sec:capture_field}

In order to be properly assessed, a capture scenario cannot neglect the fact that a stellar association, at any point of its history, is nestled in a galactic environment where most of the stars are physically unrelated to the association. Thanks to its peculiar kinematic signature -- markedly different from the one of the field --, the association can be said to experience a wind of field stars: we would like to assess here the probability of a BEAST star capturing a substellar object, either isolated or bound to a lower-mass star, from the field.

A list of bona-fide Sco-Cen members was built by considering all the 8878 stars that are listed as members in at least two among the following catalogs: \citet{rizzuto15,galli18,damiani19,luhman20,esplin20,luhman22}. The sample was merged with the 85-star BEAST sample, yielding a sample of Sco-Cen sources, $S$. Afterwards, we queried the {\it Gaia} Archive for all sources located within the box containing $S$: $152^\circ < \alpha < 268^\circ$, $-75^\circ < \delta < -10^\circ$, $4~\text{mas} < \varpi < 13~\text{mas}$. Removing the intersection with $S$ yielded a sample of field stars, $F$.

For every star in $S$, we computed the number of stars in $F$ found within 5 pc. We estimate a mean density of $n_F = 0.34$ stars pc$^{-1}$, four times higher than the corresponding estimate for Sco-Cen sources, $n_S$. We assume $n_F$ to be independent of time.

Using {\it Gaia} proper motion measurements $\mu_\alpha^*$ and $\mu_\delta$, the 2D mean differential velocity between Sco-Cen stars and field stars can be estimated as
\begin{equation}
\DvtwoD = \sqrt{\left(\langle v_{\alpha,\,s}\rangle -\langle v_{\alpha,\,f}\rangle\right)^2+\left(\langle v_{\delta,\,s}\rangle -\langle v_{\delta,\,f}\rangle\right)^2}= 11.2~\text{km s}^{-1},
\end{equation}
where $v_\alpha = 4.74~\text{km s}^{-1} \text{yr} \cdot \mu_\alpha^*/\varpi$ and $v_\delta = 4.74~\text{km s}^{-1} \text{yr} \cdot \mu_\delta/\varpi$ and the subscripts $s$ and $f$ indicate stars belonging to $S$ and $F$, respectively. Assuming the velocity is isotropic, we estimate $v_\infty \approx \DvthreeD = \sqrt{3/2}~\DvtwoD \approx 13.7~\text{km s}^{-1}$.

According to Eq.~\ref{eq:encounter_timescale}, a close encounter within 1000 pc with a $5 M_\odot$ star occurs on a timescale $\tau_{\mathrm{enc},\,f} = 26.6$ Gyr, so that $f_{p,f} = 2 \cdot 10^{-4}$ (Eq.~\ref{eq:captured_objects_per_star}) and finally $\pcapt = 3 \cdot 10^{-5}$ (Eq.~\ref{eq:captured_objects_survey}).

Due to the high relative velocity between field and Sco-Cen stars, the probability of a capture of a field object is negligible compared to the capture of a Sco-Cen object.

\section{Mean probability detection maps using various ages and models}
\label{appendix:all_mean_detmaps}

\begin{figure*}[!htbp]
    \centering
    \includegraphics[width=\textwidth]{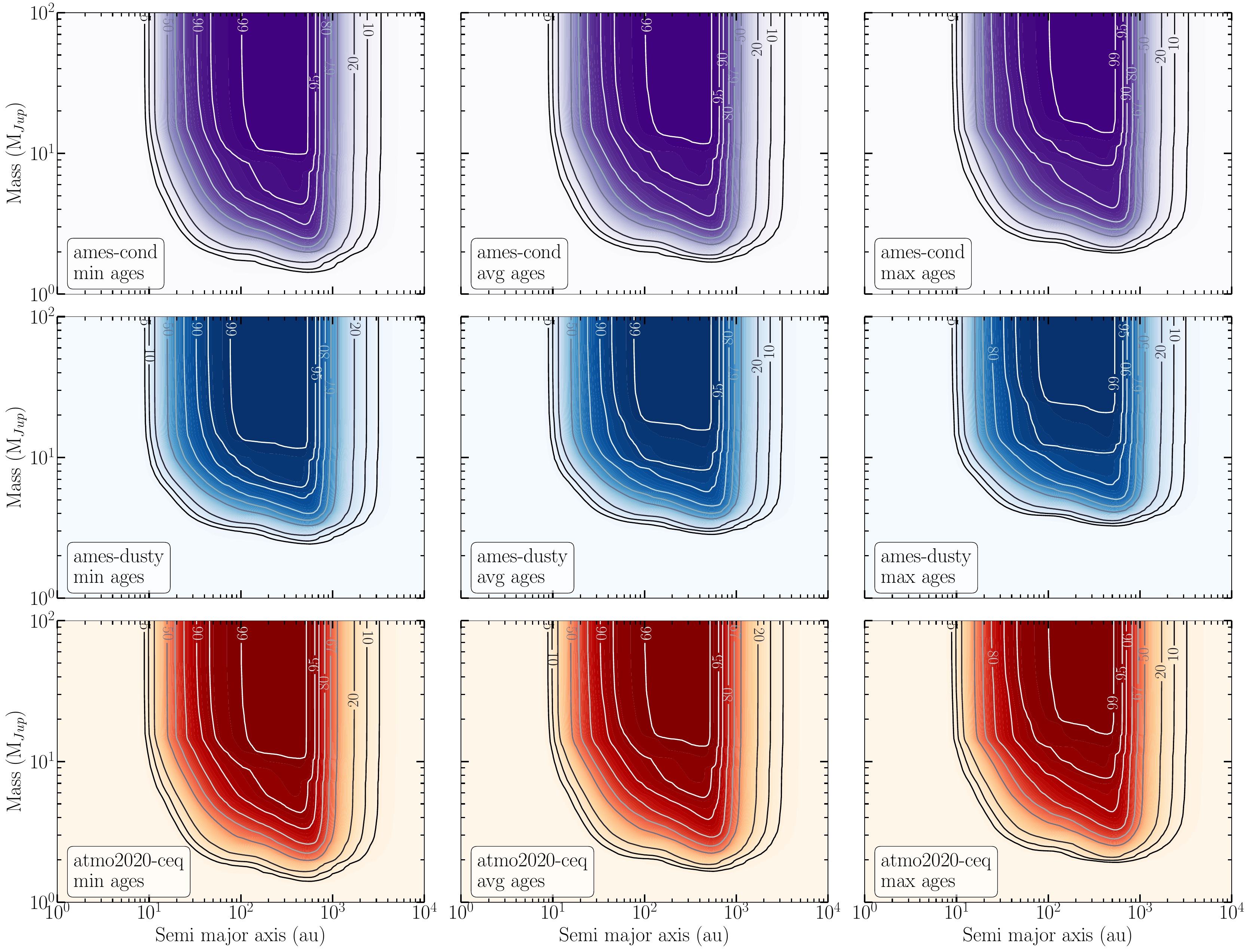}
    \caption{Mean probability detection maps computed using 3 different model and the nominal, minimum and maximum age for all stars indicated in Table~\ref{tab:star_table}.}
    \label{fig:all_mean_detmaps}
\end{figure*}

Figure~\ref{fig:all_mean_detmaps} shows the mean probability detection maps for the minimum, average and maximum age estimates of the sample's stars and the 3 distinct evolutionary models ( atmo2020 \citep{Phillips.2020}, ames-cond \citep{Baraffe.2003}, ames-dusty \citep{Chabrier.2000}) that we used to convert  detection limits in contrast into detection limits in masses.

\newpage

\section{Individual probability detection maps}
\label{appendix:individual_det_maps}

\begin{figure*}[t!]
    \centering
    \includegraphics[width=\textwidth]{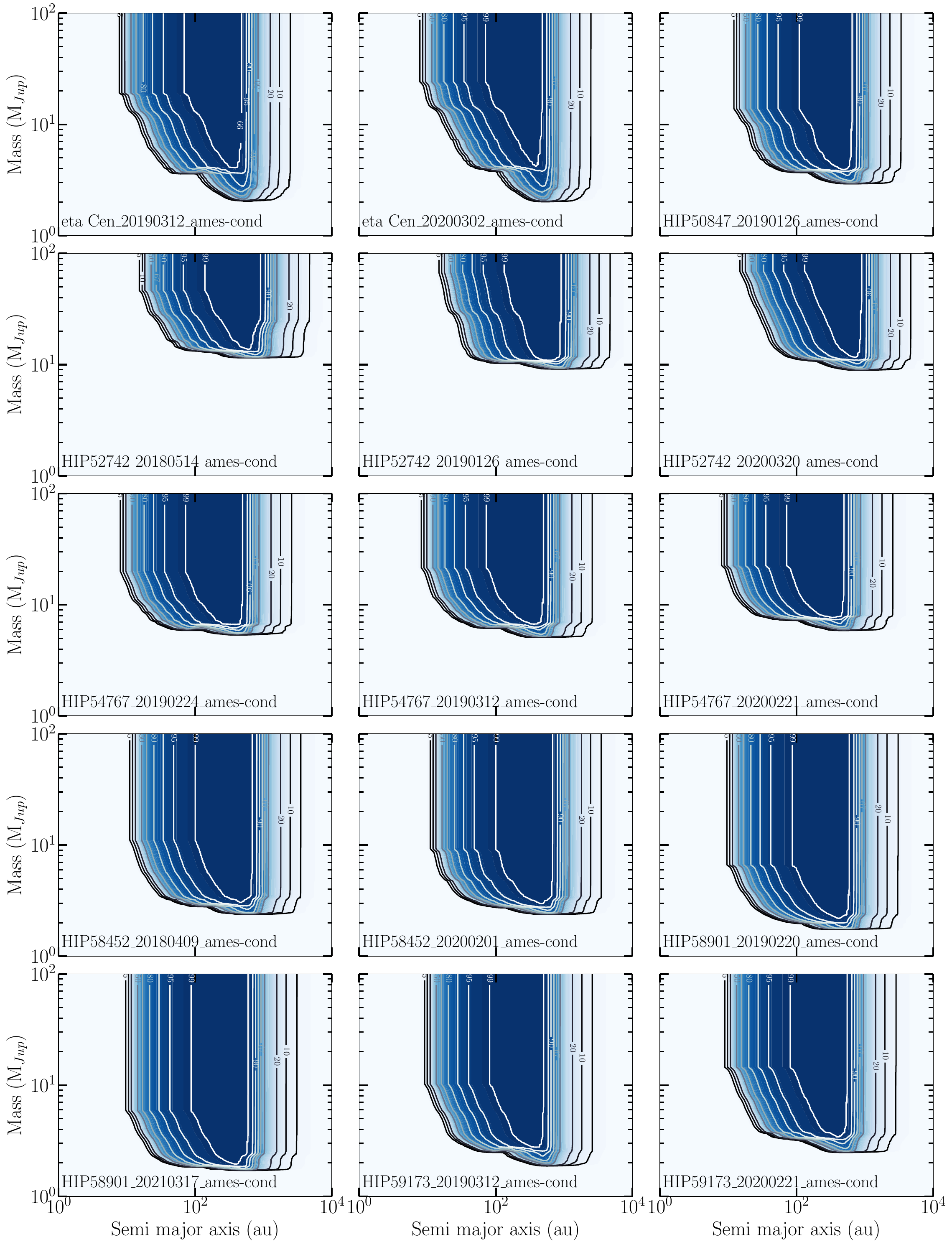}
    \caption{Probability detection maps for every observations within our sample, computed using the cond model and the nominal age.}
    \label{fig:individual_det_maps_1}
\end{figure*}

\begin{figure*}[t!]
    \centering
    \includegraphics[width=\textwidth]{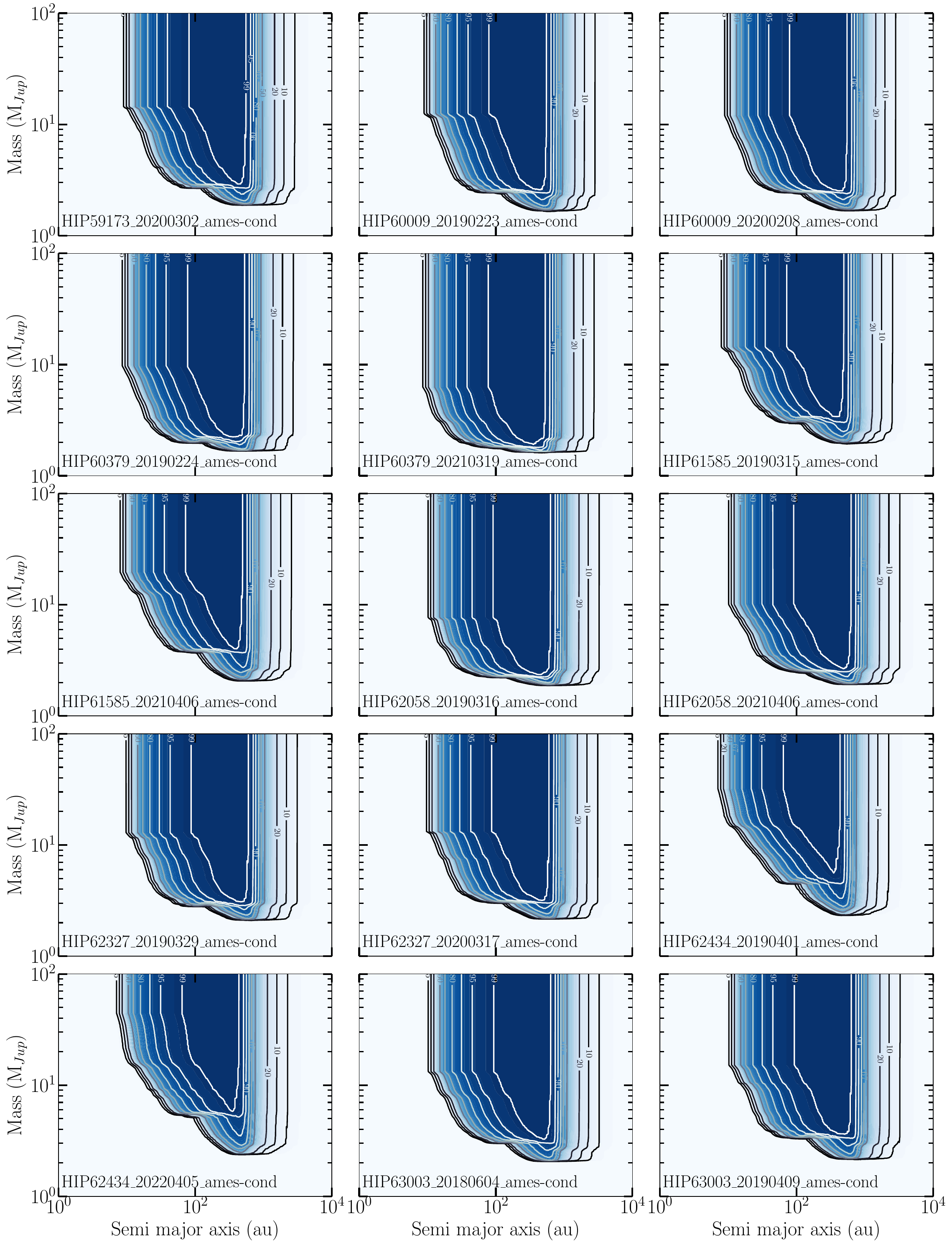}
    \caption{Probability detection maps for every observations within our sample, computed using the cond model and the nominal age.}
    \label{fig:individual_det_maps_2}
\end{figure*}

\begin{figure*}[t!]
    \centering
    \includegraphics[width=\textwidth]{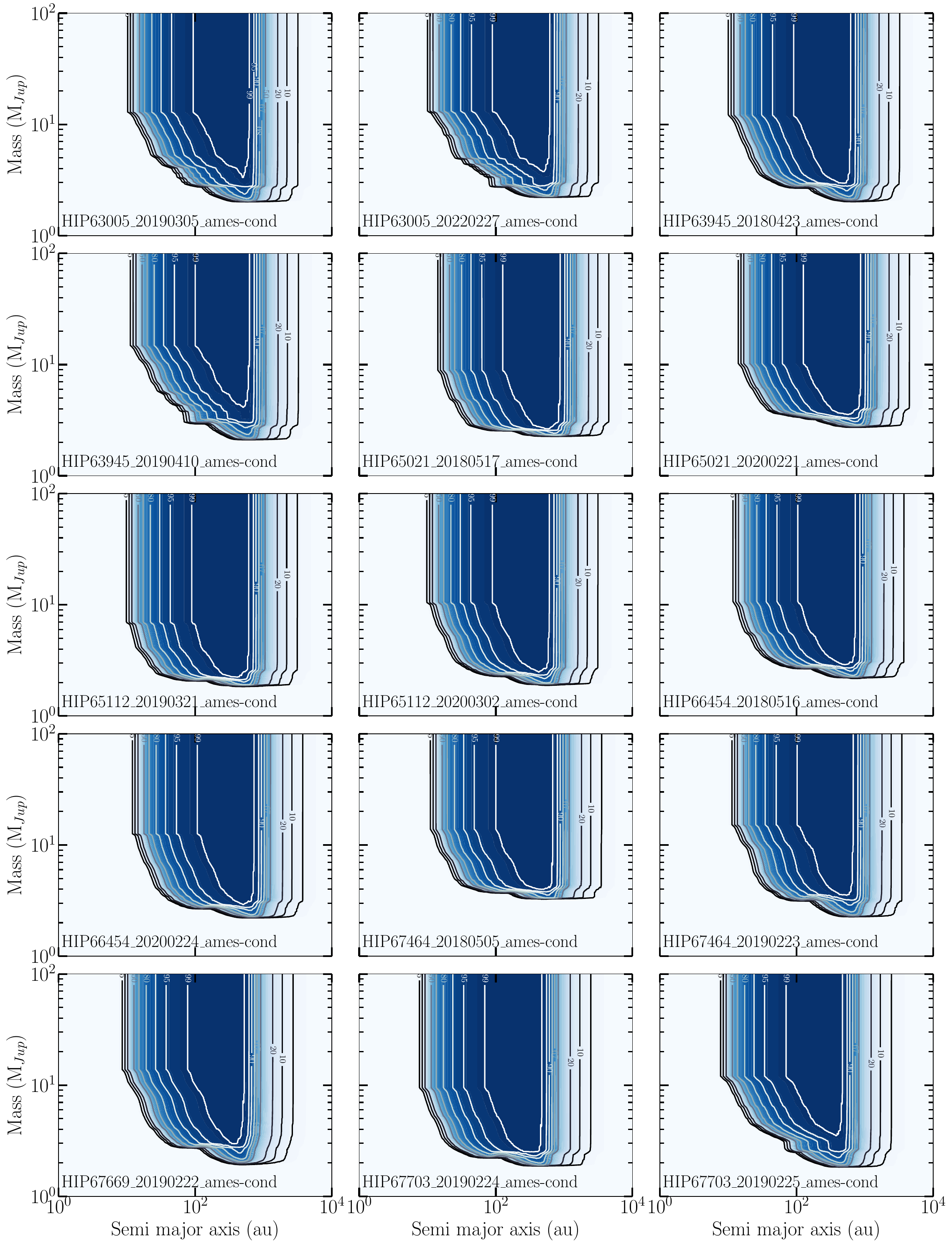}
    \caption{Probability detection maps for every observations within our sample, computed using the cond model and the nominal age.}
    \label{fig:individual_det_maps_3}
\end{figure*}

\begin{figure*}[t!]
    \centering
    \includegraphics[width=\textwidth]{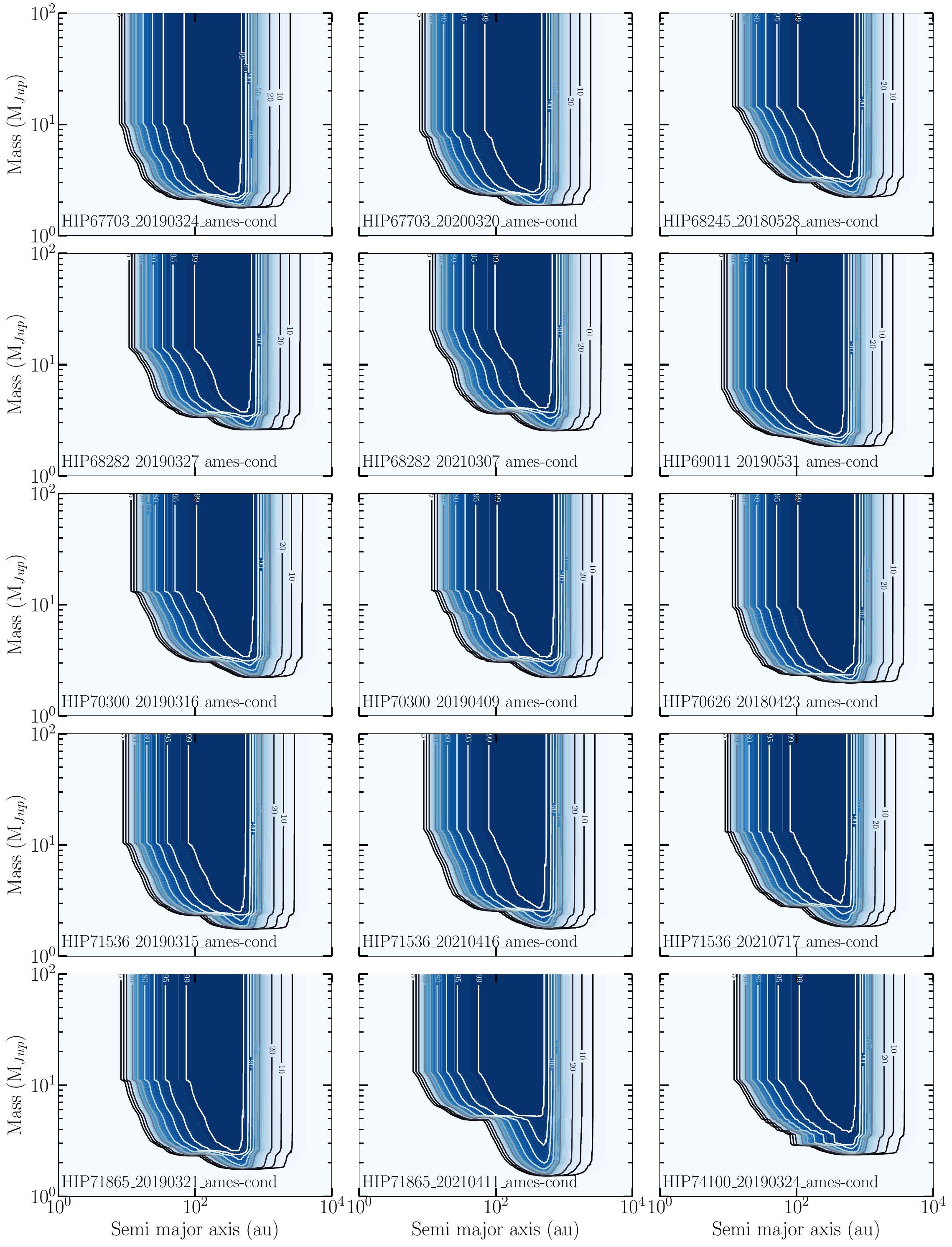}
    \caption{Probability detection maps for every observations within our sample, computed using the cond model and the nominal age.}
    \label{fig:individual_det_maps_4}
\end{figure*}

\begin{figure*}[t!]
    \centering
    \includegraphics[width=\textwidth]{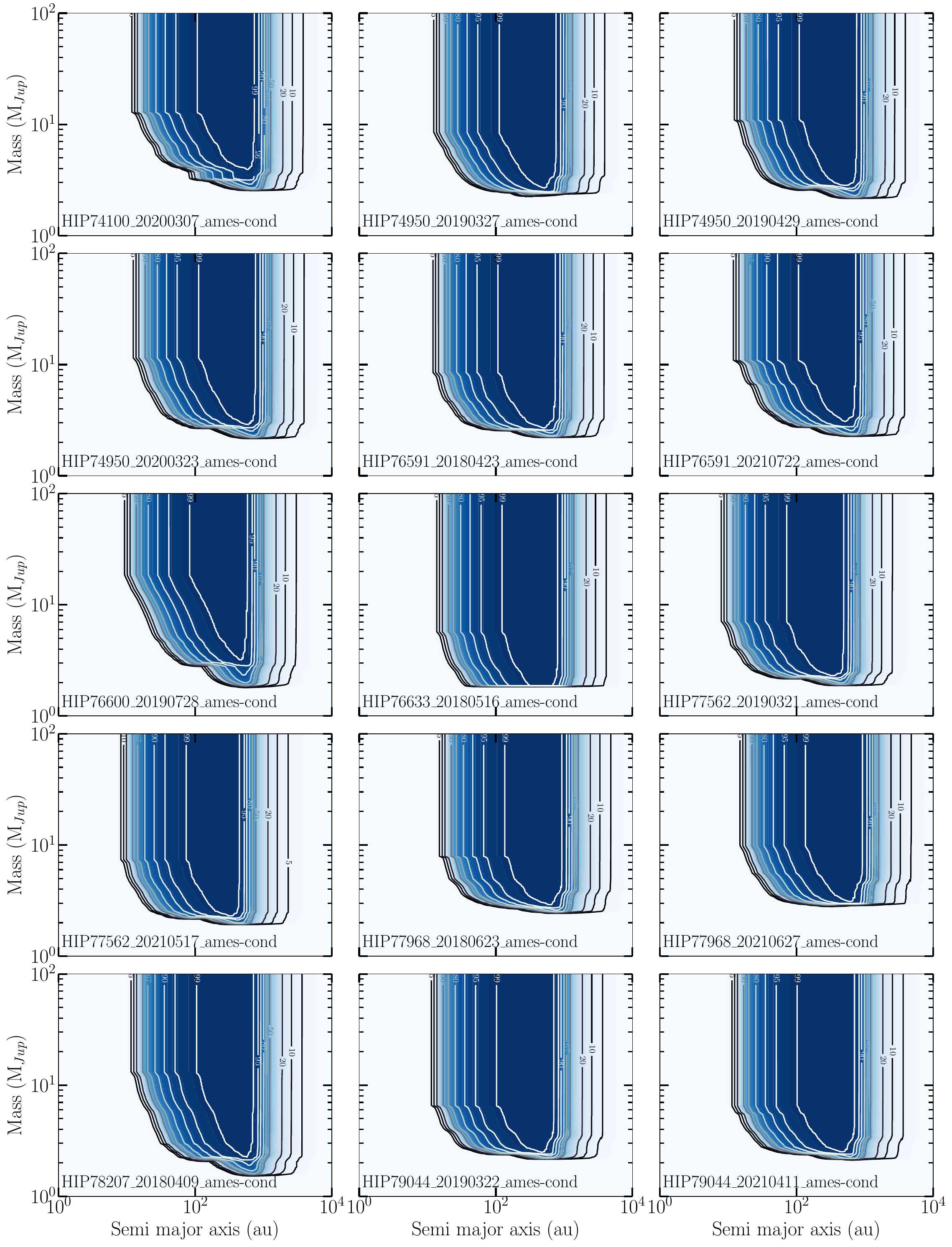}
    \caption{Probability detection maps for every observations within our sample, computed using the cond model and the nominal age.}
    \label{fig:individual_det_maps_5}
\end{figure*}

\begin{figure*}[t!]
    \centering
    \includegraphics[width=\textwidth]{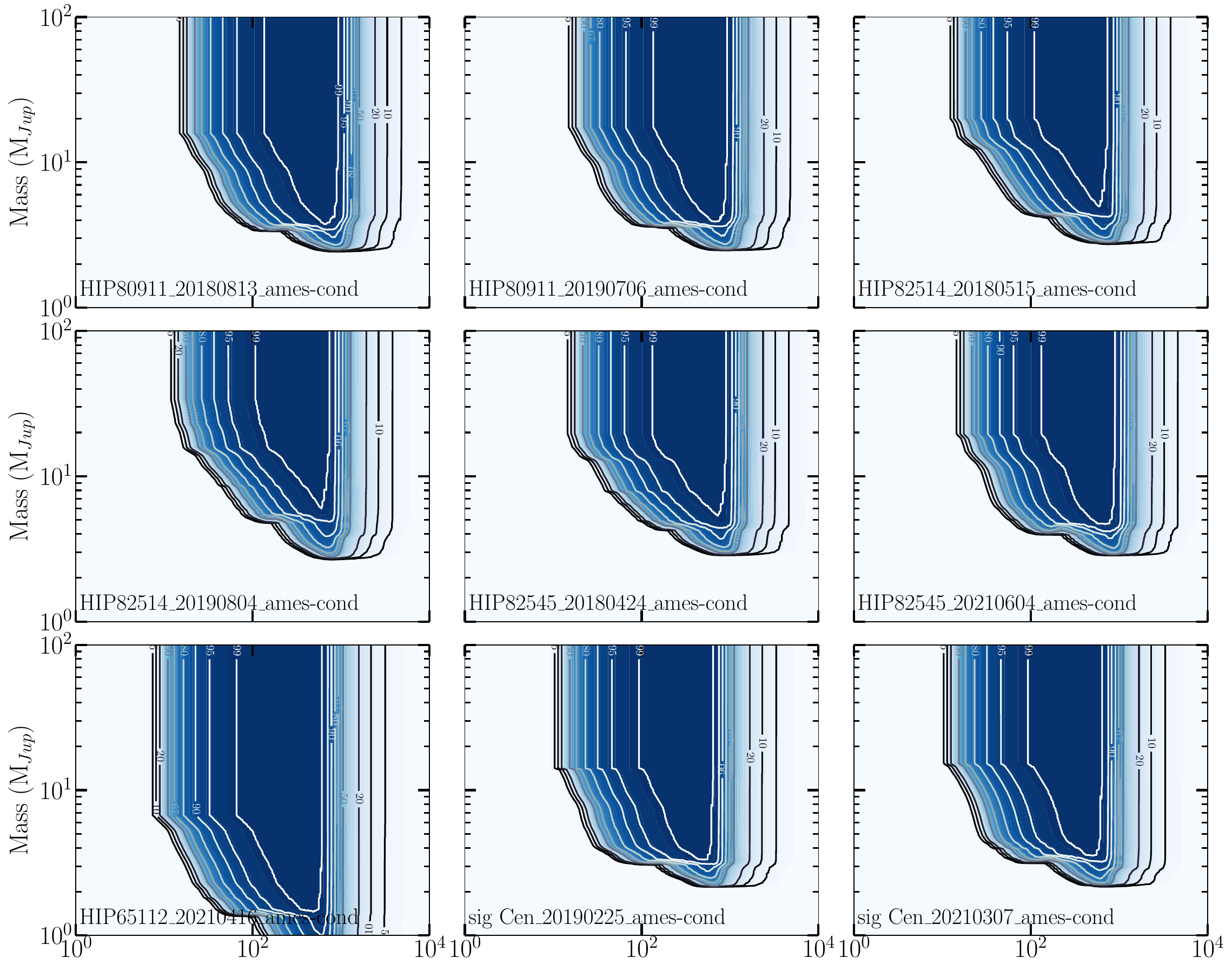}
    \caption{Probability detection maps for every observations within our sample, computed using the cond model and the nominal age.}
    \label{fig:individual_det_maps_6}
\end{figure*}

Figures~\ref{fig:individual_det_maps_1}--\ref{fig:individual_det_maps_6} show the individual probability detection maps for each star in the studied sample.

\end{appendix}

\end{document}